\documentclass[11pt]{article}

\usepackage[]{agmacros}
\usepackage{andrews_bib}
\usepackage{dirtytalk}
\title{Some algebraic and geometric computations in CH via Chow forms}
\author{
Robert Andrews~\orcidlink{0000-0002-6979-8757}\thanks{School of Computing Science, Simon Fraser University. Email: robert\_andrews@sfu.ca.} 
\and 
Abhibhav Garg\thanks{Department of Mathematics, University of Toronto. Email: abhibhav.garg@utoronto.ca.} 
\and 
Rafael Oliveira~\orcidlink{0000-0001-8917-8689}
\thanks{Cheriton School of Computer Science, University of Waterloo. Email: rafael@uwaterloo.ca.} 
\and 
Éric Schost\thanks{Cheriton School of Computer Science, University of Waterloo. Email: eschost@uwaterloo.ca.}}
\date{}

\begin{document}

\pagenumbering{roman}

\maketitle

\begin{abstract}
  The problem of decomposing an algebraic variety into more structured components (such as equidimensional, irreducible, or connected components) is a fundamental problem in computer algebra and algebraic geometry. 
  In this work, we study the computational complexity of computing equidimensional components of algebraic varieties, represented via Chow forms.

  We improve the state-of-the-art complexity bounds from $\PSPACE$ to $\CH$, the counting hierarchy, for a number of problems related to equidimensional decomposition, including deciding
  \begin{enumerate}
    \item whether the Chow form of the $k^{\text{th}}$ equidimensional component of a given projective variety vanishes at a given point,
    \item whether a given monomial is in the support of the Chow form of the $k^{\text{th}}$ equidimensional component of a given projective variety,
    \item membership in a given equidimensional component of a given algebraic variety, and
    \item membership in the closure of the image of an irreducible variety under a polynomial map.
  \end{enumerate}
  As an application of our algorithm to decide membership in the closure of the image of a variety under a polynomial map, we show that the approximate polynomial satisfiability problem can be decided in $\CH$.
  We also show that the degree of the $k^{\text{th}}$ equidimensional component of a given projective variety can be computed in $\FP^{\CH}$.

  Our main technical result is the construction of a polylogtime-uniform family of constant-depth, exponential-size threshold circuits that compute the Chow forms of the equidimensional components of a given projective variety.
  We construct these circuits by combining the machinery of geometric resolutions together with the uniform constant-depth, exponential-size threshold circuits of Andrews, Garg, and Schost (FOCS 2026) for the multivariate resultant.
\end{abstract}

\newpage
\tableofcontents

\newpage

\pagenumbering{arabic}

\section{Introduction} \label{section: introduction}

A central quest in computational algebraic geometry and algebraic complexity theory is to understand the computational complexity of deciding or computing geometric properties of algebraic varieties (common zero sets of a collection of polynomials) given the defining collection of polynomials. 
In this quest, fundamental problems include deciding whether the algebraic variety is empty (known as the Hilbert Nullstellensatz problem), computing the dimension of the variety, computing its degree, and decomposing the variety into more structured components (e.g., equidimensional, irreducible, and connected components).

Some of the above questions can also be (and have been) asked in more general settings, such as the setting of algebraic varieties defined by polynomials over arbitrary rings.
To make this introduction more concise and accessible, we will focus on the setting of algebraic varieties defined by polynomials over the complex numbers.
However, we will state our results in a more general setting.

In light of the importance of the aforementioned problems, much work has been devoted to them.
The earliest works on these problems showed that they can all be solved in doubly-exponential time, as seen in \cite{lasker1905theorie,Hermann1926,seidenberg74,gianni1988grobner}.
These works used techniques from elimination theory and Gröbner bases to reduce these problems to the task of solving doubly-exponential sized linear systems, or to factoring polynomials of doubly-exponential degree.
Together with the work of \textcite{Csanky76}, this brings the complexity of the Hilbert Nullstellensatz problem down to \EXPSPACE.
Later, improved degree bounds for the Nullstellensatz (see \cite{Jelonek05} and references therein) improved the complexity down to \PSPACE.
Subsequent works have improved the complexity of these problems, and we now give a brief overview of the history and state of the art regarding the complexity-theoretic improvements for them.

\paragraph{Hilbert Nullstellensatz (\HN) and Dimension.}
The complexity of the $\HN$ problem has been brought down to the counting hierarchy by \textcite{AGS26}, while \textcite{koiran96} proved that this problem lies in $\AM$, assuming an effective prime ideal theorem.\footnote{Recent work by \textcite{CT26}, building on OpenAI's proof of the Quasi-Riemann Hypothesis \cite{OAI:The-Quasi-Riemann-Hypothesis-September-30-2026}, claims to prove an effective version of the prime ideal theorem, which would yield an unconditional version of Koiran's result.}
Later work by \textcite{canny90,Ierardi89} reduced $\HN$ to the task of computing multivariate resultants.
\textcite{AGS26} showed that multivariate resultants can be evaluated in $\FP^{\CH}$, and as a consequence placed $\HN$ in the counting hierarchy.
Lastly, \textcite{koiran96} gave an $\AM$-protocol for the Hilbert Nullstellensatz problem inspired by Stockmeyer's protocol for approximate counting.
With regards to complexity-theoretic lower bounds, it is easy to see that arithmetization of $\SAT$ shows that this problem is \NP-hard.
Under plausible hardness assumptions, \textcite{MV05} showed that $\NP = \AM$, so Koiran's result establishes that $\HN$ is conditionally \NP-complete over $\bC$.
For a concise yet thorough account of developments on the Hilbert Nullstellensatz problem, we refer the reader to \cite{AGS26} and references therein.

On the geometric front, the computation of the dimension of algebraic varieties was shown to be in the polynomial hierarchy \cite{koiran1997randomized}, again assuming an effective prime ideal theorem, and unconditionally shown to be in the counting hierarchy by \textcite{AGS26}. 
Each of these works on computing dimension used similar techniques to those used for the Hilbert Nullstellensatz problem as described above.
We once again refer the reader to \cite{AGS26} for more details on works on the dimension of algebraic varieties.

As shown by \textcite{BC06}, the natural setting for obtaining complexity-theoretic lower bounds for several of the aforementioned geometric problems is the setting of counting problems.
Thus, before we discuss the complexity of computing the degree and decomposing varieties into equidimensional, irreducible, and connected components, we first introduce the notion of geometric counting complexity.

\paragraph{Geometric Counting Complexity.}
The study of complexity-theoretic lower bounds for more general geometric problems was initiated by \textcite{BC06}, who defined counting variants of the above problems, along with the complexity class $\GCC$ (standing for Geometric Counting Complex problems), which they show to be situated between $\sharpP$ and \FPSPACE.
Moreover, in \cite{BC06}, the authors show that the counting version of the Hilbert Nullstellensatz problem is $\GCC$-complete, and that the problem of computing the top-component degree of an algebraic variety\footnote{In \cite{BC06} the authors denote by \say{geometric degree} the degree of the top-dimensional component of an algebraic variety. 
Since we adopt the definition of degree of an algebraic variety as the sum of the degrees of its irreducible components, we will refer to the notion of degree used in \cite{BC06} as \say{top-component degree}.} is $\FP^{\GCC}$-complete with respect to Turing reductions. 
Recently, \textcite{AGS26} showed that $\GCC \subseteq \FP^{\CH}$, thereby improving the upper bound for $\GCC$ from $\FPSPACE$ to $\FP^{\CH}$.

With the above complexity class in mind, and equipped with the results of \textcite{BC06}, we are now ready to discuss the complexity of the remaining geometric problems.

\paragraph{Equidimensional Components and Degree.}
Decomposing varieties into equidimensional components was first shown to be in exponential time by \cite{chistov1983subexponential,grigoriev1986factorization,GiHeMoPa95,lecerf2000computing,jeronimo2004computational}.
Several concepts and techniques used in these works, such as geometric resolutions and the computation of Chow forms representing each equidimensional component, will be used in our work as well, see Sections~\ref{subsection: chow forms intro} and~\ref{subsection: geometric res intro}.
The work of \cite{giusti1991algorithmes} shows how to produce an exponential-size and polynomial-depth algebraic circuit that computes the equidimensional components of a given algebraic variety, therefore showing that the problem of computing equidimensional components is in $\FPSPACE$. 

Since the degree of an algebraic variety is the sum of the degrees of its equidimensional components, the above results also imply that one can compute the degree of an algebraic variety in $\FPSPACE$. 
Because these works are able to compute the equidimensional components of an algebraic variety, they are also able to compute the top-component degree of the variety, which is known to be $\FP^{\GCC}$-complete by \cite{BC06}.
This places the problem of computing the equidimensional components of an algebraic variety between $\FP^{\GCC}$ and $\FPSPACE$.

\paragraph{Irreducible Components.}
The problem of decomposing varieties into irreducible components is a fundamental problem in algebraic geometry, being a highly nontrivial generalization of the problem of factoring polynomials (the latter corresponds to the case of decomposing a hypersurface into its irreducible components).
The first complexity-theoretic improvement on decomposing varieties into irreducible components was to place this problem in single exponential time by \cite{chistov1983subexponential,grigoriev1986factorization,chistov1986algorithm,giusti1991algorithmes,GiHeMoPa95}. 
These works used geometric resolutions to construct exponential-sized algebraic circuits that compute the equidimensional components of a given algebraic variety, and then used polynomial factorization techniques to decompose each equidimensional component into irreducible components.
In \cite[Theorem 4.1 and Proposition 4.2]{burgisser2009complexity}, the authors show that the problem of counting the number of irreducible components of an algebraic variety is $\GCC$-hard and lies in $\FPSPACE$.
Later, an improvement by \cite{scheiblechner2010generalization} showed that the problem of computing a representation of the irreducible components is in $\FPSPACE$ via a generalization of Stickelberger's theorem for varieties of positive dimension.

\paragraph{Connected Components.}
The problem of counting the number of connected components of an algebraic variety was shown by \cite[Theorem 3.1]{burgisser2009complexity} to be in $\FPSPACE$.
Subsequently, \cite{scheiblechner2007complexity} showed that this counting problem is $\FPSPACE$-hard, therefore establishing $\FPSPACE$-completeness for counting the number of connected components of an algebraic variety.

The complexity of decomposing varieties into connected components was shown to be in FPSPACE by \cite[Theorem 1.1]{scheiblechner2010generalization}.
Scheiblechner's decomposition algorithm for this problem is also based on a generalization of Stickelberger's theorem for varieties of positive dimension.

\paragraph{Summary of the State of the Art.}
The Hilbert Nullstellensatz problem and the problem of computing the connected components of an algebraic variety are complexity-theoretically well-understood, with (nearly) tight upper and lower bounds for their complexity.
In contrast, over the complex numbers, the complexity of the problems of decomposing an algebraic variety into its equidimensional components or irreducible components remain unknown, with a complexity-theoretic gap between the known upper and lower bounds.
The same is true for the complexity of the counting versions of these problems.

Given existing results on decompositions of varieties into equidimensional and irreducible components, it is natural to ask whether these decompositions can be computed more efficiently.
Among these decompositions, the decomposition into equidimensional components is the simplest problem, as it is a coarser decomposition than (and can be computed from) the decomposition into irreducible components.
Moreover, it is a natural first step towards computing the decomposition into irreducible components, as most algorithms for computing irreducible components first compute the equidimensional components and then decompose each equidimensional component into irreducible components via factorization.

\subsection{Our results}

In this work, we improve the complexity-theoretic upper bounds for the problem of computing the equidimensional components of an algebraic variety, and along the way we obtain improved upper bounds for the problems of computing the degree of an algebraic variety as well as computing projections of an algebraic variety.
This also yields applications to the problem of computing coefficients of annihilators, thereby improving the complexity-theoretic upper bounds for the Approximate Polynomial Satisfiability problem, which is also known to be $\NP$-hard via arithmetization of $\SAT$.
We now give a more detailed overview of our results.

Before we state our main results, we first give a brief and informal introduction to Chow forms, which are a classical tool in algebraic geometry for representing equidimensional projective varieties.
For a more detailed introduction to Chow forms, we refer the reader to \cref{subsection: chow forms intro} and references therein.
From now on, we will refer to homogeneous polynomials as \emph{forms}, and we will refer to the common zero set of a collection of forms as a \emph{projective variety}.

If one is given an equidimensional projective variety $V \subseteq \bP^n$ of dimension $k$, it is well-known that the intersection of $V$ with the zero set of a generic choice of $k+1$ linear forms $L_i = u_{i,0}x_0 + \dots + u_{i,n}x_n$ for $i = 0, \dots, k$ is an empty set.
As it turns out, there is a unique (up to a scalar multiple) form $C_{V}(\vec{u})$, depending \emph{only on the variables} $\vec{u} \coloneqq \{u_{i,j} \}_{\substack{0 \leq i \leq k,\ 0 \leq j \leq n}}$, such that the intersection of $V$ with the zero set of the linear forms $L_i$ is nonempty if and only if $C_{V}(\vec{u}) = 0$.
The form $C_{V}$ is called the \emph{Chow form} of the equidimensional projective variety $V$, and it is also known that $\deg(C_{V}) = (k+1) \deg(V)$, where $\deg(V)$ is the degree of the variety $V$.
This gives a way to represent equidimensional projective varieties by means of a single form (the Chow form), and this form carries important information about the variety.
Although we have only defined Chow forms for projective equidimensional varieties, it is possible to use Chow forms to obtain useful information about equidimensional affine varieties as well (see \cref{remark: chow form affine} for more details).

We are now ready to state two of our main results, which improve the complexity-theoretic upper bounds for computing Chow forms of equidimensional components of algebraic varieties.
To make the statement of our results more concise, we will use the following setup.

\noindent\textbf{Setup.} Let $F_1, \dots, F_m \in \bQ[x_0, \dots, x_n]$ be forms given by constant-free algebraic circuits and let $V \coloneqq \var{F_1, \dots, F_m}$ be the projective variety defined by these forms.
For $0 \leq k \leq n$, denote by $V_k$ the $k$-equidimensional component of $V$, and by $C_{V_k}$ the Chow form of $V_k$.

\begin{theorem}[Zeros of Chow Form]\label{theorem: intro chow forms zero}
    Let $V$ be a projective variety given as in the above setup.
    Let $\vec{a} \in \bQ^{(n+1)(k+1)}$ and $k \in \bN$. 
    Given $V, k, \vec{a}$, one can decide whether $C_{V_k}(\vec{a}) = 0$ in $\CH$.
\end{theorem}

In fact, we can do more than just decide whether the Chow form vanishes at a point.

\begin{theorem}[Chow Form Coefficients]\label{theorem: intro chow form coefficients}
    Let $V$ be a projective variety given as in the above setup.
    Let $k \in \bN$, and $\vec{e} \in \bN^{(n+1)(k+1)}$ be the exponent vector of a monomial in the variables of $C_{V_k}$. 
    Given $V, k, \vec{e}$, one can decide if the coefficient of the monomial corresponding to $\vec{e}$ in $C_{V_k}$ is zero in $\CH$.
\end{theorem}

\begin{remark}
    The above two results are a consequence of our main technical result (\cref{theorem: chow form tc0}): we construct a polylogtime-uniform family of constant-depth threshold circuits that computes the Chow forms of the equidimensional components of a given projective variety.
    For more details on our main technical result and its consequences, we refer the reader to \cref{subsection: pruning}.
\end{remark}

The above two results allow us to decide membership in equidimensional components of algebraic varieties, as well as to compute the degree of the given variety.
This is the content of our next results (see \cref{corollary: degree in ch,corollary: equidimensional component membership} and \cref{remark: chow form affine} for proofs of these results).

We now state the first consequence of the above results, which allows us to compute the degree of an algebraic variety.

\begin{theorem}[Degree]\label{theorem: intro degree}
    Given a collection of polynomials $f_1, \ldots, f_m \in \bQ\bs{x_1, \ldots, x_n}$ represented as constant-free algebraic circuits and $k \in \bN$, one can compute the degree of the $k^{th}$-equidimensional component of $V \coloneqq \var{f_1, \ldots, f_m}$ in $\FP^\CH$.
    In particular, this implies that one can compute $\deg(V)$, as well as the number of isolated points of $V$ in $\FP^\CH$.
\end{theorem}

The above result improves the previous best known upper bound of $\FPSPACE$ for computing the degree of an algebraic variety.

The next consequence of the above results is on the problem of deciding membership in equidimensional components of algebraic varieties.

\begin{theorem}[Equidimensional Component Membership]\label{theorem: intro equidimensional membership}
    Given a collection of polynomials $f_1, \ldots, f_m \in \bQ\bs{x_1, \ldots, x_n}$ represented as constant-free algebraic circuits, $\valpha \in \bQ^{n}$, and $k \in \bN$, one can decide whether $\valpha$ lies in the dimension-$k$ equidimensional component of $\var{f_1, \ldots, f_m}$ in $\CH$.
\end{theorem}

The above result improves the previous best known upper bound of $\PSPACE$ for deciding membership in equidimensional components of algebraic varieties \cite{scheiblechner2010generalization}.

Another consequence of our main technical results is related to deciding membership in the closure of images of irreducible varieties under polynomial maps. 
This is a basic problem in algebraic geometry, capturing many natural questions, two special cases being the computation of the transcendence degree and the Approximate Polynomial Satisfiability problem ($\APS$) of \textcite{GSS19}.
We now state our result on membership in the closure of images of irreducible varieties under polynomial maps, and then discuss its implications for $\APS$.

\begin{theorem}[Image Closure Membership]\label{theorem: image closure membership}
    Let $p \in \bN$ and $f_1, \ldots, f_m \in \bQ\bs{x_1, \ldots, x_n}$ represented as constant-free algebraic circuits be such that the variety $\var{f_1, \ldots, f_m}$ is irreducible and of dimension $p-1$.
    Let $h_1, \ldots, h_p \in \bQ\bs{x_1, \ldots, x_n}$ be represented as constant-free algebraic circuits defining the polynomial map $\Phi_\vec{h} : \bC^n \to \bC^p$.
    Given a point $\valpha \in \bQ^p$ and polynomials $f_1, \ldots, f_m, h_1, \ldots, h_p$ as above, where the variety $\var{f_1, \ldots, f_m}$ is irreducible and of dimension $p-1$, one can decide whether $\valpha$ lies in the closure of $\Phi_\vec{h}(\var{f_1, \ldots, f_m})$ in $\CH$.
\end{theorem}

Before we discuss the implications of the above result for $\APS$, we briefly introduce the $\APS$ problem.
Given a field $\bF$, we denote by $\bA^n$ the affine $n$-space over $\overline{\bF}$, the algebraic closure of $\bF$.

\begin{definition}[Approximate Polynomial Satisfiability ($\APS$)]\label{definition: aps}
    Let $\bF$ be a field, and let $f_{1}, \dots, f_{m} \in \bF\bs{x_{1}, \dots, x_{n}}$ be polynomials.
    This set of polynomials is \emph{approximately satisfiable} if the closure of the image of the map $\phi_{\vf}: \bA^{n} \to \bA^{m}$ defined by $\phi_{\vf}(\va) = \br{f_{1}(\va), \dots, f_{m}(\va)}$ contains the point $(0, \dots, 0)$.

    The $\APS$ problem is the following: given an algebraic circuit that computes polynomials $f_{1}, \dots, f_{m} \in \bF\bs{x_{1}, \dots, x_{n}}$, decide if the polynomials $f_{1}, \dots, f_{m}$ are approximately satisfiable.
\end{definition}

\noindent Note that $\{f_{1}, \dots, f_{m}\}$ is \emph{satisfiable} in the usual sense if $(0, \dots, 0) \in \phi_{\vf}(\bA^{n})$.

\begin{remark}
    In the above, if the underlying field $\bF = \bQ$, then \cref{theorem: image closure membership} already shows us that we can solve the $\APS$ problem in $\CH$, as $\bA^n$ is an irreducible variety.
    However, since the $\APS$ problem is a very structured special case of the image closure membership problem, we can actually solve $\APS$ in $\CH$ over a larger class of fields, as we will show in \cref{section: projections}.
    Thus, below we state our result on $\APS$ in $\CH$ over a larger class of base fields $\bF$.
\end{remark}

\begin{theorem}[$\APS$ in $\CH$]\label{theorem: intro aps ch}
    Let $\bF$ be one of the fields $\bF_{p^{a}}, \bF_{p^{a}}\br{t_{1}, \dots, t_{k}}$, a number field $\bQ\bs{\alpha}$, or $\bQ\bs{\alpha}\br{t_{1}, \dots, t_{k}}$.
  Then $\APS$ over $\bF$ lies in $\CH$. 
\end{theorem}

\subsection{Proof overview}

Having stated our main results, we now give a high-level overview of the challenges and new ideas behind the proof of our main technical result (\cref{theorem: chow form tc0}), where we construct a polylogtime-uniform family of constant-depth threshold circuits that compute the Chow forms of the equidimensional components of a given projective variety.
Even though the threshold circuits we construct are exponentially large, being polylogtime-uniform and constant depth means they are highly structured.
In particular, we can decide the value of an output of such a circuit in $\CH$, the counting hierarchy, which is exactly why $\CH$ enters into our complexity bounds.
Once we have the circuits \cref{theorem: chow form tc0}, we can use them along with the framework of \cite{AGS26} to simulate such circuits in $\CH$ and other standard techniques from algebraic geometry to obtain the other results stated above.

It is a classical fact that Chow forms of equidimensional varieties can be computed from resultants (see \cref{lemma: chow form of equidimensional complete intersection} for a quick proof in the complete intersection case).
The recent work of \cite{AGS26} shows that resultants can be computed by a uniform family of algebraic circuits of constant depth and exponential size, so we can directly apply the framework of \cite{AGS26} to prove \cref{theorem: intro chow forms zero,theorem: intro chow form coefficients,theorem: intro degree} when the variety $V$ is equidimensional.
The main difficulty therefore is to reduce the general case (when the variety $V$ has more than one equidimensional component) to the equidimensional case.
We begin by asking the following question:
can we compute the Chow form of any equidimensional component of a variety $V$ via resultants?

The inspiration and starting point for our approach comes from the work of \textcite{canny90} and the subsequent algebraic proof given by \textcite{Ierardi89}. 
The work of \cite{canny90} introduced the concept of the generalized characteristic polynomial (GCP), which allows one to study the solution set of a variety $V$ via a \say{perturbed system} of equations which \say{degenerates} to the original system of equations defining $V$.
This method allows one to argue about the zero-dimensional component of $V$ by studying the zero-set of the perturbed system.
In \cite{canny90,Ierardi89}, the authors show that the GCP can be computed by exponential-size algebraic circuits.
An important consequence of their result is that they can extract from the GCP a zero-dimensional subvariety of $V$ that \emph{contains} the zero-dimensional component of $V$.
While their result does not give a way to compute the zero-dimensional component of $V$ itself, it makes some important progress towards it.
Moreover, they can compute the GCP via resultants, which is a key ingredient in our approach.

The above leads us to the following two challenges:

\begin{enumerate}
    \item Can we generalize the above GCP-based method to compute a $k$-equidimensional subvariety of $V$ that \emph{contains} the $k$-equidimensional component of $V$?
    
    The above works only deal with the zero-dimensional case (i.e., $k=0$), and it is not clear how to generalize their approach to the case of arbitrary $k$.

    \item How do we compute the zero-dimensional component of $V$ from the GCP?
    Is there an efficient way (in constant depth) to prune the extra points (i.e. components) from the zero-dimensional subvariety of $V$ that contains the zero-dimensional component of $V$?
   
    More generally, even if we solve item 1, how do we compute the $k$-equidimensional component of $V$ by pruning the $k$-equidimensional subvariety of $V$ that was obtained from item 1?
\end{enumerate}

In this work, we give a positive answer to both of the above questions, and we show that both of these steps (the higher-dimensional generalization of the GCP-based method, and the pruning step for any dimension) can be done in constant depth.
With the solution of both items 1 and 2 above and with a bit more work (see \cref{subsection: pruning}), we can compute the Chow form of any equidimensional component of a variety $V$ in constant depth, thereby proving \cref{theorem: chow form tc0}.

We now discuss the main conceptual ideas and challenges behind the solution to the above two problems, and how we overcome them.

\paragraph{Solving item 1.} 
The works of \textcite{canny90,Ierardi89} show that the GCP computes a $u$-resultant of a zero-dimensional subvariety of $V$ that contains the zero-dimensional component of $V$.
Since the $u$-resultant of a zero-dimensional variety is the Chow form of that variety, this naturally leads to the approach of using a higher-dimensional generalization of the GCP to compute the Chow form of a $k$-equidimensional subvariety of $V$ that \emph{contains} the $k$-equidimensional component of $V$.
We establish this generalization in \cref{subsection: limit subvarieties}, and it is our first main technical contribution.

\paragraph{Solving item 2.} 
Once we solved item 1, for any $k \in \{0, 1, \dots, n\}$, we can compute a $k$-equidimensional subvariety $W_k \subset V$ that contains the $k$-equidimensional component of $V$, which we will denote by $V_k$.
Let us call the irreducible components of $W_k$ that are not in $V_k$ the \emph{excess components} of $W_k$.
Because $W_k \subset V$ and $W_k$ is equidimensional, we know that \emph{all} excess components must lie inside a \emph{higher dimensional component} of $V$.
In particular, if $r$ is the dimension of $V$, then we know that $W_r = V_r$, as $W_r$ cannot have any excess components due to the dimension of $V$. 

Now that we have one equidimensional component of $V$ (the $r$-equidimensional component), by the above dimensionality argument, we can prune the excess components of $W_{r-1}$ by removing the intersection of $W_{r-1}$ with $V_r$.
Proceeding inductively, we can prune the excess components of $W_k$ by removing the intersection of $W_k$ with $V_{k+1} \cup V_{k+2} \cup \dots \cup V_r$.
This pruning can be done via a GCD computation, which can be done in constant depth.
However, note that this inductive way of pruning the excess components will not yield a constant-depth circuit, as it requires a sequential computation of the equidimensional components of $V$!

To address this sequentiality issue, we need the following important observations. 
We definitely know that the excess components of $W_k$ must lie inside a higher-dimensional component of $V$, that is, inside $V_{k+1} \cup V_{k+2} \cup \dots \cup V_r$.
Since $V_k \subset W_k$, this implies that the excess components of $W_k$ must lie inside $W_{k+1} \cup W_{k+2} \cup \dots \cup W_r$.
Moreover, note that since each $W_k \subset V$, any $k$-dimensional irreducible component $X \subset V$ \emph{cannot} lie inside $W_{k+1} \cup W_{k+2} \cup \dots \cup W_r$, as that would imply that $X$ is contained in another irreducible component of $V$ (one of dimension greater than $k$), which contradicts the fact that $X$ is an irreducible component of $V$.\footnote{The most intuitive way to see this is in the case of $k=0$, i.e. the case of isolated points of the variety $V$. 
By definition, any isolated point of $V$ cannot lie inside a higher-dimensional component of $V$, as that would contradict the fact that it is an isolated point.
Since each $W_\ell \subset V$ is equidimensional of dimension $\ell$, for any $\ell >0$, every irreducible component of $W_\ell$ is contained in an irreducible component of $V$ of dimension at least $\ell$, and thus $W_\ell$ cannot contain any isolated points of $V$.}

This gives us a way to prune the excess components of $W_k$ in constant depth: we first \say{compute} all the $W_k$'s in parallel, and then we can prune the excess components of each $W_k$ by removing the intersection of $W_k$ with $W_{k+1} \cup W_{k+2} \cup \dots \cup W_r$, which can be done in constant depth via a GCD computation.

To perform this pruning step, we need to compute both the Chow forms of $W_k$ and a geometric resolution of $W_k$ for all $k \in \{0, 1, \dots, r\}$.
But so far, step 1 only gives us a way to compute the Chow forms of $W_k$ for all $k \in \{0, 1, \dots, r\}$.
To compute the geometric resolutions of $W_k$ for all $k \in \{0, 1, \dots, r\}$, we show that we can compute a geometric resolution of $W_k$ from the Chow form of $W_k$ (and this can be done in constant depth).
The relationship between Chow forms and geometric resolutions is done in \cref{section: chow form geometric resolution}, and the pruning step is done in \cref{subsection: pruning}.
This concludes step 2, thereby giving an overview of the proof of \cref{theorem: chow form tc0}.

\subsection{Related work}

In addition to the references mentioned above, a few further comments are in order to compare the approaches used in some of these papers to our work.  
The algorithm in~\cite{giusti1991algorithmes} iteratively computes polynomials defining the equidimensional components of $V$ by a succession of projections and saturations; in effect, this reduces the problem to linear algebra in size $d^{n^{O(1)}}$. 
Follow-up papers such as~\cite{GiHeMoPa95,GiHeMoMoPa98} introduced the notion of geometric resolution, on which we rely. 
In this approach, equidimensional components are described by means of birational projections to hypersurfaces: this yields polynomials with useful algorithmic properties, but in general, they only describe open dense subsets of the components of $V$.
The algorithms in these papers, which culminate with Lecerf's equidimensional decomposition algorithm~\cite{lecerf2000computing}, are based on a formal version of Newton iteration, which makes it possible to store all polynomials by straight-line programs of size $d^{O(n)}$, and perform $d^{O(n)}$ operations altogether.

Inspired in part by these results, the algorithm in~\cite{jeronimo2004computational} computes Chow forms for the equidimensional components of $V$. 
It relies on similar techniques -- in particular, straight-line program encoding and Newton iteration and the use of geometric resolutions to perform a pruning step similar to ours -- and results in an overall $d^{O(n)}$ operations to compute straight-line programs of size $d^{O(n)}$ for the Chow forms.
The algorithms in this paper are instead based on multivariate resultant computations; the bounds on the size of the resulting circuits are slightly inferior, of order $d^{n^{O(1)}}$, but achieving constant-depth through the approach of~\cite{jeronimo2004computational} remains an open question.

\subsection{Concurrent work}

Independent of and concurrent to our work, two other sets of authors have obtained similar improvements to the complexity of the approximate polynomial satisfiability problem.
\textcite{BSTW26} proved $\APS \in \CH$ over the rational numbers and finite fields.
\textcite{AHPS26} likewise gave a proof that $\APS \in \CH$.

\subsection{AI methodology}

\begin{flushleft}
  \renewcommand{\arraystretch}{1.3}
  \begin{tabular}{@{}l@{\qquad}c@{}}
    \textbf{Activity} & \textbf{AI used} \\
    Asking the research question & No \\
    Coming up with the approach & No \\
    Proof development & No \\
    Writing and exposition & No \\
    Checking for bugs & No \\
    Other supporting tasks & No \\
  \end{tabular}
\end{flushleft}

\paragraph{Details}

No AI was used at any point in the development of this work.

\section{Preliminaries} \label{section: preliminaries}
\subsection{Notation and conventions} \label{subsection: notation}

We work over a field $\bF$ and use $\overline{\bF}$ to denote its algebraic closure.
The \emph{coefficient subfield} of $\bF$ is the largest algebraic extension of the prime field of $\bF$ within $\bF$.
We use $\bF\bs{x_1,\ldots,x_n}$ and $\bF\br{x_1,\ldots,x_n}$, respectively, to denote the ring of polynomials and field of rational functions with coefficients in $\bF$ and variables $x_1,\ldots,x_n$.
We write vectors as $\vec{x} = (x_1,\ldots,x_n)$.
Given $\vec{a} \in \bN^n$, we write $\vec{x}^{\vec{a}}$ for the monomial $x_1^{a_1} \cdots x_n^{a_n}$.

A \emph{form} of degree $d$ is a homogeneous polynomial of degree $d$.
We follow the convention that forms are written using capital letters and inhomogeneous polynomials using lowercase letters.
We frequently encounter pairs of polynomials where one is the (de-)homogenization of the other.
In such instances, we write both polynomials with the same letter to indicate this relationship.
As an example, if $F \in \bF\bs{x_0,\ldots,x_n}$ is a form, its dehomogenization $F(1,x_1,\ldots,x_n)$ will be denoted by $f$.
Similarly, when $f \in \bF\bs{x_1,\ldots,x_n}$ is an inhomogeneous polynomial of degree $d$, we use $F$ to denote the homogenization $x_0^d f(x_1 / x_0, \ldots, x_n / x_0)$.

The greatest common divisor of any two univariate polynomials with coefficients in a field $\bF$ will always be monic.
If $R$ is any unique factorization domain with field of fractions $\bF$, then the greatest common divisor of two univariate polynomials with coefficients in $R$ will be their greatest common divisor as polynomials in $\bF\bs{z}$.
For multivariate polynomials $f, g \in \bF\bs{x_{1}, \dots, x_{n}}$, we define $\gcd\br{f, g}$ to be the unique $\bF$-multiple of the greatest common divisor that has leading coefficient $1$ in the lexicographic order.
As in the univariate case, if $R$ is a UFD with field of fractions $\bF$, then the GCD of two multivariate polynomials with coefficients in $R$ is their GCD as polynomials with coefficients in $\bF$.

Given polynomials $f_1, \ldots, f_m \in \bF\bs{x_1,\ldots,x_n}$, we use $\var{f_1,\ldots,f_m}$ to denote their common zeroes in $\bA_{\overline{\bF}}^n$, which we will often denote by simply $\bA^{n}$ when the field $\bF$ is clear from context.
If these polynomials are instead forms $F_1, \ldots, F_m \in \bF\bs{x_0,\ldots,x_n}$, then $\var{F_1,\ldots,F_m}$ is taken to be their common zeroes in the projective space $\bP_{\overline{\bF}}^n$, which we will simply denote by $\bP^n$.
For us, a variety is any set of solutions to a system of polynomial equations over an algebraically closed field; in particular, varieties may be reducible.
The degree of a variety $V$, denoted $\deg V$, is defined as the sum of the degrees of all irreducible components of $V$.

We on occasion examine the trailing term and trailing coefficient, defined below, of a polynomial with respect to one of its variables.

\begin{definition} \label{definition: trailing term}
  Suppose $f \in \bF\bs{t, y_{1}, \dots, y_{m}}$ is a nonzero polynomial.
  The \emph{trailing term of $f$ with respect to $t$}, denoted by $\Tt_t f(y_1, \ldots, y_m)$, is the coefficient of the lowest-degree term in $f$ when viewed as a polynomial in $t$.
  Concretely, if $f = t^{i} f_{i}(y_{1}, \dots, y_{m}) + t^{i+1} f_{i+1}(y_{1}, \dots, y_{m}) + \cdots + t^{d} f_{d}(y_{1}, \dots, y_{m})$ where $f_i(\vy) \neq 0$, then $\Tt_{t} f = f_{i}$.

  The \emph{trailing part of $f$ with respect to $t$}, written $\TP_t f$, is the polynomial obtained by factoring out the highest power of $t$ from $f$.
  In the above setting, we have $\TP_t f = f_i + t f_{i+1} + \cdots + t^{d - i} f_d$.

  The trailing term and the trailing part of the zero polynomial are zero.
\end{definition}

The height of an integer $a \in \bZ$ is defined to be $\ceil{\log\br{\abs{a}}}$.
The height of a polynomial with integer coefficients is the maximum height of its coefficients.

\subsection{Algebraic geometry} \label{subsection: alg geom}

In this subsection, we list some basic (computational) algebraic geometry facts that we will use throughout this article.
The first two facts are standard bounds on the probability that a random hyperplane section of a variety has codimension $1$ and on the probability that the variety defined by random linear combinations of a given set of equations has certain nice behaviour.

\begin{lemma}\label{lemma: intersect dimension drop}
  Let $B \subseteq \bF$ be a finite set.
  Let $V \subseteq \bP^{n}$ be a projective variety of dimension $r$ and degree $D$.
  Suppose $\ell$ is a linear form with each coefficient picked independently and uniformly at random from $B$.
  Then with probability at least $1 - D /\abs{B}$, the intersection $V \cap \var{\ell}$ has dimension $r-1$.
  If $\ell_{1}, \dots, \ell_{r+1}$ are linear forms chosen the same way, then $V \cap \var{\ell_{1}, \dots, \ell_{r+1}} = \emptyset$ with probability at least $1 - (r+1) D / \abs{B}$.

  Let $W \subseteq \bA^{n}$ be an affine variety of dimension $r$ and degree $D$.
  Suppose $\ell$ is a linear polynomial with each coefficient picked independently and uniformly at random from $B$.
  Then with probability at least $1 - 2D /\abs{B}$, the intersection $W \cap \var{\ell}$ has dimension $r-1$.
  If $\ell_{1}, \dots, \ell_{r+1}$ are linear polynomials chosen the same way, then $W \cap \var{\ell_{1}, \dots, \ell_{r+1}} = \emptyset$ with probability at least $1 - 2(r+1) D / \abs{B}$.
\end{lemma}

\begin{lemma}
  \label{lemma: input processing}
  Let $B \subseteq \bF$ be a finite set.
  Let $f_{1}, \dots, f_{m} \in \bF\bs{x_{1}, \dots, x_{n}}$ be polynomials of degree at most $d$.
  Suppose $g_{1}, \dots, g_{n+1}$ are linear combinations of $f_{1}, \dots, f_{m}$, with each coefficient picked independently and uniformly at random from $B$.
  Then with probability at least $1 - \br{n+1} d^{n} /\abs{B}$, the following is true.
  For each $s \in [n+1]$, every irreducible component of $\var{g_{1}, \dots, g_{s}}$ of codimension less than $s$ is a component of $\var{f_{1}, \dots, f_{m}}$.
  As a consequence, $\var{g_{1}, \dots, g_{n+1}} = \var{f_{1}, \dots, f_{m}}$.
  The analogous result in the projective setting also holds.
\end{lemma}
The proof of both \cref{lemma: intersect dimension drop} and \cref{lemma: input processing} are implicit in the methods of \textcite{Heintz83}.
For an explicit proof of the latter, see for example \cite[Lemma~36]{KP96}.
The former can be proved using essentially the same arguments.

Next, we list some basic properties of ideals in polynomial rings that are preserved by certain extensions of scalars.

\begin{lemma}
  \label{lemma: extension of scalars}
  Let $\bF$ be any algebraically closed field.
  Let $\bL$ be an extension of $\bF$.
  Let $I$ and $J$ be ideals in $\bF\bs{x_{1}, \dots, x_{n}}$, and let $I'$ and $J'$ denote the extensions $I \bL\bs{x_{1}, \dots, x_{n}}$ and $J \bL\bs{x_{1}, \dots, x_{n}}$, respectively.
  The following are true.
  \begin{enumerate}[noitemsep]
    \item $I' \cap \bF\bs{x_{1}, \dots, x_{n}} = I$.
    \item $\br{I \cap J}' = I' \cap J'$.
    \item If $I$ is radical, then $I'$ is radical.
    \item If $I$ is a prime ideal of dimension $r$ and degree $D$, then $I'$ is also a prime ideal of dimension $r$ and degree $D$.
  \end{enumerate}
\end{lemma}
\begin{proof}
  Let $\bc{\alpha_{i}}_{i \in S}$ be a basis for $\bL$ as a vector space over $\bF$, with $\alpha_{s} = 1$.
  Given any $f \in I'$, we can write $f = \sum_{i \in S} \alpha_{i} f_{i}$, where $f_{i} \in I$ with only finitely many nonzero $f_{i}$.
  Further, this representation is unique since $\bc{\alpha_{i}}_{i \in S}$ is a basis for $\bL\bs{x_{1}, \dots, x_{n}}$ as a free $\bF\bs{x_{1}, \dots, x_{n}}$-module.
  The intersection $I' \cap \bF\bs{x_{1}, \dots, x_{n}}$ consists of exactly those elements which, when written in the above representation, only have nonzero $f_{s}$.
  This shows that $I' \cap \bF\bs{x_{1}, \dots, x_{n}} = I$.

  Given any element $f \in I' \cap J'$, the same argument along with uniqueness shows that we can write $f = \sum_{i \in S} \alpha_{i} f_{i}$ with $f_{i} \in I \cap J$.
  Therefore we can deduce $f \in \br{I \cap J}'$, showing $I' \cap J' \subseteq \br{I \cap J}'$.
  The reverse containment is immediate from definitions.

  If $I$ is radical, then $\bF\bs{x_{1}, \dots, x_{n}}/I$ is a reduced $\bF$-algebra.
  The algebra $\bL\bs{x_{1}, \dots, x_{n}} / I' $ is exactly $\br{\bF\bs{x_{1}, \dots, x_{n}} / I} \tensor \bL$.
  Since $\bF$ is a perfect field, the extension $\bL / \bF$ is separably generated.
  Therefore we can invoke \cite[\href{https://stacks.math.columbia.edu/tag/05DS}{Tag 05DS}]{stacks-project} to deduce that $\bL\bs{x_{1}, \dots, x_{n}} / I'$ is reduced, equivalently that $I'$ is radical.
  Similarly, if $I$ is prime, then we can deduce that $I'$ is prime \cite[\href{https://stacks.math.columbia.edu/tag/020J}{Tag 020J}]{stacks-project} .

  Suppose now that $I$ is prime.
  Any transcendence basis of the fraction field of $\bF\bs{x_{1}, \dots, x_{n}} / I$ as an $\bF$-algebra is also a transcendence basis of the field of fractions of $\bL\bs{x_{1}, \dots, x_{n}} / I'$ as an $\bL$-algebra.
  Therefore $I$ and $I'$ have the same dimension.

  Finally, when $I$ is prime and has dimension $r$, then the degree of $I$ is exactly the degree of the field extension $\bF\br{\vM, \vb} \to \bF\br{\vM, \vb}\bs{x_{1}, \dots, x_{n}} / \ideal{I, \vM \vx + \vb}$, where $\vM$ are the entries of a generic matrix of size $r \times n$, and $\vb$ are the entries of a generic vector of size $r \times 1$ (\cite[Definition~1]{Heintz83}).
  This degree is unchanged when $\bF$ is replaced by $\bL$.
  This shows that the degree of $I'$ agrees with that of $I$.
\end{proof}

The main setting in which we will invoke the above lemma is when $\bL$ is the algebraic closure of $\bF\br{u_{0}, \dots, u_{m}}$, where the $u_{i}$ are new indeterminates.

Finally, we discuss the notion of the multivariate resultant.
For $d \in \bN$, we write $M_{n, d}$ for the set of homogeneous degree-$d$ monomials in the $n+1$ variables $x_{0}, \dots, x_{n}$.
\begin{definition}[{see, e.g., \cite[Proposition 2.3]{Jouanolou91}}] \label{definition: multivariate resultant all fields}
  Fix positive integers $n, d_{0}, \dots, d_{n}$.
  Let $\vec{u} = \setbuild{u_{i, \valpha}}{i \in \bc{0, \ldots, n}, \valpha \in M_{n, d}}$ be a set of variables corresponding to the coefficients of $n+1$ homogeneous polynomials of degrees $d_0, \ldots, d_n$, respectively.
  There exists a unique polynomial $\res_{n, d_{0}, \dots, d_{n}} \in \bF\bs{\vu}$ with the following properties.
  \begin{enumerate}[noitemsep]
    \item
      \emph{Main property of the resultant:} If $F_0, \ldots, F_n \in \bF\bs{x_0, \ldots, x_n}$ are homogeneous polynomials of degrees $d_0, \ldots, d_n$, respectively, then $\res_{d_{0}, \dots, d_{n}}(F_0, \ldots, F_n) = 0$ if and only if the system of equations $F_0(\vec{x}) = \cdots = F_n(\vec{x}) = 0$ has a solution in $\bP^{n}_{\overline{\bF}}$.
      By $\res_{d_0, \ldots, d_n}(F_0, \ldots, F_n)$, we mean the evaluation of $\res_{d_0, \ldots, d_n}$ obtained by setting $u_{i, \valpha}$ equal to the coefficient of the monomial $\vec{x}^{\valpha}$ in $F_i$.
    \item
      $\res_{d_0, \ldots, d_n}(x_0^{d_0}, \ldots, x_n^{d_n}) = 1$.
    \item
      The polynomial $\res_{d_0, \ldots, d_n}$ is irreducible in $\overline{\bF}[\vu]$.
  \end{enumerate}
  Moreover, if we write $\res_{d_0, \ldots, d_n}^{\bQ}$ and $\res_{d_0, \ldots, d_n}^{\bF}$ for the resultants over $\bQ$ and $\bF$, respectively, then the former has integer coefficients and $\res_{d_0, \ldots, d_n}^{\bF}$ is the image of $\res_{d_0, \ldots, d_n}^{\bQ}$ under the natural ring homomorphism $\bZ\bs{\vec{u}} \to \bF\bs{\vec{u}}$.

  When the degrees of the input polynomials and the field of their coefficients are clear, we write $\res$ in place of $\res^{\bF}_{n, d_{0}, \dots, d_{n}}$ for brevity.
\end{definition}

A typical setting in which we invoke the above result is the following.
Suppose $F_{0}, \dots, F_{n} \in \bF\bs{x_{0}, \dots, x_{n}, y_{1}, \dots, y_{m}}$ are polynomials that are homogeneous in the $\vx$ variables.
These polynomials define a variety in $\bP^{n} \times \bA^{m}$.
We can treat $F_{0}, \dots, F_{n}$ as polynomials in $\bF\br{y_{1}, \dots, y_{m}}\bs{x_{0}, \dots, x_{n}}$.
Let $R := \res_{d_{0}, \dots, d_{n}}^{\bF\br{y_{1}, \dots, y_{m}}}\br{F_{0}, \dots, F_{n}}$, where $d_{i}$ is the degree in $\vx$ of $F_{i}$.
By the universality of the resultant, namely by the fact that the resultant over $\bF$ is the image of the resultant over $\bZ$ under the natural map $\bZ\bs{\vu} \to \bF\bs{\vu}$, we can deduce that $R \in \bF\bs{y_{1}, \dots, y_{m}}$.
From the universality, we can also deduce that the zeroes of $R$, or of its squarefree part, form the projection of $\var{F_{0}, \dots, F_{n}}$ onto $\bA^{m}$.
Indeed for any setting $\vgamma \in \overline{\bF}^{m}$ of the $\vy$ variables, the universality implies that $\res_{d_{0}, \dots, d_{n}}^{\overline{\bF}}\br{F_{0}(\vx, \vgamma), \dots, F_{m}(\vx, \vgamma)} = R(\vgamma)$.
By definition, the expression on the left vanishes exactly on the image of the projection of $\var{F_{0}, \dots, F_{n}}$ to $\bA^{m}$.
One consequence of this property of the resultant is that the projection of $\var{F_{0}, \dots, F_{n}}$ to $\bA^{m}$ is either surjective (which is equivalent to the condition that $R$ is the zero polynomial), empty (when $R$ is a nonzero constant), or has codimension one.
\subsection{Uniformity of boolean and arithmetic circuit families} \label{subsection: circuit uniformity}

This subsection describes the notions of uniformity we use for both arithmetic and boolean circuits.
We start by recalling the definition of arithmetic circuits.

\begin{definition} \label{definition: arithmetic circuit}
  Let $\bF$ be a field and let $\bF(\vec{x})$ be the field of rational functions in the variables $x_1, \ldots, x_n$ over $\bF$.
  An \emph{arithmetic circuit over $\bF$} is a directed acyclic graph.
  The vertices of this graph of in-degree zero are either called \emph{input gates} and are labeled by a variable $x_i$, or  called \emph{constant gates} and are labeled by a field element $\alpha \in \bF$.
  Vertices of positive in-degree are called \emph{internal gates} and are labeled by an operation from $\bc{+, \times, \div}$.
  Vertices of out-degree zero are called \emph{output gates}.
  Each gate of the circuit naturally computes an element of $\bF(\vec{x})$, assuming no division by zero takes place in the circuit, which we require.
  If $\bc{f_1, \ldots, f_m}$ are the functions computed by the output gates of the circuit, we say that the circuit \emph{computes} the functions $f_1, \ldots, f_m$.
  The \emph{size} of the circuit is the number of wires in the circuit.
  The \emph{depth} of the circuit is the length of the longest path from an input to an output of the circuit.

  If the only constants labeling the input gates of the circuit are $0, +1$, and $-1$, we say that the circuit is \emph{constant-free}.
  For a fixed variable $x_{i}$, if the subcircuits rooted at the denominators of each division gate in the circuit do not contain the input gate corresponding to $x_{i}$, then we say that the circuit is \emph{division-free with respect to $x_{i}$.}
  If $X'$ is a subset of variables and the circuit is division-free with respect to every variable in $X'$, we say the circuit is \emph{division-free with respect to $X'$.}
  If the circuit is division-free with respect to all variables, we say the circuit is \emph{weakly division-free}.
\end{definition}

For simplicity, we restrict our attention to constant-free arithmetic circuits whenever uniformity is a concern.
This avoids the need to impose a uniformity constraint on the field elements appearing as constants in the circuit family.
As we shall see, constant-free circuits will be sufficient for our purposes.

We now proceed to define the notion of uniformity we use for families of both boolean and arithmetic circuits.
In short, we follow \textcite{vollmer1999} and bound the uniformity of a circuit family through the complexity of its direct connection language.
To decide the direct connection language of a circuit family $\cC = (C_{k,n_1,\ldots,n_k})_{k,n_1,\ldots,n_k \in \bN}$, a Turing machine must be able to efficiently answer queries of the form ``Is gate $a$ in circuit $C_{k,n_1,\ldots,n_k}$ of type $t$?'' and ``Is gate $a$ an ancestor of gate $b$ in circuit $C_{k,n_1,\ldots,n_k}$?'' 
If this language can be decided in polynomial time, we say that the circuit family $\cC$ is polylogtime-uniform, as the length of the input to the direct connection language will be logarithmic in the size of the circuit $C_{k,n_1,\ldots,n_k}$.

To keep this work self-contained, we provide a complete definition of the direct connection language and of polylogtime-uniformity.
We start with the notion of an admissible encoding, which enforces mild constraints on the manner in which gates of a circuit are indexed.

\begin{definition}[Admissible encoding] \label{defintion: admissible encoding}
  Let $\mathcal{C} = (C_{k, n_1, \ldots, n_k})_{k, n_1, \ldots, n_k \in \bN}$ be a family of either threshold or arithmetic circuits of size $s \coloneqq s(k, n_1, \ldots, n_k)$.
  An \emph{admissible encoding of $\cC$} is a numbering of the gates of each circuit $C = C_{k, n_1, \ldots, n_k}$ in the family with the following properties.
  \begin{enumerate}[noitemsep]
    \item
      If $C$ has $m$ input gates, then they are numbered $0, \dots, m-1$.
    \item
      If $C$ has $m'$ output gates, then they are numbered $m, \dots, m + m' - 1$.
    \item
      There is a constant $c$, depending only on the circuit family, such that the binary representation of the number of any gate in $C$ has length at most $c \cdot \log\br{s}$.
  \end{enumerate}
  Further, if $\cC$ is a family of arithmetic circuits, then for each gate $a \in C$ we fix a numbering of the set of immediate predecessors of $a$ in $C$.
  We require that this numbering of predecessors is contiguous, that is, it starts at $1$ and ends at the arity of $a$.
  For division gates, we require that the numerator occurs before the denominator in the numbering of the predecessors.
\end{definition}

We now define the direct connection language of a circuit family.

\begin{definition}[Direct connection language] \label{defintion: direct connection language}
  Let $\mathcal{C} = (C_{k, n_1, \ldots, n_k})_{k, n_1, \ldots, n_k \in \bN}$ be a family of either threshold or arithmetic circuits with an admissible encoding.
  Suppose a numbering of the gate types is fixed.
  The \emph{direct connection language of $\mathcal{C}$}, denoted by $L_{DC}(\mathcal{C})$, is the language consisting of binary encodings of tuples $(k, n_1, \ldots, n_k, a, p, b)$, where
  \begin{enumerate}[noitemsep]
    \item
      $a$ is the number of a gate in $C_{k, n_{1}, \dots, n_{k}}$.
    \item
      If $p \neq \epsilon$, then 
      \begin{enumerate}
        \item
          If $\cC$ is a family of arithmetic circuits, $b$ is the number of the $p\ts{th}$ predecessor gate in $C_{k, n_{1}, \dots, n_{k}}$ of $a$.
        \item
          If $\cC$ is a family of threshold circuits, $b$ is the number of some predecessor gate in $C_{k, n_{1}, \dots, n_{k}}$ of $a$.
      \end{enumerate}
    \item
      If $p = \epsilon$, then $b$ encodes the gate type of $a$ (using the fixed numbering of gate types).
      \qedhere
  \end{enumerate}
\end{definition}

The complexity of the direct connection language provides a way to measure the uniformity of a circuit family.

\begin{definition}[Polylogtime-uniformity] \label{definition: uniform circuit}
  Let $\mathcal{C} = (C_{k, n_1, \ldots, n_k})_{k, n_1, \ldots, n_k \in \bN}$ be a family of either threshold or arithmetic circuits of size $s \coloneqq s(k, n_1, \ldots, n_k)$ with an admissible encoding.
  We say that the circuit family $\mathcal{C}$ is \emph{polylogtime-uniform} if 
  \begin{enumerate}[noitemsep]
    \item
      its direct connection language $L_{DC}(\mathcal{C})$ can be decided in time polynomial in $\log(s) + \log(k) + \sum_{i=1}^k \log(n_i)$, 
    \item
      the number of inputs and outputs of $C_{k,n_1, \ldots, n_k}$, and an upper bound on the size $s$ can be computed in time polynomial in $\log(s) + \log(k) + \sum_{i=1}^k \log(n_i)$, given $(k, n_1, \ldots, n_k)$ as input.
    \item
      when $\cC$ is a family of arithmetic circuits, the arity of a gate can be computed in time polynomial in $\log(s) + \log(k) + \sum_{i=1}^k \log(n_i)$, given $(k, n_1, \ldots, n_k)$ and the gate name as input.
      \qedhere
  \end{enumerate}
\end{definition}

\begin{remark} \label{remark: uniformity}
  We make a few remarks on the above definitions.
  For justifications of the following remarks and a more detailed discussion on the above notion of uniformity, we refer the reader to \cite[Section~2.2]{AGS26}.
  \begin{itemize}
    \item In the above definition, the first index is a count of the remaining indices.
      In the circuits we design, we extend this slightly and allow the number of indices to be some function of the first index.
    \item The freedom in numbering the gates afforded by the above definition allows us to number our gates using more than just the integers $1$ through $s$: by using appropriate pairing functions, we can for example number our gates using pairs of integers.
      To avoid confusion, when we number our gates using more than just natural numbers, we will refer to the name of a gate instead of its number, with the understanding that the actual number is obtained from the name via a pairing function.
    \item When working with a polylogtime-uniform family of circuits, given an index $(k, n_{1}, \dots, n_{k})$ and a candidate gate name $a$, we can test if $a$ is a valid gate name and compute the type of gate $a$ in time polynomial in $\log\br{s} + \log\br{k} + \sum_{i=1}^{n} \log \br{n_{i}}$.
      We can also determine if $a$ is an input or output gate. \qedhere
  \end{itemize}
\end{remark}

\begin{remark}
  \label{remark: uniformity proofs}
  Showing that a given family of circuits is polylogtime-uniform requires checking many details and is rarely insightful.
  For an example of a proof of uniformity including every detail, we refer the reader to the proofs of \cite[Lemma~3.1]{AGS26} and \cite[Lemma~3.2]{AGS26}.
  As discussed in the text following those two proofs, all of our circuits will be constructed using previously-constructed uniform circuit families combined with other simple machinery.
  The direct connection language of a circuit family constructed this way can be decided by encoding the rules for the simple machinery, and simulating the machines that decide the direct connection languages of the subcircuits.
  In \cite{AGS26}, over the course of the article, the proofs of uniformity became increasingly terse, only highlighting minor differences from earlier proofs.
  We continue this trend for the sake of readability.
\end{remark}

\subsection{Uniform circuits for basic arithmetic operations} \label{subsection: basic ops uniformity}

In this section, we collect a few results that show that basic operations can be performed by uniform families of boolean and arithmetic circuits.
These results will used as building blocks throughout our work.
Whenever we work with boolean circuits or Turing machines, integers will be represented in base two with a sign bit.
Therefore, representing an integer of height $h$ requires $h+1$ bits (accounting for the sign bit).
We represent rational numbers as pairs of integers $(p, q)$ with $q \neq 0$, where $p$ and $q$ are not necessarily relatively prime.

We start with some basic results.
\begin{theorem} \label{theorem: integer ops in tc0}
  The following functions are in polylogtime-uniform $\TC^0$.
  \begin{enumerate}
    \item 
      \textsc{Iterated Integer Addition}: Given $n$ integers $a_1, \ldots, a_n$ as input, each consisting of at most $n$ bits, compute the sum $a_1 + \cdots + a_n$.
    \item 
      \textsc{Iterated Integer Multiplication}: Given $n$ integers $a_1$, \ldots, $a_n$ as input, each consisting of at most $n$ bits, compute the product $a_1 \cdots a_n$.
    \item
      \textsc{Integer Division}: Given integers $a$ and $b$ as input, with $b\ne 0$, compute $\lfloor a/b \rfloor$, and compute an integer $r$ such that $a = \lfloor a/b \rfloor b + r$.
    \item 
      \textsc{Polynomial Division with Remainder}: given integers $a_0, \ldots, a_n$ and $b_0, \ldots, b_m$ as input, where $b_m \neq 0$ and $n \geq m$, compute $d \coloneqq b_m^{n-m+1}$ and integers $q_0, \ldots, q_{n-m}, r_0,\ldots,r_{m-1}$ such that for
      \begin{align*}
        f(x) &\coloneqq a_n x^n + \cdots + a_1 x + a_0 &
        q(x) &\coloneqq q_{n-m} x^{n-m} + \cdots + q_1 x + q_0 \\
        g(x) &\coloneqq b_m x^m + \cdots + b_1 x + b_0 &
        r(x) &\coloneqq r_{m-1} x^{m-1} + \cdots + r_1 x + r_0,
      \end{align*}
      we have $d \cdot f(x) = q(x) g(x) + r(x)$.
    \item
      \textsc{Iterated Polynomial Multiplication}: given integers $a_{1,0}, \ldots, a_{m,n}$ as input, compute integers $b_0, \ldots, b_{mn}$ such that
      \[
        b_{mn} x^{mn} + b_{mn-1} x^{mn - 1} + \cdots + b_1 x + b_0 = \prod_{i=1}^m (a_{i, n} x^n + a_{i, n-1} x^{n-1} + \cdots + a_{i, 1} x + a_{i, 0}).
      \]
    \item 
      \textsc{Polynomial Greatest Common Divisor}: given $2 \cdot (n+1)$ integers as input, representing two univariate polynomials $f, g$ of degree at most $n$, compute integers $c_{0}, \dots, c_{n}$ and integer $d$ such that for $c(x) = c_{n} x^{n} + \cdots + c_{0}$, the polynomial $c(x) / d$ is the unique monic multiple of the greatest common divisor of $f$ and $g$ in $\bQ\bs{x}$.
    \item 
      \textsc{Squarefree Part}: given $(n+1)$ integers as input, representing an univariate polynomial $f$ of degree at most $n$, compute integers $c_{0}, \dots, c_{n}$ and integer $d$ such that for $c(x) = c_{n} x^{n} + \cdots + c_{0}$, the polynomial $c(x) / d$ is the unique monic multiple of the squarefree part of $f$ in $\bQ\bs{x}$.
    \item 
      \textsc{Univariate Polynomial Interpolation}: given distinct integers $a_1, \ldots, a_n$ and not necessarily distinct integers $b_1, \ldots, b_n$ as input, compute $d \coloneqq \prod_{1 \leq i < j \leq n} \br{a_{i} - a_{j}}$ and integers $c_0, \ldots, c_{n}$ such that $f(x) \coloneqq c_n x^n + \cdots + c_1 x + c_0$  is the unique polynomial in $\bZ\bs{x}$ of degree at most $n$ that satisfies $f(a_{i}) = d b_{i}$ for $1 \leq i \leq n$.
    \item
      \textsc{Iterated Multivariate Polynomial Multiplication}: given $m \cdot \binom{d+n}{d}$ integers as input, representing the coefficients of $m$ polynomials of degree at most $d$ in $n$ variables, compute $\binom{dm + n}{dm}$ integers representing the coefficients of their product.
    \item
      \textsc{Multivariate Polynomial Division with Remainder}: given $2 \cdot \binom{d+n}{d}$ integers as input, representing the coefficients of two polynomials $f, g$ degree at most $d$ in $n$ variables, compute integer $e$ and polynomials $q, r$ with integer coefficients in $n$ variables satisfying the following:
      \begin{itemize}[noitemsep]
        \item $ef = qg + r$.
        \item $e$ is a power of the leading coefficient of $g$ in lexicographic order.
        \item The leading monomial of $r$ in lexicographic order is smaller than that of $g$.
      \end{itemize}
    \item
      \textsc{Multivariate Polynomial Interpolation}: given $\br{d+1}^{m}$ integers $b_{1}, \dots, b_{\br{d+1}^{m}}$ as input, compute $\binom{d+m}{m}$ integers representing the coefficients of an $m$-variate polynomial $f$ of degree at most $d$, and compute $D \coloneq \prod_{1 \leq i < j \leq \br{d+1}^{m}} \br{i - j}$ such that 
      \[
        f(k, k^{(d+1)}, \dots, k^{(d+1)^{m-1}}) = D b_{k}
      \]
      for every $1 \leq k \leq \br{d+1}^{m}$.
  \end{enumerate}
\end{theorem}
\begin{proof}
  Item 1 is classical, see \cite{reif92}.
  Items 2 and 3 are the main result of \cite{HAB02}.
  Items 4, 5, and 8 are from \cite[Corollary~6.5]{HAB02}.
  Item 6 is a special case of \cite[Lemma~6.13]{AGS26}.
  Item 7 follows from Items 6 and 4, and the fact that the squarefree part of a polynomial $f \in \bQ\bs{x}$ is simply $f / \gcd\br{f, f'}$.
  Item 9 follows from item 5 using Kronecker substitution (see \cite[Section~2.3]{AGS26} for details).
  Item 10 follows from item 4 using Kronecker substitution.
  Item 11 follows from item 8 using Kronecker substitution.
  For all items above, the number of input and output bits are easily computable from the index of a circuit in the corresponding family.
\end{proof}
As discussed in \cite{Eberly89} and also \cite[Section~2.3]{AGS26}, we can use the above circuit families to obtain logtime-uniform families of circuits performing iterated addition and multiplication of rational numbers.

Next, we describe a family of polylogtime-uniform boolean circuits that computes the GCD of two multivariate polynomials.
In \cite[Theorem~6.13]{AGS26}, a family of polylogtime-uniform $\TC^{0}$ circuits is constructed that computes the greatest common divisor of two polynomials in $\bZ\bs{y_{1}, \dots, y_{k}}\bs{x}$.
This circuit returns the unique monic multiple of the GCD in $\bQ\br{y_{1}, \dots, y_{k}}\bs{x}$.
For example, if queried on the inputs $y_{1} y_{2} x$ and $y_{1} x$, the algorithm returns $x$ as the GCD.
To compute the GCD of multivariate polynomials, we will invoke the above algorithm after a random linear transformation of the input variables.

\begin{lemma}
  \label{lemma: multivariate GCD}
  There exists a polylogtime-uniform family of $\TC^{0}$ circuits with the following properties.
  \begin{itemize}[noitemsep]
    \item The circuits take as input $2 \binom{d + m}{d}$ integers representing the coefficients of two $m$-variate polynomials of degree $d$ with integer coefficients $f, g$.
      The circuits also take as input $m-1$ integers, denoted $a_{2}, \dots, a_{m}$.
    \item The circuits computes $\binom{d+m}{d}$ integers representing an $m$-variate polynomial $h$ of degree at most $d$.
    \item If the integers $a_{2}, \dots, a_{m}$ are independently and uniformly sampled from a set of size $b$, and if the highest degree parts of $f$ and $g$ depend on the variable $x_{1}$, then with probability $1 - 2md^{m}/b$, the polynomial $h$ is a scalar multiple of the GCD of $f$ and $g$ in $\bQ\bs{x_{1}, \dots, x_{m}}$.
  \end{itemize}
\end{lemma}

\begin{proof}
  We use the integers $a_{2}, \dots, a_{m}$ to perform a change of basis for the polynomials $f$ and $g$.
  Let 
  \begin{align*}
    \hat{f} &\coloneqq f(x_1, x_2 - a_2 x_1, \ldots, x_m - a_m x_1) \\
    \hat{g} &\coloneqq g(x_1, x_2 - a_2 x_1, \ldots, x_m - a_m x_1).
  \end{align*}
  Let $\tilde{f}$ and $\tilde{g}$ be the univariate polynomials obtained by applying the Kronecker substitution to $\hat{f}$ and $\hat{g}$, respectively.
  Given the coefficients of $f$ and $g$ and the integers $a_{2}, \dots, a_{m}$, combined with the arithmetic described in \cref{theorem: integer ops in tc0}, we can compute the evaluations of $\tilde{f}$ and $\tilde{g}$ at the integers $1, \dots, \br{d+1}^{m}$.
  We can then compose with a circuit for multivariate interpolation (item 11 of \cref{theorem: integer ops in tc0}) to obtain the coefficients of $\bZ$-multiples of $\hat{f}$ and $\hat{g}$.
  Next, we compose with the circuit for univariate polynomial GCD in function fields (\cite[Lemma~6.13]{AGS26}), treating the polynomials $\hat{f}$ and $\hat{g}$ as univariate polynomials in $x_{1}$ with coefficients in $\bQ\br{x_{2}, \dots, x_{m}}$.
  This returns two polynomials $h_{n} \in \bZ\bs{x_{1}, \dots, x_{m}}$ and $h_{d} \in \bZ\bs{x_{2}, \dots, x_{m}}$ such that $h_n / h_d$ is the GCD of $\hat{f}$ and $\hat{g}$ in $\bQ\br{x_2,\ldots,x_m}\bs{x_1}$.
  We then compose with the circuits for multivariate polynomial division with remainder to divide every coefficient of $x_{1}$ in $h_{n}$ by $h_{d}$, and discard the remainders.
  We use integer arithmetic to bring the resulting polynomial to common denominator, and discard the denominator.
  Let the resulting polynomial be denoted by $\hat{h}$.
  We then compute the coefficients of $h \coloneq \hat{h}(x_{1}, x_{2} + a_{2} x_{1}, \dots, x_{m} + a_{m} x_{1})$ using evaluation-interpolation as above.
  The final output if the circuit are the coefficients of $h$.

  The fact that the above computation can be carried out by a polylogtime-uniform family of $\TC^{0}$ circuits follows by a simulation argument (see \cref{remark: uniformity proofs}).
  We now show that with high probability, the polynomial $h$ is a $\bQ$-multiple of the GCD of $f$ and $g$.
  To this end, observe that with probability at least $1 - 2md^{m} / b$, the leading coefficients of $\hat{f}$ and $\hat{g}$ in the variable $x_{1}$ are integers.
  If this holds, then by Gauss's lemma, the GCD of $\hat{f}$ and $\hat{g}$ in $\bQ\br{x_{2}, \dots, x_{m}}\bs{x_{1}}$ and the GCD in $\bZ\bs{x_{1}, \dots, x_{m}}$ are $\bQ$-multiples of each other.
  Specifically, every coefficient of $h_{n}$ in $x_{1}$ is divisible by $h_{d}$ in $\bQ\bs{x_{2}, \dots, x_{m}}$.
  Therefore, the division step does not produce any remainders and $\hat{h}$ is a $\bQ$-multiple of the GCD of $\hat{f}$ and $\hat{g}$.
  Finally, the last step undoes an invertible linear transformation, so the polynomial $h$ is a $\bQ$-multiple of the GCD of $f$ and $g$ as claimed.
\end{proof}

As a corollary, we also obtain a circuit family that computes squarefree parts of multivariate polynomials.
\begin{corollary}
  \label{corollary: multivariate squarefree part}
  There exists a polylogtime-uniform family of $\TC^{0}$ circuits with the following properties.
  \begin{itemize}[noitemsep]
    \item The circuits take as input $\binom{d + m}{d}$ integers representing the coefficients of an $m$-variate polynomial of degree $d$ with integer coefficients, denoted $f$.
      The circuits also take as input $m-1$ integers, denoted $a_{2}, \dots, a_{m}$.
    \item The circuits computes $\binom{d+n}{d}$ integers, representing an $m$-variate polynomial $h$ of degree at most $d$.
    \item If the integers $a_{2}, \dots, a_{m}$ are independent and uniformly sampled from a set of size $b$, and if the highest degree part of $f$ depends on the variable $x_{1}$, then with probability $1 - 2md^{m}/b$, the polynomial $h$ is a scalar multiple of the squarefree part of $f$ in $\bQ\bs{x_{1}, \dots, x_{m}}$.
  \end{itemize}
\end{corollary}

\begin{theorem}[{\cite[Corollary 5.13]{AGS26}}]
  \label{theorem: resultant uniform circuits}
  There exists a polylogtime-uniform family of weakly division-free, constant-free circuits $\cC$ indexed by parameters $n, d_{0}, \dots, d_{n}, k, D$ with the following properties.
  \begin{itemize}
    \item 
      The circuit $C_{n, d_{0}, \dots, d_{n}, k, D}$ has size bounded by $\br{PD}^{\poly\br{n, k}}$, where $P = \sum_{i=0}^{n} \prod_{j \neq i} d_{j}$, and depth bounded by a universal constant.
    \item
      The circuit has $\binom{k + D}{k} \cdot \sum_{i=0}^{n} \binom{n + d_{i} - 1}{d_{i}}$ input variables, which we denote by $u_{i, \valpha, \vbeta}$, where $\vbeta \in \bN^{k}$ satisfies $\abs{\vbeta} \leq D$, and $\valpha, i$ satisfy $\abs{\valpha} \leq d_{i}$.
      The circuit has $\binom{D D' + k}{k}$ outputs, where $D'$ is the degree of $\res_{n}\br{\vF_{n}}$.
    \item
      When evaluated at $\gamma_{i, \valpha, \vbeta}$, the output gates compute the coefficients of the monomials in $\vw$ of the resultant of the polynomials
      \[
        H_{i} = \sum_{\valpha} \sum_{\vbeta} \gamma_{i, \valpha, \vbeta} \vx^{\valpha} \vw^{\vbeta}
      \]
      with respect to the variables $\vx$, where the ranges of $\valpha, \vbeta$ are the same as above.
  \end{itemize}
\end{theorem}

\subsection{Uniformity of lifting factorizations} \label{subsection: factor lifting uniformity}

Suppose $\bF$ is a field of characteristic $0$ and $P \in \bF\bs{t, y}$ is a bivariate polynomial.
There are a number of different algorithms to compute the irreducible factors of $P$.
We refer the reader to \cite[Section~1]{gao03} for a historical account of these algorithms.

A common approach to bivariate factorization is to lift to the ring $\bF\bsd{t}\bs{y}$.
Under mild assumptions, Hensel's lemma gives us an algorithm to approximate the roots of the polynomial $P$ in this ring.
If $P$ and $P(0, y)$ are both squarefree, then each root of $P(0, y)$ lifts to a unique root of $P$ in $\bF\bsd{t}\bs{y}$.
Following this, linear algebraic methods can be used to compute which roots of $P$ in $\bF\bsd{t}\bs{y}$ combine together to produce factors of $P$ in $\bF\bs{t, y}$.

A key observation is that the coefficients of a root $\phi \in \bF\bsd{t}$ of $P$ as an element of $\bF\bsd{t}\bs{y}$ are rational functions of the coefficients of $P$ and the root of $P(0, y)$ that lifts to $\phi$.
As a consequence, we can define the generic bivariate polynomial and construct a universal power series root such that for any bivariate polynomial $P$ with coefficients in $\bF$, as long as $P$ satisfies some mild assumptions, the roots of $P$ in $\bF\bsd{t}\bs{y}$ can be obtained by specializing the universal power series root.
This same observation is used in \cite{kaltofen95} to derive irreducibility forms.

Let $F_{d}$ denote the generic bivariate polynomial of degree $d$ with indeterminate coefficients, that is
\[
  F_{d} = \sum_{i + j \leq d} c_{i, j} t^{i} y^{j} \in \bZ\bs{C_{d},  t, y},
\]
where $C_{d}$ is the set of indeterminates $\bc{c_{i, j}}_{i + j \leq d}$.
Let $F'_{d}$ denote the partial derivative $\partial_{y}F_{d}$.
Define
\[
  \tilde{F_{d}}(t, y) := F'_{d}(0, z)^{-1} \cdot F_{d}(t, y + z).
\]
Define the \emph{universal power series root} $\Phi^{(d)} \in \bZ\bs{C_{d}, z}_{F'_{d}(0, z)}\bsd{t}$ by
\begin{align*}
  \Phi^{(d)}_{N} &\coloneq \sum_{i \leq 2N} \br{-1}^{i} \mathrm{coeff}\br{y \br{\tilde{F_{d}}(t, y) - y}^{i} \partial_{y} \tilde{F_{d}}(t, y), t^{N}y^{i}} \\
                 &= \sum_{i \leq 2N} \frac{\br{-1}^{i}}{F'_{d}(0, z)^{i+1}} \mathrm{coeff}\br{y \br{F_{d}(t, y + z) - y F'_{d}(0, z)}^{i} F'_{d}(t, y+z) , t^{N}y^{i}} \\
                 &= \frac{1}{F'_{d}(0, z)^{2N+1}} \sum_{i \leq 2N} \br{-1}^{i} {F'_{d}(0, z)^{2N - i}} \cdot \mathrm{coeff}\br{y \br{F_{d}(t, y + z) - y F'_{d}(0, z)}^{i} F'_{d}(t, y+z) , t^{N}y^{i}} \\
\end{align*}
and $\Phi^{(d)} \coloneq z + \sum_{N \geq 1} \Phi^{(d)}_{N} t^{N}$.
The universal power series root satisfies the following property.
Let $\bF$ be any field of characteristic $0$, and let $\vgamma \in \bF^{\abs{C_{d}}}$.
Let $P$ denote the specialization $F_{d}^{\vgamma}$.
Suppose $\vgamma$ is such that $P(t, y)$ is a degree $d$ squarefree polynomial whose leading coefficient with respect to $y$ lies in $\bF$.
Suppose $\vgamma$ is also such that $P(0, y)$ is a squarefree polynomial.
In this setting, the specialization of $F'_{d}(0, z)$ at $\vgamma$ does not vanish, therefore the specialization of $\Phi^{(d)}$ at $\vgamma$ is well defined.
Let $P'$ denote the specialization of $F'_{d}$ at $\vgamma$, equivalently the derivative of $P$ with respect to $y$.

Since $P(0, y)$ is squarefree, $P'(0, z)$ is invertible in $\bK \coloneqq \bF\bs{z} / P(0, z)$.
Therefore, the specialization of $\Phi^{(d)}$ at $\vgamma$ is naturally an element of $\bK\bsd{t}$.
Call this element $\phi_{z}$.
Then $P(t, \phi_{z}) = 0$ in $\bK\bsd{t}$.

Essentially equivalently, suppose $\alpha \in \overline{\bF}$ is a root of $P(0, y)$.
Then $P'(0, \alpha) \neq 0$ by assumption.
Therefore, the specialization of $\Phi^{(d)}$ at $\vgamma$ and $\alpha$ is naturally an element of $\overline{\bF}\bsd{t}$.
Call this element $\phi_{\alpha}$.
Then $P(t, \phi_{\alpha}) = 0$ in $\overline{\bF}\bsd{t}$.

The proof of the above is a straightforward application of an explicit version of the implicit function theorem \cite[Proposition~20.3]{AiYu83}.
The special case above can also be deduced from the Lagrange inversion formula.
That this formula can be used to factor polynomials was first observed by \textcite{BKRRSS26a}, and our presentation closely follows theirs.
From the description above, combined with the uniform families of algebraic circuits for polynomial interpolation (\cref{theorem: integer ops in tc0}), the following lemma is immediate (see \cref{remark: uniformity proofs}).
For later applications, it will be convenient to compute the denominator and numerator in the final expression for $\Phi_{N}^{(d)}$ above separately.
\begin{lemma} 
  \label{lemma: universal root uniform}
  There exists a polylogtime-uniform family of constant-free circuits $\cC$ indexed by $N, d$ such that
  \begin{itemize}[noitemsep]
    \item The circuit $C_{N, d}$ has size polynomial in $N$ and $d$, and depth bounded by a universal constant.
    \item The circuit $C_{N, d}$ has $\binom{d+2}{2} + 1$ input gates, labeled by variables $c_{i, j}$ for $i + j \leq d$ and $z$.
      The circuit has two output gates and is weakly division free.
    \item The circuit computes the coefficient $\Phi^{(d)}_{N}$ of the universal power series root.
      Specifically, the first output gate computes the sum in the final expression above, and the second output gate computes $F'_{d}(0, z)^{2N+1}$.
  \end{itemize}
\end{lemma}

Fix an integer $e \leq d$, and let $b_{0}, \dots, b_{e}$ be new indeterminates.
Let $H_{e}$ denote the generic univariate polynomial of degree $d$ in the variable $z$, that is, $H_{e} = b_{0} + b_{1} z + \cdots + b_{e} z^{e}$.
Let the definitions of $F_{d}$ and $\Phi^{(d)}_{N}$ be as above.

Define the truncated universal power series root $\Psi^{(d)}$ as
\[
  \Psi^{(d)} \coloneqq z + \sum_{1 \leq N \leq d} \Phi^{(d)}_{N} t^{N}.
\]
From the definition of $\Phi_{N}^{(d)}$ we deduce that $F'_{d}(0, z)^{2d+1} \Psi^{(d)} \in \bZ\bs{C_{d}, t, z}$.
Further, the degree in $z$ of this polynomial is $2d^{2} - d$.
Let $R_{1}$ denote the resultant of $H(z)$ and $F'_{d}(0, z)^{2d+1}y - F'_{d}(0, z)^{2d+1}\Psi^{(d)}$ as univariate polynomials in $z$.
Let $R_{2}$ denote the resultant of $H(z)$ and $F'_{d}(0, z)^{2d+1}$ as univariate polynomials in $z$.

The polynomials above have the following properties.
Let $\vgamma$ be a specialization of the variables in $C_{d}$ with the properties described above.
Let $\vdelta$ be a specialization of the variables $b_{0}, \dots, b_{e}$ such that $Q(z) \coloneqq H^{\vdelta}(z)$ has degree $e$ and $Q(z)$ divides $P(0, z)$.

In this setting, the quotient $R_{1}^{\vgamma, \vdelta} / b_{e} R_{2}^{\vgamma, \vdelta}$ is a polynomial in $t$ and $y$ that is monic in $y$.
The assumptions above ensure that $P$ factors into linear polynomials in $\overline{\bF}\bsd{t}$, and that the quotient $R_{1}^{\vgamma, \vdelta} / b_e R_{2}^{\vgamma, \vdelta}$ is the product of those linear factors whose constant terms are roots of $H^{\vdelta}$.
Specifically, if we are guaranteed that $H^{\vdelta}$ lifts to a factor of $P$ in $\bF\bs{t, y}$, then the truncation of this quotient up to degree $d$ is the lift of $H^{\vdelta}$.
By combining uniform families of algebraic circuits for the resultant (\cref{theorem: resultant uniform circuits}) with the circuits for the universal power series roots (\cref{lemma: universal root uniform}), we obtain a uniform family of algebraic circuits that computes the polynomials $R_{1}$ and $R_{2}$ described above.

\begin{lemma} 
  \label{lemma: universal factor uniform}
  There exists a polylogtime-uniform family of constant-free circuits $\cD$ indexed by $d$ and $e$ such that
  \begin{itemize}[noitemsep]
    \item The circuit $D_{d, e}$ has size polynomial in $d$ and $e$ and depth bounded by a universal constant.
    \item The circuit $C_{N, d}$ has $\binom{d+2}{2} + e+ 3$ input gates labeled by variables $c_{i, j}$ for $i + j \leq d$ and $b_{0}, \dots, b_{e}, t, y$.
      The circuit has two output gates and is weakly division free.
    \item The circuit computes the truncation of $R_{1}$ up to degree $d$ in $t$, and also computes the polynomial $R_{2}$ described above.
  \end{itemize}
\end{lemma}

Combining the above lemma with \cite[Lemma~6.1]{AGS26}, we obtain a polylogtime uniform family of constant-depth threshold circuits that takes as input $\binom{d+2}{2} + e + 1$ integers, and computes the coefficients of $R_{1}$ and $R_{2}$ in the variables $y$ and $t$.
Using standard reductions from multivariate polynomial factorization to bivariate polynomial factorization, we can also use the above circuits to compute lifts of factors of multivariate polynomials.
We reemphasize that the reduction from multivariate polynomial factorization to univariate factorization in the following lemma is folklore, the main content of the following lemma is the uniformity statement.

\begin{lemma}
  \label{lemma: multivariate lifting}
  There exists a polylogtime-uniform family of $\TC^{0}$ circuits with the following properties.
  \begin{itemize}[noitemsep]
    \item The circuit takes as input $\binom{d+m+1}{d} \cdot \binom{e+k}{e}$ integers, representing the coefficients of a polynomial $P \in \bZ\bs{w_{1}, \dots, w_{k}}\bs{y, x_{1}, \dots, x_{m}}$ of degree at most $e$ in the variables $w_{1}, \dots, w_{k}$ and degree at most $d$ in the variables $y, x_{1}, \dots, x_{m}$.
      The circuit also takes as input $\br{d+1} \cdot \binom{e+k}{k}$ integers, representing the coefficients of a polynomial $Q \in \bZ\bs{w_{1}, \dots, w_{k}}\bs{y}$ of degree at most $e$ in $w_{1}, \dots, w_{k}$ and degree at most $d$ in $y$.
    \item The circuits computes $\binom{d+m+1}{d} \cdot \binom{e+k}{e}$ integers representing an $m$-variate polynomial $R$ of degree at most $d$ in $y, x_{1}, \dots, x_{m}$ and at most $e$ in $w_{1}, \dots, w_{k}$.
    \item Suppose the leading term of $P$ in $y$ lies in $\bZ$.
      Suppose that both $P$ and $P(y, 0, \dots, 0)$ are squarefree, and suppose $Q$ divides $P(y, 0, \dots, 0)$.
      Under these assumptions, using Gauss' lemma we can deduce that after scaling, all the factors of $P$ in the ring $\bQ\br{w_{1}, \dots, w_{k}}\bs{y, x_{1}, \dots, x_{m}}$ lie in $\bQ\bs{w_{1}, \dots, w_{k}}\bs{y, x_{1}, \dots, x_{m}}$.
      Suppose $R'$ is such a factor of $P$ such that $R'(y, 0, \dots, 0)$ is a $\bQ$-multiple of $Q$.
      Then the computed polynomial $R$ is a $\bQ$-multiple of $R'$.
  \end{itemize}
\end{lemma}
\begin{proof}
  Let $\bK \coloneqq \bQ\br{w_{1}, \dots, w_{k}, x_{1}, \dots, x_{m}}$.
  Let $\hat{P}$ denote the polynomial $P(y, tx_{1}, \dots, t x_{n}) \in \bK\bs{t, y}$.
  There exists a polylogtime-uniform family of constant-depth arithmetic circuits with $\binom{d+m+1}{d} \cdot \binom{e+k}{k}$ inputs that takes as input the coefficients of $P$ and computes the coefficients of $\hat{P}$.
  There also exists a polylogtime-uniform family of arithmetic circuits that computes the coefficients of $Q$ as a polynomial in $y$.
  We then compose these with the circuits constructed in \cref{lemma: universal factor uniform}.
  The resulting family of circuits is still polylogtime-uniform and constant depth, and it computes two polynomials in $y, t$ with coefficients in $\bK$, call them $R_{1}$ and $R_{2}$.
  Since $\hat{P}(0, t) \in \bQ\bs{y, w_{1}, \dots, w_{k}}$, we can deduce that $R_{2} \in \bQ\bs{w_{1}, \dots, w_{e}}$.
  We use \cite[Lemma~6.1]{AGS26} to obtain a polylogtime-uniform family of threshold circuits that performs the same computation.
  We compose with circuits for polynomial division to divide each coefficient of $R_{1}$ in $t$ by $R_{2}$, and discard the remainders.
  Finally, we add up all the coefficients in $t$ of the resulting polynomial (equivalently, we evaluate the resulting polynomial at $t = 1$).
  We then discard all monomials whose degree in $w_{1}, \dots, w_{k}$ is higher than $e$.
  This final polynomial is the output polynomial $R'$.
\end{proof}

\subsection{Computation and the counting hierarchy}

A key technical tool that is used in \cite{AGS26} that will also be useful in this work is that uniform families of exponentially large constant-depth threshold circuits can be evaluated in the counting hierarchy.
Since such a circuit can have exponentially many output gates, we must make precise what we mean by evaluating such a circuit.

\begin{definition}[{\cite[Definition~6.2]{AGS26}}]
  \label{definition: language of function family}
  Let $\cF \coloneqq \br{f_{\vn} : \bc{0,1}^{m'_{\vn}} \to \bc{0,1}^{m_{\vn}}}_{\vn}$ be a multi-indexed family of boolean functions.
  The \emph{language $L_{\cF}$ corresponding to $\cF$} is defined as 
  \[
    L_{\cF} \coloneqq \setbuild{(i, b, x, \vn)}{f_{\vn}(x)_{i} = b}.
  \]
  If $\cD$ is a family of circuits, then we extend notation and use $L_{\cD}$ to denote the language corresponding to the family of functions computed by $\cD$.
\end{definition}

The following simulation result was proved in \cite{AGS26}, and is inspired by a similar result in \cite{ABKM09}.
\begin{lemma}[{\cite[Lemma~6.3]{AGS26}}] \label{lemma: threshold to ch}
  Let $\cD = \br{D_{\vn}}_{\vn}$ be a polylogtime-uniform family of threshold circuits of size $s_{\vn}$ and depth $\Delta$, where $D_{\vn}$ has $m_{\vn}$ inputs.
  Let $L_{\cD}$ be the language corresponding to the circuit family $\cD$ as defined in \cref{definition: language of function family}.
  Suppose $s_{\vn} \le \exp(\poly(m_{\vn}))$.
  Then $L_{\cD} \in \CH$.
\end{lemma}

The next lemma generalizes the above, and shows that an exponential sized circuit with exponentially many inputs can be evaluated in $\CH$, as long as the input bits can themselves be computed in $\CH$.

\begin{lemma}[{\cite[Lemma~6.5]{AGS26}}]  \label{lemma: threshold with oracle to ch}
  Let $\cD = \br{D_{n}}_{n}$ be a polylogtime-uniform family of polynomial-size threshold circuits of depth $\Delta$, where $D_{n}$ has $n$ inputs.
  Let $\cF = \br{f_{\vn} : \bc{0,1}^{\ell_{\vn}} \to \bc{0,1}^{m_{\vn}}}_{\vn}$ be a family of boolean functions.
  Suppose $m_{\vn} \le \exp(\poly\br{\ell_{\vn}})$ and that a binary representation of $m_{\vn}$ can be computed in polynomial time given a unary representation of $\ell_{\vn}$ and a binary representation of $\vn$.
  Let $\cD \circ \cF = (D_{m_{\vn}} \circ f_{\vn})_{\vn}$ denote the family of composed functions.
  Then $L_{\cD \circ \cF} \in \CH^{L_{\cF}}$.
  In particular, if $L_{\cF} \in \CH$, then $L_{\cD \circ \cF} \in \CH$.
\end{lemma}

For our work, we need a slightly more general version of \cref{lemma: threshold with oracle to ch}.
\cref{lemma: threshold with oracle to ch} assumes that the circuit family $\cD$ has a single index, that $D_{n}$ takes exactly $n$ inputs, and that the circuit $D_n$ has size $n^{O(1)}$.
We need to relax the requirements of \cref{lemma: threshold with oracle to ch} to allow the family $\cD$ to be multi-indexed and for the index to the outer family $\cD$ to depend on both the index $\vn$ to the inner family $\cF$ as well as the input $x$ on which the inner family is evaluated.
We omit the proof of this lemma, as it is a straightforward adaptation of the proof of \cref{lemma: threshold with oracle to ch}.
The only change to the proof is that the machine for $L_{\cD \circ \cF}$ must decide which member of the circuit family $\cD$ to invoke.
This is easily handled by the assumption that the correct index into $\cD$ can be computed in $\CH$.

\begin{lemma} \label{lemma: generalized threshold oracle ch}
  Let $\cF = \br{f_{\vn} : \bc{0,1}^{\ell_{\vn}} \to \bc{0,1}^{m_{\vn}}}_{\vn}$ be a family of boolean functions and let $\cD = \br{D_{\vn}}_{\vn}$ be a polylogtime-uniform family of constant-depth threshold circuits.
  For every index $\vn$ and input $x \in \bc{0,1}^{\ell_{\vn}}$, let $\vp$ be an index such that $D_{\vp}$ has $m_{\vn}$ inputs.
  Let $\cD \circ \cF = (D_{{\vp}} \circ f_{\vn})_{\vn}$ denote the family of composed functions.
  Suppose that 
  \begin{enumerate}[noitemsep]
    \item
      the size of $D_{\vp}$ is at most $\exp(\poly\br{\ell_{\vn}})$, and
    \item
      given a unary representation of $\ell_{\vn}$, a binary representation of $\vn$, and an input $x \in \bc{0,1}^{\ell_{\vn}}$, the binary representation of $\vp$ can be decided in $\CH$.
  \end{enumerate}
  Then $L_{\cD \circ \cF} \in \CH^{L_{\cF}}$.
  In particular, if $L_{\cF} \in \CH$, then $L_{\cD \circ \cF} \in \CH$.
\end{lemma}

We list some functions whose corresponding languages can be computed in $\CH$.
We say that a circuit has bounded syntactic degree if the syntactic degree of every output is at most the size of the circuit.

\begin{lemma}
  \label{lemma: circuit eval in CH}
  Let $\cE$ be the family of boolean functions that take as input the description of a constant-free, weakly division-free algebraic circuit and compute the coefficient vectors of the polynomials computed by the circuit.
  Then the language $L_{\cE}$ corresponding to $\cE$ can be decided in $\CH$. 
\end{lemma}
\begin{proof}
  Let $C$ be the given algebraic circuit.
  We will compute the coefficient vectors of the polynomials computed by $C$ using evaluation-interpolation.
  To do this, we will show that (1) we can decide the bits of the binary representation of an evaluation of $C$ in $\CH$, and (2) given oracle access to the binary representations of evaluations of $C$, we can decide the bits of the coefficient vectors in $\CH$.

  We first reduce to the univariate case using the Kronecker substitution.
  Let $d \coloneqq 2^{s+1}$ so that every polynomial computed by $C$ has degree strictly less than $d$.
  Define the algebraic circuit $\hat{C}$ via
  \[
    \hat{C}(x) \coloneqq C(x, x^d, x^{d^2}, \ldots, x^{d^{n-1}}).
  \]
  By our choice of $d$, the binary representation of the coefficients of $C$ may be read off directly from that of the coefficients of $\hat{C}$, so in what remains it suffices to compute the coefficients of $\hat{C}$ in $\CH$.
  The circuit for $\hat{C}$ can be built from the circuit for $C$ using $O(n \log d) \le O(s^2)$ additional gates and the polynomials computed by $\hat{C}$ have degree bounded by $d^n \le 2^{s^2 + s}$.

  Suppose we want to evaluate $\hat{C}$ on an integer input $a$ of height at most $h$.
  We replace the input gate of $\hat{C}$ with a subcircuit that computes the value of $a$ starting from the constants $0$ and $1$.
  By writing $a$ in binary, we can implement this subcircuit using $O(h)$ gates.
  This yields an arithmetic circuit of size $O(s^2 + h)$ that has no inputs and computes $\hat{C}(a)$.
  Given indices $n, i, j \in \bN$, the task of computing the $i$\ts{th} bit of the $n$-bit binary representation of the $j$\ts{th} output of $\hat{C}(a)$ is exactly an instance of the problem $\BitSLP$, which lies in $\CH$ by \cite[Theorem 4.1]{ABKM09}.
  When our goal is to evaluate $\hat{C}$ on an integer input of height $h \le s^{O(1)}$, the circuit constructed for $\hat{C}(a)$ is of size $s^{O(1)}$, so we may decide the value of bits of the binary representation of $C(\va)$ in $\CH$.

  To recover the coefficients of $\hat{C}(x)$ from evaluations, we use the fact that polynomial interpolation can be performed by polylogtime-uniform threshold circuits of constant depth and polynomial size.
  By \cite[Lemma 3.1]{AGS26}, there is a family of polylogtime-uniform arithmetic circuits that compute the inverse of the Vandermonde matrix $\vand{1,\ldots,m}$ in constant depth and $m^{O(1)}$ size.
  Multiplying $\vand{1,\ldots,m}^{-1}$ with the evaluations $f(1),\ldots,f(m)$ of a univariate polynomial of degree less than $m$ recovers the coefficients of $m$, which can again be performed by an arithmetic circuit of constant depth and size $m^{O(1)}$.
  By \cite[Lemma 6.1]{AGS26}, this yields a polylogtime-uniform family $\cD = (D_{m, h})_{m, h \in \bN}$ of threshold circuits of constant depth and size $\poly(m, h)$, where $D_{m,h}$ receives evaluations $f(1), \ldots, f(m)$, each of height at most $h$, and computes the coefficients of the unique polynomial $f(x)$ of degree less than $m$ that agrees with these evaluations.

  We compose the circuit family $\cD$ with the evaluations $\hat{C}(1),\ldots,\hat{C}(2^{s^2+s}+1)$.
  Because each evaluation of $\hat{C}$ can be computed by a constant-free algebraic circuit of size $s^{O(1)}$, the evaluation itself has height bounded by $s^{O(1)}$, so it suffices to invoke the circuit $D_{m, h}$ for $m = 2^{s^2+s}+1$ and $h = s^{O(1)}$.
  This threshold circuit has constant depth and size $\exp(s^{O(1)})$.
  Moreover, the index of this circuit can be computed in polynomial time from the value of $s$.
  Thus, we may apply \cref{lemma: threshold with oracle to ch} to conclude that bits of the evaluation of $D_{m,h}$ on input $\hat{C}(1), \ldots, \hat{C}(2^{s^2+s}+1)$---which are precisely the coefficients of $\hat{C}$---can be decided in $\CH$, as desired.
\end{proof}

\section{Chow forms and geometric resolutions}  \label{section: chow form geometric resolution}

In this section, $\bK$ is an algebraically closed field.
Chow forms and geometric resolutions are two fundamental ways to describe equidimensional affine, or projective, varieties: the former is a single polynomial in several groups of new variables, while the latter involves a family of multivariate polynomials in the ambient coordinates. In order to establish our main result, that Chow forms can be computed in the Counting Hierarchy, we will rely on both encodings. The goal of this section is to introduce both concepts; for convenience, for the latter, existence and basic properties are derived from the ground up.

\subsection{Chow forms} \label{subsection: chow forms intro}

All facts about the Chow form that we state here appear in \cite[Section~1.2.2]{KPS01}, and proofs can be found in \cite[Chapter~1]{philipponindep}.

Let $\bK\bs{x_{0}, \dots, x_{n}}$ be the homogeneous coordinate ring of $\bP^{n}$.
Let $V \subset \bP^{n}$ be an irreducible projective variety of dimension $r$.
Let $\vu_{0}, \dots, \vu_{r}$ be $r+1$ sets of variables, with each $\vu_{i}$ consisting of $n+1$ variables $u_{i, 0}, \dots, u_{i, n}$.
We use $\vu$ to denote the union of $\vu_{0}, \dots, \vu_{r}$.
We also use $U$ and $U_{i}$ to denote the size of the sets $\vu$ and $\vu_{i}$, respectively.
For each $0 \leq i \leq r$, let $L_{i}$ denote the polynomial $u_{i, 0} x_{0} + u_{i, 1} x_{1} + \cdots + u_{i, n} x_{n}$.
For $\vgamma \in \bA^{n+1}$, we use $L_{\vgamma}$ to denote the linear form $\gamma_{0} x_{0} + \cdots + \gamma_{n} x_{n}$.

The \emph{incidence variety} of the forms $L_{0}, \dots, L_{r}$ in $V$ is defined to be the subvariety of $\bP^{n} \times \br{\bP^{n}}^{r+1}$ consisting of points $(\valpha, \vgamma_{0}, \dots, \vgamma_{r})$ such that $\valpha \in V$ and $L_{i}(\valpha, \vgamma_{i}) = 0$ for all $0 \leq i \leq r$.
We use $V_{L}$ to denote this incidence variety.
Let $\pi_{1}: \bP^{n} \times \br{\bP^{n}}^{r+1} \to \bP^{n}$ denote the projection map onto the first factor, and similarly let $\pi_{2}$ denote the projection map onto $\br{\bP^{n}}^{r+1}$.
It is straightforward to see that $\pi_{1}(V_{L}) = V$.

Consider now $W \coloneqq \pi_{2}(V_{L})$.
The variety $W$ is a closed proper subvariety of $(\bP^n)^{r+1}$ of codimension $1$.
A defining equation for $W$ is called the \emph{Chow form} of $V$.
We denote this polynomial by $\cC_{V}$.
Note that a defining equation for $W$ is only fixed up to a scalar factor.
In order to avoid this indeterminacy in the definition of $\cC_{V}$, we will fix the following normalization for $\cC_{V}$: we order the variables $u_{i, j}$ such that $u_{i, j} > u_{i', j'}$ if $i < i'$ or $i = i'$ and $j < j'$.
We normalize the Chow form so that its leading coefficient in the lexicographic order extending the variable ordering is always $1$.

The points of $W$ correspond to linear subspaces of codimension at most $r+1$ that have nonempty intersection with $V$.
If $(\vgamma_{0}, \dots, \vgamma_{r}) \in W$, then by the definitions of $V_L$ and $W$, the linear subspace defined by the equations $L_{\vgamma_{0}} = \cdots = L_{\vgamma_{r}} = 0$ has a point that lies in $V$.
The Chow form is homogeneous of degree $\deg V$ in each set of variables $\vu_{i}$.

If $V$ is a reducible variety, it is still possible to define the Chow form of $V$, assuming $V$ is equidimensional.
When $V$ is an equidimensional variety of dimension $r$, the product of Chow forms of the irreducible components of $V$ is the Chow form of $V$.

The Chow form of a zero-dimensional variety is also called its $u$-resultant.
If $V$ is a zero-dimensional variety with all points at finite distance, then the $u$-resultant of $V$ has the following factorization: $\cC_{V} = \prod_{\valpha \in V} \br{u_{0} + \valpha_{1} u_{1} + \cdots + \valpha_{n} u_{n}}$, where the points $\valpha \in \bP^{n}$ are written with $\valpha_{0} = 1$.

Chow forms behave nicely under linear changes of coordinates.
Given the Chow form of a variety $V$ and a linear change of coordinates $\phi : \bP^n \to \bP^n$, the following lemma describes how to obtain the Chow form of the variety $\phi(V)$.

\begin{lemma}
  \label{lemma: chow form linear transform}
  Let $M \in \bK^{\br{n+1} \times \br{n+1}}$ be an invertible matrix and let $\phi : \bP^{n} \to \bP^{n}$ be the linear map given by $\phi(\valpha) = M \valpha$.
  Let $V \subseteq \bP^{n}$ be an equidimensional variety.
  If $C_{V}$ is the Chow form of $V$, then the Chow form of $\phi(V)$ is obtained by performing the substitution $u_{i, j} \to u'_{i, 0} M_{0, j} + u'_{i, 1} M_{1, j} + \cdots + u'_{i, n} M_{n, j}$ in $C_V$, and making the result monic.
\end{lemma}
\begin{proof}
  Let $\bK\bs{x_{0}, \dots, x_{n}}$ and $\bK\bs{y_{0}, \dots, y_{n}}$ denote the homogeneous coordinate rings of the source and target copies of $\bP^{n}$ in the definition of $\phi: \bP^{n} \to \bP^{n}$ respectively.
  Corresponding to $\phi$, there is a map $\phi^{*}: \bK\bs{y_{0}, \dots, y_{n}} \to \bK\bs{x_{0}, \dots, x_{n}}$ defined by $\phi^{*}(y_{i}) = M_{i, 0} x_{0} + \cdots + M_{i, n} x_{n}$ that satisfies $\phi^{-1}\br{\var{f(\vy)}} = \var{\phi^{*}(f)(\vx)}$.
  
  Now suppose $V$ has dimension $r$.
  Let $\ell_{0}, \dots, \ell_{r} \in \bK\bs{y_{0}, \dots, y_{n}}$ be linear forms and let $\beta_{i, j} \in \bK$ be such that $\ell_{i} = \sum \beta_{i, j} y_{j}$.
  The variety $\phi\br{V}$ intersects $\var{\ell_{0}, \dots, \ell_{r}}$ if and only if $V$ intersects $\phi^{-1}\br{\var{\ell_{0}, \dots, \ell_{r}}}$.
  By the previous discussion, we have $\phi^{-1}\br{\var{\ell_{0}, \dots, \ell_{r}}} = \var{\phi^{*}(\ell_{0}), \dots, \phi^{*}(\ell_{r})}$.
  By the definition of $\phi^{*}$, we have $\phi^{*}(\ell_{i}) = \sum_{j} \sum_{k} M_{k, j} \beta_{i, k} x_{j}$.
  By the definition of the Chow form $\cC_{V}$, this intersection is nonempty if and only if the evaluation of $\cC_{V}$ at $u_{i, j} = \sum_{k} M_{k, j} \beta_{i, k}$ is zero.
  If $\cC'$ is the polynomial obtained by substituting $u_{i, j} \to u'_{i, 0} M_{0, j} + u'_{i, 1} M_{1, j} + \cdots + u'_{i, n} M_{n, j}$, then the above evaluation of $\cC_{V}$ is zero if and only if $\cC'$ vanishes at $u'_{i, j} = \beta_{i, j}$.
  Finally, since $\cC_{V}$ can be obtained from $\cC'$ by a similar substitution using the inverse of the matrix $M$, and since $\cC_{V}$ is squarefree, we can deduce that $\cC'$ is squarefree.
  Therefore, $\cC'$, once made monic, is the Chow form for $\phi(V)$.
\end{proof}

\subsection{Geometric resolutions} \label{subsection: geometric res intro}

In this subsection, we work with affine varieties.
A \emph{geometric resolution} of an equidimensional affine variety is a certain parametric representation of the variety that was introduced in the computer algebra literature in work by Giusti, Heintz, Pardo and their collaborators~\cite{GiHeMoPa95,GiHeMoMoPa98}. Similar objects were used in prior work by Grigoriev and his collaborators~\cite{Grigoriev1986Complexity}, but the explicit requirement to be in Noether position (see below for the definition) and its consequences appear in the former references. For the convenience of the reader, we provide self contained proofs of all statements about geometric resolutions that we will need, including their existence.

Below, we work with equidimensional varieties of dimension $m$,
for which it will be convenient to divide the variables into two sets.
Let $S \coloneqq \bK\bs{x_{1}, \dots, x_{m}, y_{1}, \dots, y_{n}}$ be the coordinate ring of the affine space $\bK^{m+n}$ and let $\pi_{\vx} : \bK^{m + n} \to \bK^{m}$ be the projection onto the first $m$ coordinates.
We make the following assumption.

\begin{definition} \label{definition: projective noether position}
  We say that an equidimensional variety $V$ of dimension $m$ is in \emph{projective Noether position} if there exists a point $\valpha \in \bK^{m}$ such that the fiber $\pi^{-1}_{\vx}(\valpha) \cap V$ has cardinality exactly $\deg V$.
  We say that \emph{$y_{1}$ is a primitive element for $V$} if there is a point $\valpha$ such that the fiber $\pi^{-1}_{\vx}(\valpha) \cap V$ has cardinality exactly $\deg V$, and all points in this fiber have distinct $y_{1}$ coordinates.
\end{definition}

Note that if $y_{1}$ is a primitive element for $V$, then $V$ is in particular in projective Noether position.

\begin{definition}
  \label{definition: geometric resolution}
  Let $V \subset \bA^{m+n}$ be an equidimensional variety of dimension $m$ such that $y_{1}$ is a primitive element for $V$ (\cref{definition: projective noether position}).
  A \emph{geometric resolution} for $V$ is a set of polynomials $q, r_{2}, \dots, r_{n} \in \bK\bs{\vx, y_{1}}$ of degrees at most $\deg V$ with the following properties.
  \begin{itemize}[noitemsep]
    \item The polynomial $q$ is monic in $y_{1}$.
      Further, $q$ defines the projection of $V$ onto the first $m+1$ coordinates, that is $\pi_{\vx, y_{1}}(V) = \var{q}$.
    \item The discriminant $\delta$ of $q$ with respect to $y_{1}$ is nonzero.
    \item Let $q'$ denote the derivative of $q$ with respect to $y_{1}$.
      Let $J$ denote the ideal generated by $q, q'y_{2} - r_{2}, q'y_{3} - r_{3}, \dots, q' y_{n} - r_{n}$.
      Then $J : q'^{\infty} = I(V)$. \qedhere
  \end{itemize}
\end{definition}
If $y_{1}$ is not a primitive element for $V$, then we can apply a change of coordinates to ensure that $y_{1}$ is a primitive element in the new basis.
Consequently, a geometric resolution for an arbitrary variety usually consists of a change of coordinates, along with the above data.
For our application, we will not require this more general definition.
The rest of this section is devoted to showing that geometric resolutions exist, and along the way, showing some of their useful properties.

\textbf{
For the rest of this subsection, fix an equidimensional variety $V$ of dimension $m$ such that $y_{1}$ is a primitive element for $V$.
}

Before we begin our proof, we quickly remark on why a geometric resolution is a parametric representation for $V$.
Given a geometric resolution for $V$ as above, we have a map $\phi: \bA^{m+1} \setminus \var{q'} \to \bA^{m+n}$ defined by
\[
   \phi(a_{1}, \dots, a_{m}, b) = (a_{1}, \dots, a_{m}, b, r_{2}(\va, b) / q'(\va, b), \dots, r_{n}(\va, b) / q'(\va, b)).
\]
The conditions that $q, r_{2}, \dots, r_{n}$ are a geometric resolution imply that the image of $\var{q}$ under $\phi$ is a dense open subset of $V$.
Therefore, a geometric resolution allows us to parametrize a dense open subset of $V$ using a hypersurface.
Since we do not use this fact, we do not pursue this further.

Let $W_{1}, \ldots, W_{t}$ be the irreducible components of $V$.
Let $I$ denote the ideal of $V$, and $I_{j}$ denote the ideal of $W_{j}$ for each $j$.
Define $X_{j} \coloneqq \overline{\pi_{\vx}(W_{j})}$.

\begin{lemma}
  \label{lemma: component assumption}
  The element $y_{1}$ is a primitive element for each of $W_{1}, \dots, W_{t}$.
  Further, $X_{j} = \bK^{m}$ for all $j \in [t]$.
\end{lemma}
\begin{proof}
  Let $\valpha \in \bK^m$ be a point witnessing the fact that $y_{1}$ is primitive element for $V$.
  By assumption, the fiber $\pi_\vx^{-1}(\valpha) \cap V$ has cardinality exactly $\deg V = \sum_{j=1}^t \deg W_j$.
  This implies that each fiber $\pi_\vx^{-1}(\valpha) \cap W_j$ is finite and therefore of cardinality at most $\deg W_j$.
  Since
  \[
    \sum_{j=1}^t \deg W_j = \deg V = |\pi_\vx^{-1}(\valpha) \cap V| \le \sum_{j=1}^t |\pi_\vx^{-1}(\valpha) \cap W_j| \le \sum_{j=1}^t \deg W_j,
  \]
  it follows that the fiber $\pi_{\vx}^{-1}(\valpha) \cap W_j$ has cardinality precisely $\deg W_j$, so $W_j$ is in projective Noether position.
  Since the points in the fiber $\pi_{\vx}^{-1}(\valpha) \cap V$ have distinct $y_{1}$ coordinates, so do the points in each of the fibers $\pi_{\vx}^{-1}(\valpha) \cap W_{j}$, which shows that $y_{1}$ is primitive for each $W_{j}$.

  The fact that $X_j = \bK^m$ follows from the theorem on the dimension of fibers \cite[Chapter~1, Theorem~1.25]{shafarevich1994basic} as follows:
  Since the map $\pi_{\vx} : W_j \to X_j$ is dominant and has a fiber of finite cardinality, the image $X_j$ must have dimension $m$.
  Since $X_j \subseteq \bK^m$, it follows that $X_j = \bK^m$.
\end{proof}

Since $y_{1}$ is a primitive element for $W_{1}, \dots, W_{t}$, it also follows that each $W_{i}$ is in projective Noether position (\cref{definition: geometric resolution}).
From \cite[Proposition~1]{Heintz83} we can deduce that the set of points $\valpha \in \bK^m$ that witness the fact that $V$ is in projective Noether position (and likewise for each $W_i$) is a nonempty open subset of $\bK^{m}$.

Let $u_{1}, \dots, u_{n}$ be new indeterminates and let $\bL$ denote the algebraic closure of the field $\bK\br{u_{1}, \dots, u_{n}}$.
Let $I'$ and $I'_{j}$ denote the extensions of the ideals $I$ and $I_{j}$ to $\bL\bs{\vx, \vy}$.
Let $V'$ and $W'_{j}$ denote the zeroes of these ideals in $\bL^{m+n}$.
From \cref{lemma: extension of scalars} we can deduce that $I'_{j}$ is prime for all $j$ and that $I' = \bigcap_j I_{j}'$.
In particular, the varieties $W_{j}'$ are irreducible and $V'$ is their union.
Further, $W_{j}'$ is a variety of dimension $m$ and degree $\deg W_{j}$.
Since $y_{1}$ is a primitive element for $V$ (\cref{definition: projective noether position}), it is also a primitive element for $V'$ and $W'_{j}$, and any witness for the former is a witness for the latter.
By the previous lemma, the maps $\pi_{\vx} : W'_{j} \to \bL^{m}$ are still dominant.

Let $\ell$ denote the linear form $u_{1} y_{1} + \cdots + u_{n} y_{n}$.
Let $\phi: \bL^{m+n} \to \bL^{m+1}$ be defined by $(\vp, \vq) \mapsto (\vp, \ell(\vq))$.
The following lemma captures some basic properties of the polynomial that defines the image of each irreducible component of $V'$ under this map.

\begin{lemma}
  \label{lemma: component image base field}
  For each $j$, the close of the image $\overline{\phi(W'_{j})}$ is an irreducible hypersurface that is defined by an irreducible polynomial $Q_{j} \in \bK\bs{\vu, \vx, T}$ of degree at most $\deg W_{j}$ in $\vx$ and $T$.
\end{lemma}
\begin{proof}
  Since $W'_{j}$ is irreducible and $\phi$ is continuous, $\phi(W'_{j})$ is irreducible.
  The projection map $\pi_{\vx}$ onto the $\vx$ coordinates factors through $\phi$.
  Combined with the theorem on the dimension of fibers, we deduce that $\dim W'_{j} \geq \dim \phi(W'_{j}) \geq \dim \pi_{\vx}(W'_{j})$.
  Since $\dim W'_{j} = \dim \pi_{\vx}(W'_{j})$, we can deduce $\dim \phi(W'_{j}) = m$.
  This shows that the closure of the image is an irreducible hypersurface.

  Let $T := \sum u_{i} y_{i}$.
  After a change of basis, we can treat $x_{1}, \dots, x_{m}, T, y_{2}, \dots, y_{n}$ as the set of variables in our polynomial ring.
  Since $I'_{j}$ has generators in $\bK\bs{\vx, \vy}$, it admits generators in $\bK\br{\vu}\bs{\vx, \vy}$ in the new coordinates, performing the change of variables $y_1 \gets (T-u_2 y_2 - \cdots - u_n y_n)/u_1$. 
A defining equation for $\overline{\phi(W'_{j})}$ is then obtained by starting from these new equations and constructing a Gröbner basis with any lexicographic order with $y_{n}, \dots, y_{2}$ occurring before $x_{1}, \dots, x_{m}, T$, and eliminating $y_{2}, \dots, y_{n}$.
 
Since the construction of the Gröbner basis only involves operations with $\bK\br{\vu}$-coefficients, the resulting Gröbner basis and elimination ideal have generators in $\bK\br{\vu}\bs{\vx, T}$.
  After clearing denominators and factoring out any content, we can assume that the image is defined by an irreducible polynomial $Q_{j} \in \bK\bs{\vu, \vx, T}$ as claimed.
  The degree of $Q_{j}$ in $\vx, T$ is exactly the degree of $\overline{\phi(W'_{j})}$, which is at most $\deg W_{j}$ by \cite[Lemma~2]{Heintz83}.
\end{proof}

The following lemma further refines the properties of the polynomial $Q_{j}$ defined above.

\begin{lemma}
  \label{lemma: component image monic}
  Suppose $Q_{j} = p_{j, 0} + p_{j, 1}T + \cdots + p_{j, d_{j}}T^{d_{j}}$ with $p_{j, k} \in \bK\bs{\vu, \vx}$, $p_{j, d_{j}}$ nonzero, and $d_{j} \leq \deg W_{j}$.
  Then $d_{j} = \deg W_{j}$ and $p_{j, d_{j}} \in \bK$. 
\end{lemma}
\begin{proof}
  Pick $\valpha \in \bK^{m}$ that is witness to $W_{j}$ being in projective Noether position (\cref{definition: projective noether position}), such that $p_{j, d_{j}}$ does not vanish when evaluated at $\valpha$.
  Since the set of all such witnesses is a nonempty open set, and since the set of points in $\bK^{m}$ where $p_{j, d_{j}}$ does not vanish is also a nonempty open set, such an $\valpha$ exists.
  This choice of $\valpha$ ensures that $Q_{j}(\vu, \valpha, T)$ is nonzero.

  The fiber $\pi^{-1}_{\vx}(\valpha) \cap W_{j}$ has $\deg W_{j}$ points, say $(\valpha, \vbeta_{1}), \dots, (\valpha, \vbeta_{\deg W_{j}})$, with coordinates in $\bK$.
  Considered as points in $\bL^{m+n}$, these points have distinct images $(\valpha, t_{1}), \dots, (\valpha, t_{\deg W_{j}})$ under $\phi$. %
  The polynomial $Q_{j}$ vanishes on each $(\valpha, t_{i})$, since $(\valpha, \beta_{i}) \in W_{j}'$.
  Therefore the nonzero polynomial $Q_{j}(\vu, \valpha, T)$ has at least $\deg W_{j}$ many roots as a polynomial in $\bK\br{\vu}\bs{T}$, hence its degree in $T$ is at least $\deg W_{j}$.
  This gives $d_{j} = \deg W_{j}$.
  Since the total degree of $Q_{j}$ in $\vx$ and $T$ is at most $\deg W_{j}$, we can also deduce $p_{j, d_{j}} \in \bK\bs{\vu}$.

  Now we show that $p_{j, d_{j}}$ in fact lies in $\bK$.
  Consider an irreducible factor of $p_{j, d_{j}}$, say $f(\vu)$.
  Since $Q_{j}$ is irreducible in $\bK\bs{\vu, \vx, T}$, there is at least one other $p_{j, i}$ such that $f$ does not divide $p_{j, i}$.
  Therefore, there is an open set of points $\valpha \in \bK^{m}$ such that $f$ does not divide $p_{j, i}(\vu, \valpha)$.
  Since each of these conditions are open, we can find a point $\valpha \in \bK^{m}$ that is a witness to $W_{j}$ being in projective Noether position, and such that for each factor $f$ of $p_{j, d_{j}}$, the polynomial $Q_{j}(\vu, \valpha, T)$ is not divisible by $f$.
  In particular, for any such $\valpha$, the polynomial $Q_{j}(\vu, \valpha, T)$ does not have an irreducible factor that depends only on $\vu$.
  Fix such an $\valpha$, and consider again the images of the fiber of $\valpha$ under $\phi$.
  Let these be denoted $(\valpha, s_{1}), \dots, (\valpha, s_{\deg W_{j}})$.
  We have the factorization $Q_{j}(\vu, \valpha, T) = p_{j, d_{j}} \prod_{k \leq d_{j}} \br{T - s_{i}}$.
  This, combined with the choice of $\valpha$, implies that $p_{j, d_{j}}$ must lie in $\bK$.
\end{proof}

We normalize $Q_{j}$ such that $p_{j, d_{j}} = 1$ for all $j$.
Pick $\valpha$ such that the fibers $\pi^{-1}_{\vx}(\valpha) \cap W_{i}$ and $\pi^{-1}_{\vx}(\valpha) \cap W_{j}$ are disjoint.
The polynomials $Q_{i}(\vu, \valpha, T)$ and $Q_{j}(\vu, \valpha, T)$ are coprime.
Therefore we can deduce that $Q_{i}$ and $Q_{j}$ are themselves coprime for all pairs $Q_{i}, Q_{j}$.
The variety $\overline{\phi(V')}$ is the union of $\overline{\phi(W'_{j})}$ over all $j$, therefore it is a hypersurface in $\bL^{m+1}$ that is defined by the polynomial $Q = Q_{1} \cdots Q_{t} \in \bK\bs{\vu, \vx, T}$.
The polynomial $Q$ has total degree $\deg V$ in $\vx$ and $T$ and is monic in $T$.
We say that the polynomial $Q$ defines the general projection of the variety $V$.

Suppose now that $\gamma_{1}, \dots, \gamma_{n} \in \bK$ are elements, and consider the map $\bA^{m+n} \to \bA^{m+1}$ defined by $(\vp, \vq) \to (\vp, \sum_{i=1}^{n} \gamma_{i} q_{i})$.
The following lemma states that the specialization of the general projection $Q$ at $\vu = \vgamma$ vanishes on the image of the above linear map.
Since $Q$ is monic, we are guaranteed that this specialization is nonzero.
Consequently, we can also deduce that the map $\pi_{\vx}$ is a finite map when restricted to $V$.

\begin{lemma}
  \label{lemma: generic projection specialisation}
  For every $\vmu \in \bK^{n}$, the polynomial $Q(\vmu, \vx, \ell(\vmu))$ lies in $I$.
  Consequently, $\ell(\vmu)$ is integral over $\bK\bs{\vx}$ in the coordinate ring of $V$.
  In particular, the map $\pi_{\vx}: V \to \bK^{m}$ is finite.
  For every $\valpha$, and for every $(\valpha, \vbeta)$ in the fiber $\pi^{-1}_{\vx}(\valpha) \cap W$, the element $\ell(\vmu, \vbeta)$ is a root of $Q(\vmu, \valpha, T)$.
\end{lemma}
\begin{proof}
  The polynomial $Q(\vu, \vx, \ell)$ lies in $I'$.
  As in the proof of \cref{lemma: extension of scalars}, since $I'$ is the extension of an ideal in $\bK\bs{\vx, \vy}$, every coefficient of $Q(\vu, \vx, \ell)$ written as a polynomial in $\vu$ belongs to the ideal $I$.
  Therefore we can write $Q(\vu, \vx, \ell) = \sum a_{i} f_{i}$ where $a_{i} \in \bK\bs{\vu}$ and $f_{i} \in I$.
  Evaluating this expression at any $\vmu$ shows $Q\br{\vmu, \vx, \ell(\vmu)} \in I$.
  The polynomial $Q(\vmu, \vx, T)$ is a monic equation satisfied by $\ell(\vmu)$, which shows that $\ell(\vmu)$ is integral over $\bK\bs{\vx}$ in the coordinate ring of $V$.
  Invoking this for different choices of $\vmu$ we can deduce that $y_{i}$ is integral over $\bK^{m}$ in the coordinate ring of $V$ for all $i$, which shows that map $\pi_{\vx} : V \to \bK^{m}$ is finite.
  The final statement follows as in the proof of \cref{lemma: component image monic}.
\end{proof}

Let $\vnu_{0} \in \bK^{n}$ denote the point $(1, 0, \dots, 0)$.
Set
\begin{equation} \label{equation: geometric resolution from projection}
  q \coloneqq Q(\vnu_{0}, \vx, y_{1}), \qquad r_{j} \coloneqq - \frac{\partial Q}{\partial u_{j}} \br{\vnu_{0}, \vx, y_{1}}, \quad j = 1, \dots, n.
\end{equation}
These are polynomials in $\bK\bs{\vx, y_{1}}$ of degree at most $\deg V$.
Let $\delta \in \bK\bs{\vx}$ and $q' \in \bK\bs{\vx,y_1}$ denote the discriminant and derivative, respectively, of $q$ with respect to $y_{1}$.
In the rest of this section, we prove that $q, r_{2}, \dots, r_{n}$ are a geometric resolution of $V$.
We break the proof into a series of smaller lemmas.

\begin{lemma}
  \label{lemma: squarefree image}
  The discriminant $\delta$ is nonzero.
  Equivalently, the polynomial $q$ has no multiple factor.
\end{lemma}
\begin{proof}
  Let $\valpha$ be a witness to $y_{1}$ being a primitive element for $V$.
  The proof of \cref{lemma: component image monic} shows that $q(\valpha, y_{1}) = Q(\vnu_{0}, \valpha, y_{1})$ factors as $\prod_{j \leq \deg V} (y_{1} - \beta_{j})$, where $\beta_{1}, \dots, \beta_{\deg V}$ are the distinct $y_{1}$-coordinates of the fiber above $\valpha$.
  This shows that $q'(\valpha) \neq 0$, and since $q$ is monic in $y_{1}$ this further shows that $\delta \neq 0$.
\end{proof}

\begin{lemma}
  \label{lemma: shape basis in ideal}
  The polynomials $q, q'y_{2} - r_{2}, \dots, q'y_{n} - r_{n}$ are in the ideal $I$.
\end{lemma}
\begin{proof}
  As in the proof of \cref{lemma: generic projection specialisation}, we can write $Q(\vu, \vx, \ell) = \sum_i a_{i} f_{i}$, for $a_{i} \in \bK\bs{\vu}$ and $f_{i} \in I$.
  Evaluating this expression at $\vnu_{0}$ shows that $q \in I$.
  If we take the partial derivative of this expression with respect to $u_{j}$ we obtain
  \[
    \frac{\partial Q}{\partial T} \br{\vu, \vx, \ell} y_{j} + \frac{\partial Q}{\partial u_{j}} \br{\vu, \vx, \ell} = \sum_i \frac{\partial a_{i}}{\partial u_{j}} f_{i}.
  \]
  Evaluating this expression at $\vnu_{0}$ shows that $q'y_{j} - r_{j} \in I$ for all $j \geq 1$.
\end{proof}

Let $J \subset \bK\bs{\vx, \vy}$ be the ideal generated by $q, q'y_{2} - r_{2}, \dots, q'y_{n} - r_{n}$.
Since $J \subset I$, we have $\var{J} \supset V$.
The next two lemmas will show that the excess components, that is, those components of $\var{J}$ that are not components of $V$, are exactly those contained in $\var{q'}$.

\begin{lemma}
  \label{lemma: first coordinate projection}
  The polynomial $q$ defines the projection of $V$ onto the first $m+1$ coordinates, equivalently $\pi_{\vx, y_{1}}(V) = \var{q}$.
\end{lemma}
\begin{proof}
  The map $\pi_{\vx}: V \to \bK^{m}$ is finite.
  If $Y$ is the Zariski closure of $\pi_{\vx, y_{1}}(V)$ then the map $\pi_{\vx, y_{1}} : V \to Y$ is also finite, hence surjective.
  Therefore, $Y$ is a hypersurface of degree at most $\deg V$.
  Suppose this hypersurface is defined by $P \in \bK\bs{\vx, y_{1}}$.
  The total degree of $P$ is at most $\deg V$.
  For any $\valpha$ that is a witness to $y_{1}$ being a primitive element for $V$, the polynomial $P(\valpha, y_{1})$ has $\deg V$ distinct roots.
  Therefore $P$ has degree $\deg V$ in $y_{1}$, and we can scale by an element of $\bK$ to make it monic in $y_{1}$.
  
  Since $q \in I$, and since $q$ vanishes on $Y$, the polynomial $P$ must divide $q$ in $\bK\bs{\vx, y}$.
  Since $q$ and $P$ are both monic in $y_{1}$, and since $q$ has degree at most $\deg V$, we must have $P = q$.
\end{proof}

Note that the irreducible components of $V$ are in bijection with the irreducible factors of the polynomials $q$.

\begin{lemma}
  \label{lemma: components of triangular basis}
  The following equality of ideals holds: $J \colon q'^{\infty} = I$.
\end{lemma}

\begin{proof}
  First suppose $f \in J \colon q'^{\infty}$.
  Then $q'^{k} f \in J$ for some $k > 0$.
  Since $q$ and $q'^{k}$ are coprime in $\bK\br{\vx}\bs{y_{1}}$, there exists a polynomial $\delta' \in \bK\bs{\vx}$ and polynomials $A, B \in \bK\bs{\vx, y_{1}}$ such that $A q  + B q'^{k} = \delta'$.
  Multiplying this equation by $f$ and using the fact that $J \subset I$ we deduce $\delta' f \in I$.
  No minimal prime of $I$ contains any element of $\bK\bs{\vx}$, since $\pi_{\vx}(W_{j}) = \bK^{m}$ for all $j$.
  Therefore, every minimal prime, and hence $I$ itself, contains $f$.

  Conversely, suppose $f \in I$.
  Define $\tilde{f} := f(\vx, y_{1}, r_{2} / q', \dots, r_{n} / q') \in \bK\bs{\vx, y_{1}}_{q'}$ and write $\tilde{f} = F / q'^{t}$ with $F \in \bK\bs{\vx, y_{1}}$.
  Pick a point $(\valpha, \beta_{1}) \in \pi_{\vx, y_{1}}(V)$ such that $\delta(\valpha) \neq 0$.
  There exists $(\beta_{2}, \dots, \beta_{n})$ such that $(\valpha, \vbeta) \in V$.
  Since the polynomials $q' y_{j} - r_{j}$ vanish on $V$ and since $q'(\valpha) \neq 0$, we can deduce that $\beta_{j} = -r_{j}(\valpha) / q'(\valpha)$ for all $j \geq 2$.
  We also have $f(\valpha, \vbeta) = 0$.
  Therefore we can deduce $F(\valpha, \beta_{1}) = 0$.

  We have shown that $F$ vanishes on every point in $\var{q} \setminus \var{q'}$.
  No component of $\var{q}$ is contained in $\var{q'}$, therefore $F$ must vanish on $\var{q}$.
  Equivalently, $q$ divides $F$.
  To summarize, in the ring $\bK\bs{\vx, y_{1}}_{q'}$ we have shown that $f - cq \in \ideal{q'y_{2} - r_{2}, \dots, q'y_{n} - r_{n}}$ for some $c \in \bK\bs{\vx, y_{1}}_{q'}$.
  Writing $f = cq + \sum_{j \geq 2} c_{j}  \br{q'y_{2} - r_{2}}$ for $c_{j} \in \bK\bs{\vx, y_{1}}_{q'}$, and multiplying by a high enough power of $q'$ shows that $f \in J \colon q'^{\infty}$.
\end{proof}

\subsection{The relationship between Chow forms and geometric resolutions}
\label{subsection: chow geom relation}

In this section, we show how Chow forms and geometric resolutions are related.
We continue to use the same notation as in \cref{subsection: geometric res intro}.
First we show how the polynomial defining the general projection, denoted $Q$, can be computed given the Chow form of the projective closure of an equidimensional variety.
As a consequence, the geometric resolution can also be computed from the Chow form (as long as $y_{1}$ is a primitive element).

A similar procedure for the same task appears in \cite[Section~3]{jeronimo2004computational}. 

\begin{lemma}
  \label{lemma: chow form to resolution}
  Let $V \subset \bK^{m+n}$ be an equidimensional variety such that $y_{1}$ is a primitive element for $V$.
  Let $\overline{V} \subset \bP^{m+n}$ denote its projective closure (say we homogenize by adding a new variable $z$) and let $C_{\overline{V}}$ be the Chow form of $\overline{V}$.
  Let $\ell_{i}$ denote the linear form $w_{i} z - x_{i} \in \bK\bs{\vw}\bs{z, \vx, \vy}$ for $i = 1, \dots, m$.
  Let $\ell_{m+1}$ denote the linear form $T z - \sum_{j=1}^{n} u_{i} y_{i} \in \bK\bs{T, \vu}\bs{z, \vx, \vy}$.
  Then $C_{\overline{V}}\br{\coeff{\ell_{1}}, \dots, \coeff{\ell_{m+1}}} = Q(\vu, \vw, T)$.
\end{lemma}
\begin{proof}
  Let $\valpha$ be a witness to the fact that $V$ is in projective Noether position.
  Consider the linear forms $x_{1} - \alpha_{1} z, \dots, x_{m} - \alpha_{m} z$.
  The variety $\overline{V} \cap \var{x_{1} - \alpha_{1} z, \dots, x_{m} - \alpha_{m} z}$ contains $\deg V$ points on the chart $z \neq 0$.
  The projective Bézout's theorem shows that this intersection has degree at most $\deg V$, therefore it cannot contain any other points.
  In the rest of this proof, $\vbeta_{1}, \dots, \vbeta_{\deg V} \in \bK^{n}$ will be such that $(1, \valpha, \vbeta_{i})$ are the points in $\overline{V} \cap \var{x_{1} - \alpha_{1} z, \dots, x_{m} - \alpha_{m} z}$.

  Define $C' := C_{\overline{V}}\br{\coeff{\ell_{1}}, \dots, \coeff{\ell_{m+1}}}$.
  We now show that $C' \neq 0$.
  In $C'$, set $w_{1} = \alpha_{1}, \dots, w_{m} = \alpha_{m}$.
  Set $\vu$ to an element $\vgamma \in \bK^{n}$ such that $\sum_{j} \gamma_{j} \beta_{i, j}$ are distinct for all $i$.
  Let this resulting polynomial be denoted by $C''(T)$.
  Then $C''$ is nonzero at any $\tau$ that is not of the form $\sum_{j} \gamma_{j} \beta_{i, j}$ for some $i$.
  Indeed by the first step above, $\overline{V} \cap \var{x_{1} - \alpha_{1} z, \dots, x_{m} - \alpha_{m} z}$ only contains finitely many points, and these choices of $\tau$ and $\vgamma$ ensure that the hyperplane $z \tau - \sum_{j} \gamma_{j} y_{j}$ avoids these points.
  Since the Chow form is a homogeneous polynomial of degree $\deg V$ in each set of variables, we can deduce that $C''$ is monic in $T$ of degree $\deg V$.
  From this, we can deduce that $C'$ is nonzero, and also that $C'$ has degree $\deg V$ in $T$.

  Next we show $Q(\vu, \vw, T)$ divides $C'$.
  For a general choice of $\valpha' \in \bK^{m}$ and a general choice of $\vgamma' \in \bK^{n}$, both $Q(\vgamma', \valpha', T)$ and $C'(\vgamma', \valpha', T)$ have the same roots.
  Indeed for a general choice, the fiber $\pi^{-1}_{\vx}(\valpha')$ is finite, and the roots of both $Q(\vgamma', \valpha', T)$ and $C('\vgamma, \valpha', T)$ are $\sum_i \gamma'_{i} \beta'_{i}$ where $(\valpha, \vbeta')$ is a point in the fiber.
  Along with the statement on the degrees, this shows that $C' = Q \cdot P$ for some polynomial $P \in \bK\bs{\vw}$.
  We show $P \in \bK$, which will conclude the proof.
  To this end, assume that $P \not\in \bK$ and let $\vmu$ be a zero of $P$.
  The polynomial $C'(\vu, \vmu, T)$ is zero.
  By \cref{lemma: generic projection specialisation}, the map $\pi_{\vx} : V \to \bK^m$ is finite, so the fiber $\pi_{\vx}^{-1}(\vmu) \cap V$ consists of a finite number of points.
  Further, these are the only points in $\overline{V} \cap \var{x_{1} - \mu_{1} z, \dots, x_{m} - \mu_{m} z}$, since
  \[
    \overline{V} \cap \var{x_{1} - \mu_{1} z, \dots, x_{m} - \mu_{m} z} \cap \var{z} = \overline{V} \cap \var{x_{1} - \alpha_{1} z, \dots, x_{m} - \alpha_{m} z} \cap \var{z} = \emptyset.
  \]
  We can therefore find a $\tau$ and $\vgamma$ such that $\tau z - \sum_i y_{i} \gamma_{i}$ is nonzero at all of these points.
  Then the intersection of $\overline{V}$ with $\var{x_{1} - \mu_{1} z, \dots, x_{m} - \mu_{m}z, \tau z - \sum y_{i} \gamma_{i}}$ is empty, so $C'(\vgamma, \vmu, \tau) \neq 0$, which contradicts the fact that $C'(\vu, \vmu, T)$ is zero.
\end{proof}

That $V$ is in projective Noether position is crucial in the proof of \cref{lemma: chow form to resolution}.
Without this condition, the specialization of the Chow form vanishes identically.

Finally, we show how a geometric resolution of an affine variety can be used to compute which of its components lie on another variety given by a Chow form.
\begin{lemma}
  \label{lemma: resolution prune}
  Let $V \subset \bK^{m+n}$ be an $m$-equidimensional variety such that $y_{1}$ is a primitive element for $V$.
  Let $q, r_{2}, \dots, r_{n} \in \bK[\bm x,y_1]$ be a geometric resolution for $V$ as above.
  Let $W \subset \bP^{m+n}$ be another equidimensional variety of dimension $b$, and suppose $C_{W}$ is the Chow form of $W$, in the variables $u_{0, 0}, \dots, u_{0, m+n}, u_{1, 0}, \dots, u_{b, m+n}$.
  
  Consider new variables $w_{i, j}$ (for $0 \leq i \leq b$ and $1 \leq j \leq n+m$), $t_j$ (for $1 \le j \le m$) and $s$, and for $i = 0, \dots, b$ let $g_{i}$ denote the following polynomial, which is linear in $\vx, \vy, z$:
  \[
    g_{i} \coloneqq \sum_{j = 1}^{m} w_{i, j} \br{x_{j} - t_{j} z} + w_{i, m+1} \br{y_{1} - s z} + \sum_{j'=2}^{n} w_{i, m+j'} \br{y_{j'} + \frac{r_{j'}(\vt, s)}{q'(\vt, s)}z}.
  \]
  Let $\frac{P(\vt, s, \vw)}{q'(\vt, s)^{c}} = C_{W}\br{\coeff{g_{0}}, \dots, \coeff{g_{b}}}$, expressed so that $P \in \bK\bs{\vt, s, \vw}$ and so that $P$ and $q'$ are relatively prime.
  Let $P' \in \bK\bs{\vt, s}$ denote the GCD of the coefficients of $P$ as a polynomial in $\vw$.
  Then $\gcd\br{q(\vt, s), P'(\vt, s)}$ consists of those factors of $q$ that correspond to components of $V$ that lie in $W$. 
\end{lemma}

\begin{proof}
  Let $q_{1}$ denote an irreducible factor of $q$, and let $V_{1}$ be the corresponding component of $V$.

  Suppose first that $V_{1} \subset W$.
  We show $q_{1}(\vt, s)$ divides $P'$.
  To this end, let $(\valpha, \beta_{1}) \in \bK^{m+1}$ be a point in $\var{q_{1}} \setminus \var{q'}$.
  There exists $\beta_{2}, \dots, \beta_{n}$ such that $(\valpha, \vbeta) \in V_{1}$ (because the restriction of $\pi_{\bm x,y_1}$ to $V_1$ is closed).
  Further $\beta_{i} = -r_{i}(\valpha, \beta_{1}) / q'(\valpha, \beta_{1})$ for all $i \geq 2$.
  After substituting $\valpha$ and $\beta_{1}$ for $\vt$ and $s$, each of the linear forms $x_{1} - \alpha_{1} z, x_{2} - \alpha_{2} z, \dots, x_{m} - \alpha_{m} z, y_{1} - \beta_{1} z, y_{2} + \frac{r_{2}(\valpha, \beta_{1})}{q'(\valpha, \beta_{1})} z, \dots, y_{n} + \frac{r_{n}(\valpha, \beta_{1})}{q'(\valpha, \beta_{1})} z$ vanishes at $(1, \valpha, \vbeta) \in \bP^{m+n}$.
  Therefore $g_{0}, \dots, g_{b}$ also vanish at this point.
  The evaluation $C_{W}\br{\coeff{g_{0}}, \dots, \coeff{g_{b}}}$ must also vanish, which implies $P(\valpha, \beta_{1}, \vw) = 0$.
  Since this happens for every $(\valpha, \beta_{1}) \in \var{q_{1}} \setminus \var{q'}$, and since $q'$ and $q_{1}$ are relatively prime, we can deduce that $q_{1}(\vt, s)$ divides every coefficient of $P$ as a polynomial in $\vw$, and specifically that $q_{1}(\vt, s)$ divides $P'(\vt, s)$.

  Now suppose $V_{1} \not\subset W$.
  Pick a point $(\valpha, \beta_{1}) \in \var{q_{1}} \setminus \var{q'}$ such that the point $(\valpha, \vbeta) \in V$ above $(\valpha, \beta_{1})$ lies in $V_{1} \setminus W$.
  Such a point exists since $W$ has codimension at least $1$ inside $V_{1}$.
  As in the previous case, after substituting $\valpha$ and $\beta_{1}$ for $\vt$ and $s$, each of the linear forms $x_{1} - \alpha_{1} z, x_{2} - \alpha_{2} z, \dots, x_{m} - \alpha_{m} z, y_{1} - \beta_{1} z, y_{2} + \frac{r_{2}(\valpha, \beta_{1})}{q'(\valpha, \beta_{1})} z, \dots, y_{n} + \frac{r_{n}(\valpha, \beta_{1})}{q'(\valpha, \beta_{1})} z$ vanishes at $(1, \valpha, \vbeta) \in \bP^{m+n}$.
  In fact this point is exactly the vanishing set of this collection of linear forms.
  Since this point is not in $W$, we can find a specialization $\vmu$ of the variables $\vw$ such that $W \cap \var{g_{0}(\vmu), \dots, g_{b}(\vmu)} = \emptyset$.
  Thus $P(\valpha, \beta_{1}, \vmu) \neq 0$.
  Consequently, $q_{1}(\vt, s)$ does not divide $P'(\vt, s)$.
\end{proof}

\begin{remark} \label{remark: resolution prune randomness}
  When using \cref{lemma: resolution prune} to perform computation, we can directly use random linear combinations instead of general linear combinations when defining the polynomials $g_{0}, \dots, g_{b}$.
  In this case, $\gcd\br{q(\vt, s), P(\vt, s)}$ is the factor of $q$ with the claimed properties, with high probability.
  Indeed as long as the coefficients in the random linear combinations are picked from a set $\cB \subset \bK$ of size $\poly\br{n + m, \deg V, \deg W}$, then the required property fails only with probability $1 / \poly\br{n+ m, \deg V, \deg W}$.
  This follows from the above proof by considering the degree of $P$ as a polynomial in $\vw$, and also considering the size of $\cB$ to ensure that a specialisation $\vmu$ with the claimed properties can be found in the second step.
\end{remark}

\section{Computing Chow forms of equidimensional components} \label{section: chow form computation}
In this section, we start by showing how resultant computations can be used to compute the Chow form of each equidimensional component of a given projective variety.
We then show how this can be used to solve the same problem in the affine case

The construction of the Chow form for equidimensional components proceeds in two steps.
In the first step, from the defining equations of a projective variety $V$, we construct the Chow forms for certain subvarieties $V_{k} \subseteq V$.
These are equidimensional subvarieties of $V$ that are obtained via a deformation process.
This step generalizes the construction of \textcite[Section~2]{Ierardi89}, where only the isolated points were considered.
The second step involves pruning the Chow forms of the varieties $V_{k}$ to obtain Chow forms for the equidimensional components.
This step uses geometric resolutions and the results in \cref{subsection: chow geom relation}.

\subsection{Chow form of limit subvarieties}
\label{subsection: limit subvarieties}

Given a sequence of forms $G_{1}, \dots, G_{k} \in \bF\bs{x_{0}, \dots, x_{n}}$ with $k \leq n$, and $d_{i} \coloneqq \deg G_{i}$, we use $\perturbedvar{G_{1}, \dots, G_{k}}$ to denote the variety in $\bP^{n} \times \bA^{1}$ defined by $(1-t) G_{1} + t x_{1}^{d_{1}}, \dots, (1 - t) G_{k} + t x_{k}^{d_{k}}$.
Since projective varieties are proper, the projection to $\bA^{1}$ of each irreducible component of $\perturbedvar{G_{1}, \dots, G_{k}}$ is an irreducible closed set \cite[Chapter~1, Theorem~1.11]{shafarevich1994basic}.
Therefore, the projection is either all of $\bA^{1}$ or a single point.
Let $W_{1}, \dots, W_{a}$ be those irreducible components of $\perturbedvar{G_{1}, \dots, G_{k}}$ whose projection to $\bA^{1}$ is surjective.
We use $\paramvar{G_{1}, \dots, G_{k}}$ to denote $\bigcup_{i=1}^{a} W_{i}$.

The intersection of $\bP^{n} \times \bA^{1}$ with the hyperplane defined by $t = 0$ can be naturally identified with $\bP^{n}$.
The image of the intersection of $\paramvar{G_{1}, \dots, G_{k}}$ with the hyperplane defined by $t = 0$ under this identification is a subvariety of $\bP^{n}$, which we denote by $\limitvar{G_{1}, \dots, G_{k}}$.
More generally, for $\tau \in \bF$, we write $\limitvart{G_{1}, \dots, G_{k}}{\tau}$ for the subvariety of $\bP^{n}$ corresponding to the intersection of $\paramvar{G_{1}, \dots, G_{k}}$ with the hyperplane defined by $t = \tau$.

The following lemma establishes some basic structural facts about the varieties $\perturbedvar{G_1, \ldots, G_k}$ and $\limitvart{G_1, \ldots, G_k}{\tau}$.
In particular, we compute the dimensions of the varieties $W_1, \ldots, W_a$ and $\limitvart{G_1, \ldots, G_k}{\tau}$, and show how the variety $\limitvar{G_1, \ldots, G_k}$ is related to $\var{G_1, \ldots, G_k}$.
In the rest of this subsection, we use $\pi_{t}$ and $\pi_{\vx}$ to denote the projection maps $\bP^{n} \times \bA^{1} \to \bA^{1}$ and $\bP^{n} \times \bA^{1} \to \bP^{n}$, respectively.

\begin{lemma}
  \label{lemma: structure of limit variety}
  Every component of $\perturbedvar{G_{1}, \dots, G_{k}}$ has codimension at most $k$ and the components $W_{1}, \dots, W_{a}$ have codimension exactly $k$.
  For every $\tau \in \bF$, the variety $\limitvart{G_{1}, \dots, G_{k}}{\tau}$ is an equidimensional variety of codimension $k$.
  Further,
  \[ 
    \limitvar{G_{1}, \dots, G_{k}} \subseteq \var{G_{1}, \dots, G_{k}},
  \]
  and every codimension-$k$ component of the latter is an irreducible component of the former.
\end{lemma}
\begin{proof}
  That every component of $\perturbedvar{G_{1}, \dots, G_{k}}$ has codimension at most $k$ follows from the fact that $\perturbedvar{G_{1}, \dots, G_{k}}$ is defined by $k$ equations.
  Consider now the component $W_{1}$.
  Since $\pi_{t}\br{W_{1}}$ is surjective, $W_{1} \cap \var{t-1} \neq \emptyset$ and this intersection has codimension $1$ within $W_{1}$.
  Further, $W_{1} \cap \var{t-1} \subseteq \perturbedvar{G_{1}, \dots, G_{k}} \cap \var{t-1} = \var{t-1, x_{1}^{d_{1}}, \dots, x_{k}^{d_{k}}}$, and the latter has codimension $k+1$ in $\bP^{n} \times \bA^{1}$.
  From this we get
  \[
    k+1 \geq \codim{W_{1}} + 1 = \codim{W_{1} \cap \var{t-1}} \geq k+1,
  \]
  which shows that $\codim{W_{1}} = k$ in $\bP^{n} \times \bA^{1}$.

  Since $W_{1}$ is not contained in any hyperplane defined by $t = \tau$, the intersection of $W_{1}$ with the hyperplane defined by $t = \tau$ for any $\tau$ has codimension $k+1$ in $\bP^{n} \times \bA^{1}$.
  The same holds for the components $W_{2}, \dots, W_{a}$.
  This proves that $\limitvart{G_{1}, \dots, G_{k}}{\tau}$ is an equidimensional variety of codimension $k$ for every $\tau$.

  The inclusion $\limitvar{G_{1}, \dots, G_{k}} \subseteq \var{G_{1}, \dots, G_{k}}$ follows by construction.
  Let $Y$ be an irreducible component of the latter of codimension $k$.
  Since $Y \times \bc{0} \subseteq \perturbedvar{G_{1}, \dots, G_{k}}$, there must exist a component $Z$ of $\perturbedvar{G_{1}, \dots, G_{k}}$ such that $Y \times \bc{0} \subseteq Z$.
  The component $Z$ has codimension at most $k$ in $\bP^{n} \times \bA^{1}$ by the first paragraph.
  We will show that $\pi_{t}(Z) = \bA^{1}$, which implies $Z = W_{i}$ for some $i$.
  As a consequence, we have the inclusion $Z \subseteq \paramvar{G_{1}, \dots, G_{k}}$, and the final statement of the lemma follows.

  Towards a contradiction, assume that $\pi_{t}(Z) \neq \bA^{1}$.
  Since $\pi_{t}(Z)$ is closed and irreducible, and since $0 \in \pi_{t}(Z)$, we can deduce $\pi_{t}(Z) = \bc{0}$.
  Then $Z = Z \cap \var{t}$ in $\bP^{n} \times \bA^{1}$, and this variety has dimension at least $n + 1 - k$.
  Further, this variety is contained in $\var{G_{1}, \dots, G_{k}} \times \bc{0}$ by construction, so it must be contained in one of the irreducible components of $\var{G_{1}, \dots, G_{k}} \times \bc{0}$ of dimension at least $n + 1 - k$, say $Z \subseteq Y' \times \bc{0}$.
  We now have $Y \times \bc{0} \subsetneq Y' \times \bc{0}$ whence $Y \subsetneq Y'$, where these containments are strict since $Y'$ has dimension strictly larger than that of $Y$.
  This is a contradiction to the fact that $Y$ is an irreducible component of $\var{G_{1}, \dots, G_{k}}$.
\end{proof}

If a set of forms $G_{1}, \dots, G_{k}$ defines an equidimensional variety of codimension $k$, then the Chow form is easily constructed using the resultant as follows.
Let $\setbuild{u_{i, j}}{0 \le i \leq n-k, 0 \le j \le n}$ be a set of fresh variables.
We use $\vu_{i}$ to denote the variables $u_{i, 0}, \dots, u_{i, n}$ and $\vu$ to denote all the $u_{i, j}$.
Let $L_{i}$ denote the general linear form $L_{i} \coloneqq u_{i, 0} x_{0} + u_{i, 1} x_{1} + \cdots + u_{i, n} x_{n}$.
If $\gamma_{i, j}$ is a set of elements from $\bF$ for $0 \leq i \leq n-k$ and $0 \leq j \leq n$ then we use $L_{i}^{\vgamma}$ to denote the specialization of the linear form $L_{i}$.

\begin{lemma}
  \label{lemma: chow form of equidimensional complete intersection}
  If $G_{1}, \dots, G_{k}$ are homogeneous forms such that $V = \var{G_{1}, \dots, G_{k}}$ is an equidimensional variety of codimension $k$, then the squarefree part of $\res\br{G_{1}, \dots, G_{k}, L_{0}, \dots, L_{n-k}}$ (where we treat the polynomials as forms in $x_{0}, \dots, x_{n}$) is a nonzero scalar multiple of the Chow form of $V$.
\end{lemma}
\begin{proof}
  This is a straightforward consequence of the main property of the resultant (Item 1 of \cref{definition: multivariate resultant all fields}) as follows.
  Let $R \in \bK\bs{\vu}$ denote the resultant of $G_1,\ldots,G_k,L_0,\ldots,L_{n-k}$ as forms in the variables $\vx$.
  Suppose $\gamma_{i, j}$ for $0 \leq i \leq n-k$ and $0 \leq j \leq n$ is a set of elements such that $R(\vgamma) = 0$.
  Then by the main property of the resultant, $\var{G_{0}, \dots, G_{k}, L_{0}^{\vgamma}, \dots, L_{n-k}^{\vgamma}}$ is nonempty, whence $\vgamma$ is a zero of the Chow form of $V$.
  Conversely, if $\vgamma$ is a root of the Chow form, then by definition of the Chow form, $\var{G_{0}, \dots, G_{k}, L_{0}^{\vgamma}, \dots, L_{n-k}^{\vgamma}}$ is nonempty, and so $R(\vgamma) = 0$ by the main property of the resultant.
  Since $R$ and the Chow form of $V$ have the same zero set, their squarefree parts must be the same (up to a multiple of $\bK$).
  Since the Chow form is squarefree, normalizing the squarefree part of $R$ must yield the Chow form.
\end{proof}

If $\var{G_{1}, \dots, G_{k}}$ is not equidimensional, then it has a component of codimension less than $k$.
In this case, $\res\br{G_{1}, \dots, G_{k}, L_{0}, \dots, L_{n-k}}$ is the zero polynomial, since $\var{G_{0}, \dots, G_{k}, L_{0}^{\vgamma}, \dots, L_{n-k}^{\vgamma}}$ will be nonempty for every choice of $\vgamma$.

As we now prove, even if $\var{G_{1}, \dots, G_{k}}$ is not equidimensional, the resultant
\[
  \res\br{(1-t)G_{1} + t x_{1}^{d_{1}}, \dots, (1-t)G_{k} + t x_{k}^{d_{k}}, L_{0}, \dots, L_{n-k}}
\]
remains nonzero.
While we cannot recover the Chow forms of the equidimensional components of $\var{G_{1}, \dots, G_{k}}$ directly from this polynomial, we can recover the Chow forms of the varieties $\limitvart{G_{1}, \dots, G_{k}}{\tau}$.

We now prove some intermediate results towards the above facts.
The first of these is that certain incidence varieties are irreducible.
This is essentially a version of Bertini's theorem.

\begin{lemma}
  \label{lemma: generic subspace irreducible}
  Suppose $Z \subseteq \bP^{n} \times \bA^{1}$ is an irreducible variety.
  Then the incidence variety
  \[
    Z^{H} \coloneqq \br{Z \times \br{\bP^{n}}^{n-k+1}} \cap \var{L_{0}, \dots, L_{n-k}}
  \]
  is irreducible in $\bP^{n} \times \bA^{1} \times \br{\bP^{n}}^{n-k+1}$.
\end{lemma}
\begin{proof}
  Let $\pi$ denote the projection map from $\bP^{n} \times \bA^{1} \times \br{\bP^{n}}^{n-k+1}$ to $\bP^{n} \times \bA^{1}$, restricted to $Z^{H}$.
  The fibre of $\pi$ at a point $(\valpha, \tau) \in Z$ consists exactly of points of the form $(\valpha, \tau, \vgamma) \in \bP^{n} \times \bA^{1} \times \br{\bP^{n}}^{n-k+1}$ such that $L_{0}^{\vgamma}, \dots, L_{n-k}^{\vgamma}$ vanish at $\valpha$.
  These fibres are therefore all nonempty and isomorphic to $\br{\bP^{n-1}}^{n-k+1}$.
  In particular, they are also all irreducible.
  To summarize, the map $\pi$ is surjective, has irreducible image, and every fibre is irreducible of the same dimension.
  Therefore we can invoke \cite[Chapter~1, Theorem~1.26]{shafarevich1994basic} to deduce that $Z^{H}$ is irreducible.
  We remark that the statement of \cite[Chapter~1, Theorem~1.26]{shafarevich1994basic} assumes that the source and target of the map are projective; however, inspection of the proof shows that it suffices for the map to be closed.
\end{proof}

We now focus on incidence varieties of the above form arising from components of $\perturbedvar{G_{1}, \dots, G_{k}}$.
We use $\pi_{t, \vu}$ to denote the projection from $\bP^{n} \times \bA^{1} \times \br{\bP^{n}}^{n-k+1}$ onto the last two components.
Based on whether the projection to $\bA^{1}$ of a given component $Z \subseteq \perturbedvar{G_{1}, \dots, G_{k}}$ is all of $\bA^{1}$ or a single point, we can deduce information about $\pi_{t, \vu}(Z^{H})$.

\begin{lemma}
  \label{lemma: bad components projection}
  Let $Z$ be an irreducible component of $\perturbedvar{G_{1}, \dots, G_{k}}$ such that the projection of $Z$ to $\bA^{1}$ consists of a single point $\tau$.
  Then $\pi_{t, \vu}(Z^{H}) = \var{t - \tau}$.
\end{lemma}
\begin{proof}
  By \cref{lemma: structure of limit variety}, the component $Z$ has codimension at most $k$ in $\bP^{n} \times \bA^{1}$.
  Further $Z$ lies in $\var{t - \tau}$, therefore $Z \cap \var{t - \tau}$ also has codimension at most $k$ in $\bP^{n} \times \bA^{1}$.
  If we identify $\bP^{n} \times \bc{\tau}$ with $\bP^{n}$, then the image of $Z$ under this identification has codimension at most $k-1$ in $\bP^{n}$.
  As a consequence, this image meets every linear subspace defined by $n-k+1$ equations.
  This shows that the fibre of $(\tau, \vgamma)$ under the map $\pi_{t, \vu}$ meets $Z^{H}$ for every choice of $\vgamma$, proving the claim.
\end{proof}

\begin{lemma}
  \label{lemma: good components projection}
  Let $Z$ be an irreducible component of $\perturbedvar{G_{1}, \dots, G_{k}}$ such that the projection of $Z$ onto $\bA^{1}$ is surjective.
  Then $\pi_{t, \vu}(Z^{H})$ has codimension $1$ and is not contained in any hyperplane defined by an equation of the form $t = \tau$.
\end{lemma}
\begin{proof}
  By \cref{lemma: structure of limit variety}, the component $Z$ has codimension exactly $k$ in $\bP^{n} \times \bA^{1}$, so it has dimension exactly $n+1-k$.
  Therefore, $Z \times \br{\bP^{n}}^{n-k+1}$ has dimension $n+1-k + n(n-k+1)$, and $Z^{H} = \br{Z \times \br{\bP^{n}}^{n-k+1}} \cap \var{L_{0}, \dots, L_{n-k}}$ has dimension at least $ \dim \br{Z \times \br{\bP^{n}}^{n-k+1}} -  n + k - 1 = n(n-k+1)$.

  Since $\pi_{t}(Z) = \bA^{1}$, and since 
  \[
    Z \cap \var{t-1} \subset \perturbedvar{G_{1}, \dots, G_{k}} \cap \var{t-1} = \var{t, x_{1}, \dots, x_{k}},
  \]
  and since $\codim Z = k$, we have $Z \cap \var{t - 1} = \var{x_{1}, \dots, x_{k}}$.
  Pick $\vgamma$ such that $L_{0}^{\vgamma} = x_{0}, L_{1}^{\vgamma} = x_{k+1}, L_{2}^{\vgamma} = x_{k+2}, \dots, L_{n-k}^{\vgamma} = x_{n}$.
  Then the linear subspace defined by these linear forms does not meet $Z \cap \var{t - 1}$.
  This shows that $\pi_{t, \vu}(Z^{H}) \neq \bA^{1} \times \br{\bP^{n}}^{n-k+1}$.
  Since $\pi_{t, \vu}(Z^{H})$ is closed (\cite[Chapter~1, Theorem~1.11]{shafarevich1994basic}), we can deduce also that $\dim \pi_{t, \vu}(Z^{H}) \leq \dim \br{\bP^{n}}^{n-k+1} \times \bA^{1} - 1 = n(n-k+1)$.

  Now we show that there is a point $(\tau, \vgamma) \in \pi_{t, \vu}(Z^{H})$ whose fibre meets $Z^{H}$ at finitely many points.
  For this, we can pick $\tau = 1$ and $\vgamma$ such that $L_{0}^{\vgamma} = L_{1}^{\vgamma} = x_{k+1}$ and $L_{2}^{\vgamma} = x_{k+2}, \dots, L_{n-k}^{\vgamma} = x_{n}$.
  The fibre only meets $Z^{H}$ at the point $((1, 0, \dots, 0), \tau, \vgamma)$.
  By the fibre dimension theorem, since the target and the source are irreducible, we can deduce that $\dim Z^{H} = \dim \pi_{t, \vu}(Z^{H})$.
  Combined with the bounds on $\dim Z^{H}$ and $\dim \pi_{t, \vu}(Z^{H})$ shown above, we can deduce that both $Z^{H}$ and $\pi_{t, \vu}(Z^{H})$ have dimension exactly $n \br{n-k+1}$.
  The final statement, that $\pi_{t, \vu}(Z^{H})$ is not contained in a hypersurface of the form $\var{t - \tau}$ follows from the fact that for any choice of $\tau$, the intersection $Z \cap \var{t - \tau}$ is nonempty, and therefore we can always find a set of $n-k+1$ linear equations that meet it.
\end{proof}

We can combine the above observations to obtain a defining equation of $\pi_{t, \vu}\br{\paramvar{G_{1}, \dots, G_{k}}}$.

\begin{lemma}
  \label{lemma: image of perturbed variety}
  Let $G_{1}, \dots, G_{k}$ be forms. 
  Define
  \[
    R' \coloneqq \res\br{(1-t)G_{1} + t x_{1}^{d_{1}}, \dots, (1-t)G_{k} + t x_{k}^{d_{k}}, L_{0}, \dots, L_{n-k}},
  \]
  where the resultant is computed with respect to $x_{0}, \dots, x_{n}$.
  Let $R$ be the squarefree part of $R'$.
  Then the squarefree part of $\Tt_{t}R$ is a nonzero scalar multiple of the Chow form for $\limitvar{G_{1}, \dots, G_{k}}$.
\end{lemma}
\begin{proof}
  Define $Y \coloneqq \br{\perturbedvar{G_{1}, \dots, G_{k}} \times \br{\bP^{n}}^{n-k+1}} \cap \var{L_{0}, \dots, L_{n-k}}$.
  By the main property of the resultant (Item 1 of \cref{definition: multivariate resultant all fields}), the roots of $\var{R'}$ consist of exactly those $(\tau, \vgamma)$ such that $(1 - \tau) G_{1} + \tau x_{1}^{d_{1}}, \dots, (1 - \tau) G_{k} + \tau x_{k}^{d_{k}}, L_{0}^{\vgamma}, \dots, L_{n-k}^{\vgamma}$ have a common zero.
  These are exactly the points in $\pi_{t, \vu}(Y)$.
  Thus $\pi_{t, \vu}(Y) = \var{R'} = \var{R}$, where the last equality follows since taking squarefree parts does not change the zero set.
  Let $R_{1} \in \bF\bs{t}$ and $R_{2} \in \bF\bs{t, \vu}$ be such that $R = R_{1} R_{2}$ and no factor of $R_{2}$ depends only on $t$ (equivalently, $R_{2}|_{t = \tau} \neq 0$ for all $\tau$).

  We show the following three facts regarding $R_{1}$ and $R_{2}$.
  \begin{enumerate}[noitemsep]
    \item $R_{1}(\tau) = 0$ if and only if $\perturbedvar{G_{1}, \dots, G_{k}}$ has a component whose projection to $\bA^{1}$ consists of the point $\tau$.
    \item $\var{R_{2}} = \pi_{t, \vu}\br{\br{\paramvar{G_{1}, \dots, G_{k}} \times \br{\bP^{n}}^{n-k+1}}  \cap \var{L_{0}, \dots, L_{n-k}}}$.
    \item For each $\tau$, the squarefree part of the specialization $R_{2}|_{t = \tau}$ is a nonzero scalar multiple of the Chow form of $\limitvart{G_{1}, \dots, G_{k}}{\tau}$.
  \end{enumerate}

  The image $\pi_{t, \vu}(Y)$ is the union of the images of the irreducible components of $Y$ under the map $\pi_{t, \vu}$.
  By construction, the irreducible components of $Y$ are exactly the irreducible components of $Z^{H}$, where $Z$ is an irreducible component of $\perturbedvar{G_{1}, \dots, G_{k}}$.
  By \cref{lemma: generic subspace irreducible}, the varieties $Z^{H}$ are themselves irreducible, therefore the irreducible components of $Y$ are exactly $Z^{H}$.

  \cref{lemma: bad components projection} and \cref{lemma: good components projection} show that the $\codim{\pi_{t, \vu}(Z^{H})} = 1$, and therefore each such image either lies in $\var{R_{1}}$ or $\var{R_{2}}$.
  \cref{lemma: good components projection} also shows that if $Z$ is a component of $\paramvar{G_{1}, \dots, G_{k}}$ then the image lies in $\var{R_{2}}$.
  Similarly, \cref{lemma: bad components projection} also shows that if $Z$ is not a component of $\paramvar{G_{1}, \dots, G_{k}}$ then the image lies in $\var{R_{1}}$.
  These two facts prove items $1$ and $2$.

  To prove item $3$, note that from item $2$ we can deduce 
  \begin{align*}
    \var{R_{2}|_{t = \tau}} &= \pi_{\vu}\br{\br{\paramvar{G_{1}, \dots, G_{k}} \times \br{\bP^{n}}^{n-k+1}} \cap \var{L_{0}, \dots, L_{n-k}} \cap  \var{t - \tau}} \\
                            &= \pi_{\vu}\br{\br{\limitvart{G_{1}, \dots, G_{k}}{\tau} \times \br{\bP^{n}}^{n-k+1}} \cap \var{L_{0}, \dots, L_{n-k}}}.
  \end{align*}
  By the main property of the resultant (item 1 of \cref{definition: multivariate resultant all fields}), the zeroes of $\var{R_{2}|_{t = \tau}}$ are exactly those $\vgamma$ such that $L_{0}^{\vgamma}, \dots, L_{n-k}^{\vgamma}$ have a common root on $\limitvart{G_{1}, \dots, G_{k}}{\tau}$, therefore the squarefree part of $R_{2}|_{t = \tau}$ is a nonzero scalar multiple of the Chow form of $\limitvart{G_{1}, \dots, G_{k}}{\tau}$.
  The conclusion of the lemma now follows from the fact that $R_{2}|_{t = 0}$ and $\Tt_{t}R$ are scalar multiples of each other, which can be seen as follows.
  We have $\Tt_{t}R = \Tt_{t}R_{1} \cdot \Tt_{t} R_{2}$.
  The first of these terms lies in $\bF \setminus \bc{0}$, since $R_{1} \in \bF\bs{t} \setminus \bc{0}$.
  The second term is simply $R_{2}|_{t = 0}$, since no factor of $R_{2}$ depends only on $t$, therefore $R_{2}$ is not a multiple of $t$.
\end{proof}

Using the above results, given any projective variety defined by $m$ equations, and given any $k$, we can compute a Chow form for some codimension-$k$ equidimensional subvariety of the input that contains every irreducible component of codimension $k$.
\begin{lemma}
  \label{lemma: chow form without pruning}
  Let $F_{1}, \dots, F_{m} \in \bF\bs{x_{0}, \dots, x_{n}}$ be forms of degrees $d_{1}, \dots, d_{m}$ with $d_{1} \geq \cdots \geq d_{m}$, and let $V \coloneqq \var{F_{1}, \dots, F_{m}}$.
  Let $B \subseteq \bF \setminus \bc{0}$ be a subset of the coefficient field of $\bF$.
  Let $k$ be an integer between $1$ and $n$, inclusive.
  Let $G_{1}, \dots, G_{k}$ and $H_{1}, \dots, H_{k}$ be random linear combinations of $\bc{x_{j}^{d_{1} - d_{i}} F_{i}}_{i, j}$, where each coefficient is picked independently and uniformly from $B$.
  Define
  \[
    R_{G} \coloneqq \Tt_{t} \res\br{(1-t)G_{1} + t x_{1}^{d_{1}}, \dots, (1-t)G_{k} + t x_{k}^{d_{1}}, L_{0}, \dots, L_{n-k}},
  \]
  and
  \[
    R_{H} \coloneqq \Tt_{t} \res\br{(1-t)H_{1} + t x_{1}^{d_{1}}, \dots, (1-t)H_{k} + t x_{k}^{d_{1}}, L_{0}, \dots, L_{n-k}}.
  \]
  Then with probability at least $1 - 4 n d^{n} / \abs{B}$, the squarefree part of the polynomial $\gcd\br{R_{G}, R_{H}}$ is a nonzero scalar multiple of the Chow form for some $V_{k} \subseteq V$ where $V_{k}$ is equidimensional of codimension $k$ and contains every component of $V$ of codimension $k$.
\end{lemma}
\begin{proof}
  Let $V_{G} \coloneqq \var{G_{1}, \dots, G_{k}}$ and $V_{H} \coloneqq \var{H_{1}, \dots, H_{k}}$.
  Via a union bound, we can deduce that the following hold with probability at least $1 - 4 n d^{n} / \abs{B}$.
  By \cref{lemma: input processing}, every component of $V_{G}$ and $V_{H}$ of codimension less than $k$ is a component of $V$, and every component of $V$ of codimension $k$ is a component of $V_{G}$ and $V_{H}$.
  Every codimension-$k$ irreducible variety that is a component of both $V_{G}$ and $V_{H}$ is also a component of $V$: indeed if $Z$ is any irreducible component of $V_{G}$ that does not lie in $\var{F_{1}, \dots, F_{m}}$ then already $Z$ does not lie in $\var{H_{1}}$.

  Assume now that all the above properties hold.
  By \cref{lemma: image of perturbed variety}, the squarefree parts of $R_{G}$ and $R_{H}$ are nonzero scalar multiples of the Chow forms of $\limitvart{G_{1}, \dots, G_{k}}{0}$ and $\limitvart{H_{1}, \dots, H_{k}}{0}$ respectively.
  \cref{lemma: structure of limit variety} says the following about $\limitvart{G_{1}, \dots, G_{k}}{0}$: it is an equidimensional variety of codimension $k$ that contains every codimension $k$ component of $\var{G_{1}, \dots, G_{k}}$.
  Further, since $\limitvart{G_{1}, \dots, G_{k}}{0} \subseteq \var{G_{1}, \dots, G_{k}}$, every component of $\limitvart{G_{1}, \dots, G_{k}}{0}$ that is not an irreducible component of $\var{G_{1}, \dots, G_{k}}$ actually lies in some higher dimensional component of $\var{G_{1}, \dots, G_{k}}$.
  In the setting above, these higher dimensional components of $\var{G_{1}, \dots, G_{k}}$ are exactly the components of $\var{F_{1}, \dots, F_{m}}$ of codimension less than $k$.
  The same holds for $\limitvart{H_{1}, \dots, H_{k}}{0}$.
  
  Now since the Chow form of an equidimensional variety is the product of the Chow forms of its components, the squarefree part of $\gcd\br{R_{G}, R_{H}}$ is a nonzero multiple of the Chow form of the union of the those components that are both in $\limitvart{G_{1}, \dots, G_{k}}{0}$ and in $\limitvart{H_{1}, \dots, H_{k}}{0}$.
  From the description of these limit varieties above, and the properties guaranteed by the random choice, every such component lies in some irreducible component of $\var{F_{1}, \dots, F_{m}}$, and further every component of codimension $k$ of $\var{F_{1}, \dots, F_{m}}$ is one of these common components.
  This completes the proof.
\end{proof}

\subsection{Constant-depth circuits for the Chow form}
\label{subsection: pruning}

The final step to obtaining the Chow form of the equidimensional components of a given variety is to prune the extra components.
This will be carried out using \cref{lemma: chow form to resolution} and \cref{lemma: resolution prune}, relying on the interplay between geometric resolutions and Chow forms.
These results require in particular that we find a primitive element for the varieties whose geometric resolution we wish to compute (defined in \cref{subsection: geometric res intro}), which we can do by applying a randomly-chosen change of coordinates.
The proof of the following lemma is standard, see for example \cite[Section~3.5]{jeronimo2004computational}.

\begin{lemma}
  \label{lemma: random transform firstcoordinate}
  Let $W \subseteq \bP^{n}$ be an equidimensional projective variety of degree $D$.
  Let $B \subseteq \bF \setminus \bc{0}$ be a subset of the coefficient subfield of $\bF$.
  Let $M \in \bF^{\br{n+1} \times \br{n+1}}$ be a matrix where each entry is picked uniformly and independently from $B$.
  Suppose $M$ is invertible.
  Let $\phi: \bP^{n} \to \bP^{n}$ be the invertible linear map defined by matrix $M$, that is, $\phi(\valpha) = M \valpha$.
  
  Let $W'$ be the restriction of $\phi(W)$ to the affine chart of points at finite distance.
  With probability at least $1 - 5nD/\abs{B}$, the variety $W'$ is in projective Noether position, and $x_{r+1}$ is a primitive element for $\phi(W')$, where $r = \dim W$.
  Further, the projective closure of $W'$ is exactly $\phi(W)$.
\end{lemma}

We now construct a family of circuits that compute the Chow forms of the equidimensional components of a variety, given set-theoretic defining equations for the variety in the dense representation.
We will treat part of the input as random bits, but the final output of the circuit is a $\bZ$-multiple of the Chow forms, where the exact scaling factor depends in an unspecified manner on these random bits.

The computations of the Chow forms involve multivariate GCD computations and factor-lifting as subroutines, which are only developed over characteristic $0$ fields.
Consequently, even though the results of \cref{subsection: limit subvarieties} hold over arbitrary fields, the following results are only stated over $\bQ$.

\begin{theorem} \label{theorem: chow form tc0}
  There exists a polylogtime-uniform family of constant-depth threshold circuits $\cC^{\mathcal{CHOW}}$ indexed by $n, m, d_{1}, \dots, d_{m}, h$ with the following properties.
  \begin{itemize}[noitemsep]
    \item The circuit $C^{\mathcal{CHOW}}_{n, m, d_{1}, \dots, d_{m}, h}$ take as input a collection of forms $F_{1}, \dots, F_{m} \in \bQ\bs{x_{0}, \dots, x_{n}}$ of degrees $d_{1}, \dots, d_{m}$ and height at most $h$ in the dense representation.
      The circuit also takes as input $5 \cdot m \cdot \br{n+1}^{3}$ integers of height at most $h$, which we denote by $a_{1}, \dots, a_{r}$.
    \item The output of $C^{\mathcal{CHOW}}_{n, m, d_{1}, \dots, d_{m}, h}$ is a set of $n + 1$ polynomials in the dense representation.
      If the inputs $a_{1}, \dots, a_{r}$ are independently and uniformly sampled from a set of size $b$, then with probability at least $1 - 100 n^2 (n d^n)^{n^2} / b$, the computed polynomials are $\bZ$-multiples of the Chow forms of the equidimensional components of $\var{F_{1}, \dots, F_{m}}$.
      Here, $d$ denotes $\max_{i} d_{i}$.
    \item The size of $C^{\mathcal{CHOW}}_{n, m, d_{1}, \dots, d_{m}, h}$ is $\poly\br{h, n, m, \br{d+1}^{\poly(n)}}$.
  \end{itemize}
\end{theorem}

\begin{proof}
  Let $F_1, \ldots, F_m \in \bQ\bs{x_0, \ldots, x_n}$ be the given forms of degrees $d_1, \ldots, d_m$.
  By rearranging the inputs, we can assume $d_1 \ge \cdots \ge d_m$.
  For convenience, we let $d \coloneqq d_1$.
  Let $V \coloneqq \var{F_1, \ldots, F_m}$ be the projective variety defined by $F_1, \ldots, F_m$.
  If all the $F_{i}$ are identically zero then $V = \bP^{n}$, and the Chow form of every equidimensional component of $V$ except $n$ is just $1$.
  The Chow form of the component of dimension $n$ is $0$.
  In the rest of this argument, we will assume that at least one of the inputs $F_{i}$ is nonzero, so $V \neq \bP^{n}$.
  The Chow form of the component of dimension $n$ is then $1$.
  The rest of the proof will compute the Chow form of the other equidimensional components of $V$.
  Let $V = V_0 \cup V_1 \cup \cdots \cup V_{n-1}$ be the equidimensional decomposition of $V$.

  We will first compute, for every $0 \le k \le n-1$, the Chow form of a variety $V_{k}^\star \subseteq V$ that is equidimensional of dimension $k$ and contains $V_{k}$, the dimension-$k$ equidimensional component of $V$.
  Because $V_{k} \subseteq V_{k}^\star$, the Chow form of $V_{k}$ divides that of $V_{k}^\star$.
  Any points in the difference $V_{k}^\star \setminus V_{k}$ must lie on a higher-dimensional component of $V$.
  In particular, they lie on $V_\ell \subseteq V_\ell^\star$ for some $\ell > k$.
  We will use the Chow forms of $V_{k+1}^\star, \ldots, V_{n-1}^\star$ to remove these extraneous factors from the Chow form of $V_{k}^\star$, resulting in the Chow form of $V_{k}$.

  \paragraph*{The varieties $V_0^\star, \ldots, V_{n-1}^\star$ and their Chow forms.}
  We begin by computing Chow forms of the varieties $V_0^\star, \ldots, V_{n-1}^\star$ using \cref{lemma: chow form without pruning}.
  As in \cref{lemma: chow form without pruning}, we define forms $G_1, \ldots, G_{n-k}$ and $H_1, \ldots, H_{n-k}$ each to be linear combinations of the polynomials $\bc{x_j^{d_1 - d_i} F_i : 1 \le i \le m, 0 \le j \le n}$, where the coefficients of this linear combination are the first $n(n+1)^{2}m/2$ of the integers $a_{e}$.
  Let $\setbuild{u_{i, j}}{0 \le i \leq k, 0 \le j \le n}$ be a set of fresh variables and define $L_i \coloneqq u_{i,0} x_0 + \cdots + u_{i,n} x_n$.
  Let $t$ be a new variable and define the polynomials
  \begin{align*}
    R_{G} &\coloneqq \Tt_{t} \res\br{(1-t)G_{1} + t x_{1}^{d_{1}}, \dots, (1-t)G_{n-k} + t x_{n-k}^{d_{1}}, L_{0}, \dots, L_{k}} \\
    R_{H} &\coloneqq \Tt_{t} \res\br{(1-t)H_{1} + t x_{1}^{d_{1}}, \dots, (1-t)H_{n-k} + t x_{n-k}^{d_{1}}, L_{0}, \dots, L_{k}},
  \end{align*}
  where the resultants are taken with respect to the variables $x_0, \ldots, x_n$.
  By \cref{lemma: chow form without pruning}, under the assumption that the integers $a_{1}, \dots, a_{r}$ are picked uniformly and independently, with high probability the squarefree part of $\gcd(R_G, R_H)$ is a nonzero scalar multiple of the Chow form of an equidimensional dimension-$k$ variety.
  We assume for the rest of the proof that this event occurs.
  We let $V_{k}^\star$ be this variety.

  We now bound the cost of computing the Chow form of $V_k^\star$.
  Computing the coefficients of the forms $G_i$ and $H_i$ can be done using items 1 and 2 of \cref{theorem: integer ops in tc0} to multiply the coefficients of the polynomials $x_j^{d_1 - d_i} F_i$ by the coefficients $a_{e}$ and then add the resulting polynomials together.
  The threshold circuits obtained by invoking \cref{theorem: integer ops in tc0} are of constant depth and size $\poly(h, m, n, \br{d+1}^{\poly(n)})$.
  By \cref{theorem: resultant uniform circuits} and \cite[Lemma 6.1]{AGS26}, the coefficients of $R_G$ and $R_H$ can be computed by a polylogtime-uniform family of threshold circuits of constant depth and size $\poly(h, n, \br{d+1}^{\poly(n)})$.
  To compute $\gcd(R_G, R_H)$, we apply the circuit family of \cref{lemma: multivariate GCD}, which is a polylogtime-uniform family of $\TC^0$ circuits to compute GCDs of multivariate polynomials.
  These circuits in turn require ``random inputs'', for which we use the next few $a_{e}$'s.
  In our application, these circuits have size $\poly(h, n, \br{d+1}^{\poly(n)})$.
  This follows from the fact that both $R_G$ and $R_H$ are polynomials in $O(nk) \le O(n^2)$ variables of degree at most $n d^n$, and hence have at most $\poly(n, \br{d+1}^{\poly(n)})$ nonzero coefficients; together with the fact that the coefficients of $R_G$ and $R_H$ were obtained by simulating a constant-depth arithmetic circuit of size $\poly(n, \br{d+1}^{\poly(n)})$ and thus have height at most $h \cdot \poly(n, \br{d+1}^{\poly(n)})$.
  The failure probability is at most $2 n d^n / b$ for this step.
  We then apply \cref{corollary: multivariate squarefree part} to compute the squarefree part of $\gcd(R_G, R_H)$, which can be implemented by a polylogtime-uniform family of threshold circuits of constant depth and size $\poly(h, n, \br{d+1}^{\poly(n)})$, again using the next few $a_{e}$'s as the random bits.
  Similar to computing the GCD, the squarefree part computation fails with probability at most $2 n^2 (n d^n)^{n^2} / b$.
  Finally, we normalize the coefficients of the squarefree part of $\gcd(R_G, R_H)$ so that it has leading coefficient $1$ in the lexicographic order, and thus is the Chow form of $V_k^\star$.
  In total, this part of the computation can be carried out by a threshold circuit of constant depth and size $\poly(h, n, \br{d+1}^{\poly(n)})$.

  \paragraph*{Changing coordinates to projective Noether position.}
  Next, we will change coordinates so that the points of $V_k^\star$ at finite distance are in projective Noether position and have $y_1$ as a primitive element.
  To do this, we apply a linear change of coordinates using the next few $a_{e}$'s as the coefficients of the transformation.
  Let $W_k^\star$ be the image of $V_k^\star$ under this change of coordinates, and similarly let $W_k$ be the image of $V_k$.
  Let $U_k^\star$ and $U_k$ be the points of $W_k^\star$ and $W_k$, respectively, at finite distance (that is, the points on the open set defined by $x_0 \neq 0$).
  By \cref{lemma: random transform firstcoordinate}, if the $a_{e}$'s are sampled uniformly and independently, then with probability at least $1 - 5nd^k/b$, the variety $U_k^\star$ is in projective Noether position and the variable $x_{k+1}$ is a primitive element for $U_k^\star$.
  Furthermore, the variety $W_k^\star$ is precisely the projective closure of $U_k^\star$, and likewise $W_k$ is the projective closure of $U_k$.
  By \cref{lemma: chow form linear transform}, the Chow form of $W_k^\star$ is obtained by applying the same change of coordinates to the Chow form of $V_k^\star$. 

  We can compute the coefficients of the Chow form of $W_k^\star$ from the Chow form of $V_k^\star$ by evaluating the Chow form of $V_k^\star$ at a sufficient number of points and then interpolating the coefficients of the Chow form of $W_k^\star$.
  The Chow form of $V_k^\star$ is a polynomial of degree at most $nd^n$ in $n^2$ variables, so $(n d^n + 1)^{n^2} \le \poly(n, \br{d+1}^{\poly(n)})$ evaluations of the Chow form suffice to recover its coefficients.
  These evaluations can be performed using a polylogtime-uniform family of threshold circuits of constant depth and size $\poly(h, n, \br{d+1}^{\poly(n)})$ using items 1 and 2 of \cref{theorem: integer ops in tc0}.
  We then interpolate back the coefficients of the Chow form of $W_k^\star$ using the threshold circuits in item 11 of \cref{theorem: integer ops in tc0}, which again can be implemented in constant depth and size $\poly(h, n, \br{d+1}^{\poly(n)})$.
  Overall, this step can be carried out by threshold circuits of constant depth and $\poly(h, n, \br{d+1}^{\poly(n)})$ size.

  \paragraph*{Computing a geometric resolution of $U_k^\star$.}
  Our next goal is to use Chow forms of the varieties $W_k^\star, W_{k+1}^\star, \ldots, W_{n-1}^\star$ to compute the Chow form of $W_k$.
  This proceeds in three steps.
  First, we compute a geometric resolution of the variety $U_k^\star$.
  We then use the Chow forms of $W_{k+1}^\star, \ldots, W_{n-1}^\star$ to refine this to obtain (part of) a geometric resolution for $U_k$.
  Lastly, we lift this (partial) geometric resolution back to a Chow form of $W_k$.

  We start by computing a geometric resolution of $U_k^\star$.
  Rename the variables $x_{k+1}, \ldots, x_n$ to $y_1, \ldots, y_{n-k}$.
  Let $u_1, \ldots, u_{n-k}$ be new indeterminates, let $\bL = \overline{\bK\br{u_1,\ldots,u_n}}$, and let $\phi : \bK^n \to \bK^{k+1}$ be the linear map given by 
  \[
    \phi(x_1,\ldots,x_k,y_1,\ldots,y_{n-k}) = (x_1,\ldots,x_k,u_1y_1 + \cdots + u_{n-k} y_{n-k}).
  \]
  By \cref{lemma: component image base field}, the closure of the image $\overline{\phi(U_k^\star)}$ is a hypersurface defined by a single polynomial $Q_k^\star \in \bK\bs{\vu}\bs{\vx, T}$.
  Because $W_k^\star$ is the projective closure of $U_k^\star$, the polynomial $Q_k^\star$ can be obtained as a specialization of the Chow form of $W_k^\star$ by \cref{lemma: chow form to resolution}.
  Because this specialization of the Chow form only involves setting variables to $0$, $1$, or $-1$, we can compute the coefficients of $Q_k^\star$ by adding and subtracting the appropriate coefficients of the Chow form of $W_k^\star$.
  This can be carried out by threshold circuits of constant depth and size $\poly(h, n, \br{d+1}^{\poly(n)})$ by invoking item 1 of \cref{theorem: integer ops in tc0}.

  From the polynomial $Q_k^\star$, we compute a geometric resolution of the variety $U_k^\star$, following \cref{subsection: geometric res intro}.
  Let $\vnu_0 \in \bK^k$ be the point $(1, 0, \ldots, 0)$ and define the polynomials
  \begin{align*}
    q^\star(\vx, y_1) &\coloneqq Q_k^\star(\vnu_0, \vx, y_1) \\
    r_j^\star(\vx, y_1) &\coloneqq -\frac{\partial Q_k^\star}{\partial u_j}(\vnu_0, \vx, y_1) & j \in \bc{1,\ldots,n_k}.
  \end{align*}
  Together, \cref{lemma: squarefree image,lemma: shape basis in ideal,lemma: first coordinate projection,lemma: components of triangular basis} imply that $q^\star, r_1^\star, \ldots, r_{n-k}^\star$ are a geometric resolution of $U_k^\star$.
  These polynomials are obtained by specializing the polynomial $Q_k^\star$ and its first derivatives, so as was the case with $Q_k^\star$, we can compute their coefficients using threshold circuits of constant depth and size $\poly(h, n, \br{d+1}^{\poly(n)})$ thanks to item 1 of \cref{theorem: integer ops in tc0}.

  \paragraph*{Removing excess components of $U_k^\star$ in $q^\star$.}
  We now execute the second step of our construction, refining the polynomial $q^\star$ to a polynomial $q$ that corresponds to the components of $U_k^\star$ that lie on $W_k$.
  For each $\ell > k$, we will use the Chow form of $W_\ell^\star$ to prune away the factor of $q^\star$ that corresponds to components of $W_k^\star$ that lie on $W_\ell^\star$.
  To do this, we apply \cref{lemma: resolution prune}.
  Let $w_{0,1}, \ldots, w_{\ell, n}, t_1, \ldots, t_k, s$ be fresh variables.
  For $0 \le i \le \ell$, let $g_i$ be the polynomial
  \[
    g_{i} \coloneqq \sum_{j = 1}^{k} w_{i, j} \br{x_j - t_{j} x_0} + w_{i, k+1} \br{y_{1} - s x_0} + \sum_{j=2}^{n-k} w_{i, k+j} \br{y_{j} + \frac{r_{j}(\vt, s)}{q'(\vt, s)}x_0},
  \]
  where $q'(\vx,y_1)$ is the derivative of $q^\star(\vx, y_1)$ with respect to $y_1$.
  Since $g_0,\ldots,g_\ell$ are linear in the $\vx$ and $\vy$ variables, we may evaluate the Chow form of $W_\ell^\star$ at the coefficients of $g_0,\ldots,g_\ell$.
  Letting $C_\ell$ denote the Chow form of $W_\ell^\star$, write the evaluation $C_\ell(\coeff{g_0}, \ldots, \coeff{g_\ell})$ as
  \[
    C_\ell(\coeff{g_0}, \ldots, \coeff{g_\ell}) = \frac{P_\ell(\vt, s, \vw)}{q'(\vt, s)^{c_\ell}},
  \]
  where the polynomials $P_\ell(\vt, s, \vw)$ and $q'(\vt, s)$ are relatively prime.
  Letting $\hat{P}_\ell(\vt, s)$ be the GCD of the coefficients of $P_\ell(\vt,s,\vw)$ viewed as a polynomial in $\vw$, \cref{lemma: resolution prune} implies that 
  \[
    q_\ell(\vt, s) \coloneqq \gcd(q^\star(\vt, s),\hat{P}_\ell(\vt, s))
  \]
  consists exactly of those factors of $q^\star$ that correspond to components of $U_k^\star$ lying on $U_\ell^\star$.
  Equivalently, the polynomial $\frac{q^\star(\vt, s)}{q_\ell(\vt, s)}$ consists of those factors of $q^\star$ that correspond to components of $U_k^\star$ that do not lie on $W_\ell^\star$.

  Consider the polynomial
  \[
    q(\vt, s) \coloneqq \gcd\br{\frac{q^\star(\vt, s)}{q_{k+1}(\vt, s)}, \ldots, \frac{q^\star(\vt, s)}{q_{n-1}(\vt,s)}}.
  \]
  The factors of $q$ are precisely the factors of $q^\star$ that correspond to components of $U_k^\star$ not contained in any of the varieties $W_{k+1}^\star, \ldots, W_{n-1}^\star$.
  Recall that the components of $W_k^\star \setminus W_k$ are contained in $W_{k+1} \cup \cdots \cup W_{n-1} \subseteq W_{k+1}^\star \cup \cdots \cup W_{n-1}^\star$.
  Since we have removed factors of $q^\star$ that correspond to components lying on $W_{k+1}^\star \cup \cdots \cup W_{n-1}^\star$, it follows that $q$ consists of exactly those factors of $q^\star$ that lie on $U_k$.
  Thus, the polynomial $q(\vt, s)$ defines the projection of $U_k$ to the first $k+1$ coordinates, as desired.

  To compute $q$, we first compute each of the polynomials $q_\ell$ in parallel for $k < \ell < n$.
  For each value of $\ell$, we evaluate $C_\ell$ at the coefficients of $g_0, \ldots, g_\ell$, which results in a rational function $f(\vt, s, \vw) / g(\vt, s, \vw)$.
  We then compute $\gcd(f(\vt, s, \vw), g(\vt, s, \vw))$ and divide both $f$ and $g$ by this GCD, which yields the polynomials $P_\ell(\vt, s, \vw)$ and $q'(\vt, s)^{c_\ell}$.
  Next, we compute $\hat{P}_\ell(\vt, s)$ as the GCD of all the $\vw$-coefficients of $P_\ell(\vt, s, \vw)$.
  Finally, we have $q_\ell(\vt, s) = \gcd(q^\star(\vt, s), \hat{P}_\ell(\vt, s))$.
  Each of these steps can be implemented by a polylogtime-uniform family of threshold circuits of constant depth and size $\poly(h, n, \br{d+1}^{\poly(n)})$: evaluating $C_\ell$ and the divisions are carried out using the multivariate polynomial arithmetic of \cref{theorem: integer ops in tc0}, and GCDs are computed using \cref{lemma: multivariate GCD}.
  For the multivariate GCD computations, we use the $a_{e}$'s for the random inputs.
  Further, to compute the GCD of more than two multivariate polynomials at a time, we can invoke \cref{lemma: multivariate GCD} with two distinct formal linear combinations of the polynomials.

  After computing the polynomials $q_{k+1},\ldots,q_{n-1}$, we compute $q$ as $\gcd\br{\frac{q^\star}{q_{k+1}}, \ldots, \frac{q^\star}{q_{n-1}}}$.
  The divisions are carried out in parallel, each implemented with \cref{theorem: integer ops in tc0}.
  Finally, we compute the GCD of these polynomials once again using \cref{lemma: multivariate GCD}.
  In total, this step can be executed by a polylogtime-uniform family of threshold circuits of constant depth and exponential size.

  \paragraph*{Lifting $q$ to the Chow form of $W_k$.}
  Finally, we use the polynomial $q(\vt, s)$ to compute the Chow form of $W_k$.
  If $q(\vt, s)$ is a constant
  Because $W_k$ and $W_k^\star$ are the projective closures of $U_k$ and $U_k^\star$, respectively, we can relate $q$ and $q^\star$ to the Chow forms of $W_k$ and $W_k^\star$.
  Let $C$ be the Chow form of $W_k$ and recall that $C_k$ is the Chow form of $W_k^\star$.
  For $1 \le i \le k$, let $\ell_i$ be the linear form $t_i x_0 - x_i \in \bK\bs{\vt, s}\bs{\vx, \vy}$, and let $\ell_{k+1}$ be the linear form $s x_0 - y_1 \in \bK\bs{\vt, s}\bs{\vx, \vy}$.
  Then \cref{equation: geometric resolution from projection,lemma: chow form to resolution} imply that
  \begin{align*}
    C_k(\coeff{\ell_1}, \ldots, \coeff{\ell_{k+1}}) &= q^\star(\vt, s) \\
    C(\coeff{\ell_1}, \ldots, \coeff{\ell_{k+1}}) &= q(\vt, s).
  \end{align*}
  So far, we have computed the coefficients of $C_k$, $q^\star$, and $q$.
  We know that $C$ is a factor of $C_k$ and likewise that $q$ is a factor of $q^\star$.
  To compute the coefficients of $C$, we will invoke \cref{lemma: multivariate lifting} to lift $q$ to a factor of $C_k$, which will be precisely the Chow form $C$.
  
  To use \cref{lemma: multivariate lifting}, we need to work with a specialization of the polynomials $C_k$ and $C$ where variables are only set to zero, whereas the specialization above sets some variables to $-1$.
  Let $\setbuild{u_{i, j}}{1 \le i \leq k+1, 0 \le j \le n}$ be the variables of $C_k$ and $C$.
  Define the shifted polynomials $\hat{C}_k$ and $\hat{C}$ to be $C_k$ and $C$, respectively, under the change of coordinates that performs the shift $u_{i,i} \mapsto u_{i,i} - 1$ for $1 \le i \le k+1$ and acts as the identity on the remaining variables.
  Then we have
  \begin{align*}
    \hat{C}_k((t_1, 0, \ldots, 0), \ldots, (t_k, 0, \ldots, 0), (s, 0, \ldots, 0)) &= q^\star(\vt, s) \\
    \hat{C}((t_1, 0, \ldots, 0), \ldots, (t_k, 0, \ldots, 0), (s, 0, \ldots, 0)) &= q(\vt, s).
  \end{align*}
  To apply \cref{lemma: multivariate lifting}, we must verify that $\hat{C}_k$ and $q^\star$ are squarefree, and that $\hat{C}_k$ is monic as a polynomial in $s$.
  The squarefreeness of $\hat{C}_k$ follows from that fact that it is the Chow form of $W_k^\star$, hence necessarily squarefree, while the squarefreeness of $q^\star$ follows from \cref{lemma: squarefree image}.
  It is not necessarily the case that $\hat{C}_k$ is monic as a polynomial in $s$, so to guarantee this, we apply a linear change of variables $t_i \mapsto t_i + \alpha_i s$, where the scalars $\alpha_i$ are the next few $a_{e}$'s.
  The image of $\hat{C}_k$ under this change of coordinates will be monic in $s$ with probability at least $1 - nd^n/b$.

  In this new coordinate system, we can apply \cref{lemma: multivariate lifting}.
  Let $R(\vu)$ be the polynomial obtained by applying \cref{lemma: multivariate lifting} and then undoing the preceding linear change of variables.
  The conclusion of \cref{lemma: multivariate lifting} guarantees that $R$ is a $\bQ$-multiple of $C$, the Chow form of $W_k$.
  By normalizing $R$ to have leading coefficient $1$, we obtain the coefficients of the Chow form $C$.

  To carry out this computation, we shift and perform a linear change of variables on the polynomials $\hat{C}_k$ and $q$.
  We then apply \cref{lemma: multivariate lifting} to $\hat{C}_k$ and $q$ in the new coordinate system.
  Finally, we change back to the original coordinate system to obtain the coefficients of $C$, the Chow form of $W$.
  The changes of coordinates can both be carried out by threshold circuits of constant depth and size $\poly(h, n, \br{d+1}^{\poly(n)})$ by evaluating the polynomial in the new coordinate system at sufficiently many points and then interpolating out its coefficients, both of which follow from the arithmetic of \cref{theorem: integer ops in tc0}.
  The lifting procedure of \cref{lemma: multivariate lifting} can likewise be carried out by threshold circuits of constant depth and $\poly(h, n, \br{d+1}^{\poly(n)})$ size, so this entire step can be implemented by threshold circuits of exponential size and constant depth.

  \paragraph*{Computing the Chow form of $V_k$.}
  The variety $W_k$ is the image of $V_k$ under a previously-sampled linear change of coordinates.
  Using \cref{lemma: chow form linear transform}, we can compute the Chow form of $V_k$ by composing the Chow form of $W_k$ with the inverse of this change of coordinates.
  This allows us to compute the coefficients of the Chow form of $V_k$ from the coefficients of the Chow form of $W_k$.
  Once more, this change of coordinates can be carried out by evaluating the Chow form of $V_k$ at a sufficient number of points and then interpolating its coefficients, both using the results of \cref{theorem: integer ops in tc0}.
  This can be done by a threshold circuit of constant depth and $\poly(h, n, \br{d+1}^{\poly(n)})$ size.

  In total, we have computed the coefficients of the Chow form of $V_k$ using a randomized threshold circuit of constant depth and size $\poly(h, m, n, \br{d+1}^{\poly(n)})$.
  By our guarantees on the inputs $a_{1}, \dots, a_{r}$, the resulting circuit errs with probability less than $100 n^2 (n d^n)^{n^2} / b$.
  Finally, the uniformity of the circuit family follows from the uniformity of each of the subcircuit used throughout the construction, together with the uniformity of the algorithm itself.
\end{proof}

As a corollary of the preceding theorem, we deduce that the coefficients of the Chow form of a variety can be computed in the counting hierarchy.

\begin{corollary} \label{corollary: chow form in ch}
  The language $L_{\mathcal{CHOW}}$ corresponding to the circuit family $\cC^{\mathcal{CHOW}}$ defined in \cref{theorem: chow form tc0} can be decided in $\CH$.
\end{corollary}

\begin{proof}
  This is an immediate consequence of \cref{theorem: chow form tc0} together with \cref{lemma: threshold to ch}.
\end{proof}

\cref{corollary: chow form in ch} shows that the Chow forms of the equidimensional components of a projective variety $V$ can be computed in $\CH$ when a set of equations that set-theoretically define $V$ are given in the dense representation.
Our next corollary extends this result to the case where we are instead given arithmetic circuits that compute equations that set-theoretically define $V$.
This relies on \cref{lemma: circuit eval in CH}, which shows that the coefficients of a polynomial given in the circuit representation can computed in the counting hierarchy.

\begin{corollary} \label{corollary: chow form in ch circuit representation}
  There exists a family of boolean functions $\cG^{\mathcal{CIRCCHOW}}$ indexed by $s$ with the following properties.
  \begin{itemize}[noitemsep]
    \item The function $G^{\mathcal{CIRCCHOW}}_{s}$ takes as input the description of a constant-free, weakly division-free arithmetic circuit of description length at most $s$, computing a collection of forms $F_{1}, \dots, F_{m} \in \bQ\bs{x_{0}, \dots, x_{n}}$.
      The function also takes as input $5(s+1)^{4}$ integers of height at most $s^{5}$, which we denote by $a_{1}, \dots, a_{r}$.
    \item The output of $G^{\mathcal{CIRCCHOW}}_{s}$ is a set of polynomials in the dense representation.
      If the inputs $a_{1}, \dots, a_{r}$ are independently and uniformly sampled from a set of size $b$, then with probability at least $1 - 100\cdot 2^{s^{4}}/b$, the polynomials computed are $\bZ$-multiples of the Chow forms of the equidimensional components of $\var{F_{1}, \dots, F_{m}}$, in the dense representation.
    \item The language $L_{\mathcal{CIRCCHOW}}$ corresponding to $\cG^{\mathcal{CIRCCHOW}}$ can be decided in $\CH$.
  \end{itemize}
\end{corollary}

\begin{proof}
  We first describe the function $G^{\mathcal{CIRCCHOW}}_{s}$, which we denote by $G_{s}$ for brevity.
  Suppose the input to $G_{s}$ is a circuit $C$ that computes forms $F_{1}, \dots, F_{m} \in \bQ\bs{x_{0}, \dots, x_{n}}$.
  We can bound both the number of variables $n$ and the number of polynomials $m$ by $s$.
  Let $d_{1}, \dots, d_{m}$ denote the degrees of $F_{1}, \dots, F_{m}$.
  A straightforward induction on $s$ shows that the degrees and heights of the polynomials $F_1, \ldots, F_m$ are both bounded by $2^{s}$.
  Define $h \coloneqq s^{5} \cdot 2^{s}$.
  The function $G_{s}$, on input $C, a_{1}, \dots, a_{r}$, is defined to be the evaluation of $C^{\mathcal{CHOW}}_{n, m, d_{1}, \dots, d_{m}, h}$ at the coefficients of $F_{1}, \dots, F_{m}$ and the first $5 \cdot m \cdot (n+1)^{3}$ of the integers $a_{1}, \dots, a_{r}$.
  From this description, the fact that $G_{s}$ computes the Chow forms of the equidimensional components of $\var{F_{1}, \dots, F_{m}}$ follows from the guarantees in \cref{theorem: chow form tc0}.

  We now show that the language $L_{\mathcal{CIRCCHOW}}$ can be decided in $\CH$.
  First note that given the circuit $C$, we can compute the integers $n$, $m$, and $h$ in polynomial time.
  We can compute the integers $d_{1}, \dots, d_{m}$ in $\CH$ as follows: we nondeterministically guess each $d_{i}$, a monomial in $F_{i}$ of degree $d_{i}$ with nonzero coefficient, and a bit in this coefficient witnessing the fact that it is nonzero.
  These guesses can be verified in $\CH$ using \cref{lemma: circuit eval in CH}, so we can compute $d_1, \ldots, d_m$ in $\NP^{\CH} = \CH$.

  Let $\cH$ be the family of boolean functions that takes an arithmetic circuit $C$ as input and computes the coefficients of the polynomials computed by $C$. %
  \cref{lemma: circuit eval in CH} shows that the language $L_{\cH}$ is in $\CH$.
  By \cref{lemma: generalized threshold oracle ch}, the description of $\cG^{\mathcal{CIRCHOW}}$, and \cref{theorem: chow form tc0}, we deduce that the language $L_{\mathcal{CIRCCHOW}}$ is in $\CH^{L_{\cH}} \subseteq \CH^{\CH} = \CH$.
\end{proof}

We end with two applications of \cref{corollary: chow form in ch}.
Our first application is to computing the degree of a projective variety.
Because the degree of an equidimensional projective variety can easily be computed from the degree of its Chow form, \cref{corollary: chow form in ch circuit representation} immediately allows us to compute the degree of both a projective variety and its equidimensional components in $\CH$. 

\begin{corollary} \label{corollary: degree in ch}
  Given a collection of forms $F_1, \ldots, F_m \in \bQ\bs{x_0,\ldots,x_n}$ represented as constant-free, weakly division-free arithmetic circuits, the degree of the projective variety $\var{F_1, \ldots, F_m}$ can be computed in $\FP^{\CH}$.
\end{corollary}

\begin{proof}
  The degree of $\var{F_1, \ldots, F_m}$ is the sum of the degrees of its equidimensional component.
  For any equidimensional variety of dimension $r$, the degree of its Chow form as a polynomial is $r+1$ times the degree of the variety.

  Let $s$ be the description length of the input circuit $C$.
  We sample $5\br{s+1}^{4}$ random integers each between $0$ and $2^{s^{5}}$, call these $a_{1}, \dots, a_{r}$.
  Consider the boolean function $G_{s}^{\mathcal{CIRCCHOW}}$ defined in \cref{corollary: chow form in ch circuit representation}, which we denote by $G$ for the sake of brevity.
  By \cref{corollary: chow form in ch circuit representation}, with high probability $G(C, a_{1}, \dots, a_{r})$ computes the Chow forms of the equidimensional components of $\var{F_{1}, \dots, F_{m}}$.
  Assume for the rest of this proof that this event holds.

  Our algorithm now nondeterministically guesses the degree of each of the Chow forms, a monomial with nonzero coefficient of the guessed degree, and a bit witnessing the fact that this coefficient is nonzero. 
  Each equidimensional component of $\var{F_1, \ldots, F_m}$ has degree at most $\exp\br{s^{O(1)}}$, and therefore the Chow forms also have degrees bounded by $\exp\br{s^{O(1)}}$.
  Further, the algorithm to compute the Chow form also shows that the heights of the Chow forms are bounded by $\exp\br{s^{O(1)}}$.
  Consequently, all the guesses that the algorithm makes have polynomial length.
  The algorithm then verifies these guesses using the fact that $L_{\mathcal{CIRCCHOW}}$ can be decided in $\CH$.

  Once the degrees of the Chow form are obtained, we can divide the degrees by the respective dimensions and add them up to obtain the degree of the variety.
  The problem of computing the degree of the variety thus lies in the function version of $\BPP^{\CH}$, which is just $\FP^{\CH}$.
\end{proof}

Our second application is to deciding membership in the equidimensional components of a variety.
Suppose we are given a point $\valpha \in \bF^{n+1}$ and a list of equations that set-theoretically define a projective variety $V \subseteq \bP^n$.
Naturally, the equations defining $V$ can be used to test if $\valpha \in V$.
Letting $V = V_0 \cup V_1 \cup \cdots \cup V_{n-1}$ be the equidimensional decomposition of $V$, a natural next question is to decide if $\valpha \in V_k$ for some given $k$.
Using the Chow forms of the equidimensional components of $V$, we can decide these membership queries in $\CH$.

\begin{corollary} \label{corollary: equidimensional component membership}
  Given a collection of forms $F_1, \ldots, F_m \in \bQ\bs{x_0, \ldots, x_n}$ represented as constant-free, weakly division-free arithmetic circuits, a point $\valpha \in \bQ^{n+1}$, and a natural number $k \in \bN$, whether $\valpha$ lies in the dimension-$k$ equidimensional component of $\var{F_1, \ldots, F_m}$ can be decided in $\CH$.
\end{corollary}

\begin{proof}
  If $k > n$ or if $\valpha \notin \var{F_1, \ldots, F_m}$, the algorithm immediately rejects.
  The latter condition can be checked in randomized polynomial time by evaluating the circuit modulo a random prime \cite{IM83}.
  Otherwise, we first solve the case $k = 0$.
  Let $V_0$ be the zero-dimensional component of $\var{F_1, \ldots, F_m}$.
  The variety $V_0$ is a finite set.
  Let $V_0 = \bc{\vbeta_1, \ldots, \vbeta_d}$.
  Then the Chow form $C(u_0, \ldots, u_n)$ of $V_0$ is a scalar multiple of the polynomial
  \[
    \prod_{i=1}^d (u_0 \beta_{i,0} + u_1 \beta_{i,1} + \cdots + u_n \beta_{i,n}).
  \]
  Thus, to test if $\valpha \in V_0$, it suffices to determine if the Chow form $C(\vu)$ and the linear form $u_0 \alpha_0 + \cdots + u_n \alpha_n$ have a nontrivial GCD.
  \cref{corollary: chow form in ch circuit representation} allows us to compute the coefficients of the Chow form $C(\vu)$ in $\CH$.
  Together with the polylogtime-uniform $\TC^0$ circuits for multivariate GCD of \cref{lemma: multivariate GCD}, we conclude by \cref{lemma: generalized threshold oracle ch} that the coefficients of the GCD can be computed in $\CH$.
  In particular, we can decide in $\CH$ if the GCD is nontrivial, and hence if $\valpha \in V_0$.
  If $\valpha \in V_0$, we are done and the algorithm accepts.

  If $\valpha \notin V_0$, then we will iteratively intersect the variety $\var{F_1, \ldots, F_m}$ with hyperplanes containing $\valpha$ one-by-one until $\valpha$ lies on the zero-dimensional component of the resulting variety.
  If $\valpha$ is on the dimension-$k$ component of the original variety, then with high probability, $\valpha$ will lie on the zero-dimensional component after $k$ steps.

  We now make this argument precise.
  By performing an invertible linear change of coordinates, we may assume that $\valpha = (1, 0, \ldots, 0)$.
  Let $B \subseteq \bQ \setminus \bc{0}$ be a set of size $10 n 2^{sn}$.
  Let $L = a_1 x_1 + \cdots + a_n x_n$ be a linear form where $a_1,\ldots,a_n$ are chosen independently and uniformly at random from $B$.
  We claim that with probability at least $1 - 2^{sn}/|B|$, if $\valpha$ lies on the dimension-$k$ component of $\var{F_1, \ldots, F_m}$, then $\valpha$ lies on the dimension-$(k-1)$ component of $\var{F_1,\ldots,F_m,L}$.
  To prove this, it suffices to show that $L$ properly intersects each positive-dimensional component of $\var{F_1, \ldots, F_m}$, as this implies that the dimension of each such component drops by $1$ on intersecting with $L$.

  Let $W$ be such a component.
  Because $W$ is positive-dimensional, there is a point $\vbeta \in W$ different from $\valpha$.
  The point $\vbeta$ lies on $\var{L}$ with probability at most $1/|B|$.
  Since the variety $\var{F_1, \ldots, F_m}$ has degree at most $2^{sn}$, it has at most $2^{sn}$ positive-dimensional components. 
  By a union bound, $L$ properly intersects every positive-dimensional component with probability at least $1 - 2^{sn}/|B|$.

  This yields an algorithm to decide on which equidimensional component of $\var{F_1, \ldots, F_m}$ the point $\valpha$ lies.
  We iteratively sample a linear form $L$ with coefficients in $B$, add $L$ to our list of equations, and check if $\valpha$ lies on the zero-dimensional component of the resulting variety.
  With probability at least $1 - n 2^{sn}/|B| \ge 9/10$, this procedure decrements the dimension of the component on which $\valpha$ lies at every step.
  If $\valpha$ was a point on the dimension-$\ell$ component of $\var{F_1, \ldots, F_m}$, then this procedure will stop after $\ell$ steps.
  At this point, we accept if $\ell = k$ and reject otherwise.
\end{proof}

Finally, we close with two remarks. 
The first is a remark regarding computing Chow forms of equidimensional components of affine varieties, and the second is regarding the further special case where the affine variety in question is zero-dimensional.

\begin{remark}
  \label{remark: chow form affine}
  Suppose we are given polynomials $f_{1}, \dots, f_{m} \in \bQ\bs{x_{1}, \dots, x_{n}}$, and suppose $V \coloneq \var{f_{1}, \dots, f_{m}}$.
  The Chow form of the projective closure $\overline{V} \subset \bP^{n}$ can be computed as follows.
  We consider the equations $F_{i} := x_{0} f_{i}^{h}$, where we homogenize with respect to the variable $x_{0}$.
  Because of the additional factor $x_{0}$ in each of these equations, the irreducible components of $\var{F_{1}, \dots, F_{m}}$ consist of the hyperplane at infinity $\var{x_{0}}$, and the projective closures of the irreducible components of $\var{f_{1}, \dots, f_{m}}$.
  Consequently, for every $k > 1$, the Chow form of the codimension-$k$ equidimensional component of $\var{F_{1}, \dots, F_{m}}$ is exactly the Chow form of the projective closure of the codimension-$k$ equidimensional component of $\var{f_{1}, \dots, f_{m}}$.
  The Chow form of the codimension-$1$ equidimensional component of $\var{F_{1}, \dots, F_{m}}$ contains the Chow form of $\var{x_{0}}$ as a factor, and once we divide by this, we obtain the Chow form of the projective closure of the codimension-$1$ equidimensional component of $\var{f_{1}, \dots, f_{m}}$.
  The Chow form of $\var{x_{0}}$ is simply the determinant of the matrix with $i, j$ entry $u_{i, j}$, therefore it can be computed easily in $\CH$.
\end{remark}

\begin{remark}
  \label{remark: zero dim affine}
  Suppose we are given polynomials $f_{1}, \dots, f_{m} \in \bQ\bs{x_{1}, \dots, x_{n}}$, and suppose $V \coloneq \var{f_{1}, \dots, f_{m}}$.
  Suppose further that $V$ is promised to have dimension $0$, and that our goal is to compute the $u$-resultant of $V$.
  This can be done using the method discussed in \cref{remark: chow form affine}.
  However in this case, we can further simplify the calculation as follows.
  We first compute the Chow form of $\overline{V}_{0}^{\star}$ as described in the proof \cref{theorem: chow form tc0} (which is the same as its $u$-resultant, since $\overline{V}_{0}^{\star}$ has dimension $0$).
  The only points in $\overline{V}_{0}^{\star}$ that are not in $V$ are those that lie in $\var{x_{0}}$.
  Therefore, all we have to do to obtain the $u$-resultant of $V$ is to identify those factors of $\overline{V}_{0}^{\star}$ that depend on $u_{0}$.
  By randomly setting $u_{0}$ to two different values and computing a GCD, we can identify the factors that do not depend on $u_{0}$, and a final division step gives us the $u$-resultant of $V$.
  In this way, we can avoid the pruning and factor lifting steps that are used in \cref{theorem: chow form tc0}.
\end{remark}

\section{Projections}  \label{section: projections}
In this section, we solve the following problem: given an equidimensional variety $V \subset \bA^{n}$ of dimension $r$, and a linear map $\psi: \bA^{n} \to \bA^{r+1}$ described by a matrix $M \in \bF^{\br{r+1} \times n}$ such that $\psi(V')$ has codimension $1$ for every irreducible component $V'$ of $V$, compute a polynomial $C$ such that $\var{C} = \overline{\psi(V)}$.
We call this the projection problem.
Our treatment of geometric resolutions gave us an answer to this question in the special case when $\psi$ is the projection to the first $r+1$ coordinates and $x_{r+1}$ is a primitive element: this is the content of \cref{lemma: first coordinate projection}.
The computation of the polynomial $q$ described in the statement of \cref{lemma: first coordinate projection} is given in \cref{section: chow form computation}.
In this section,  we will deal with the more general case described above.

As an application of the projection problem, we give an algorithm for the problem of approximate polynomial satisfiability.
While we only give a solution to the projection problem in characteristic zero, our solution to the approximate polynomial satisfiability problem will short-circuit some of the intermediate computations and will in fact be applicable even in fields of finite characteristic.

\subsection{Projections of equidimensional varieties}

The general case of the projection problem can be reduced to the following case: the map $\psi: \bA^{n} \to \bA^{r+1}$ is the projection to the first $r+1$ coordinates, and for every component $V'$ of $V$, the projection of $V'$ to the first $r$ coordinates is dominant.
This is achieved as follows.
If $M$ is not full rank, then $\psi(\bA^{n}) \subsetneq \bA^{r+1}$.
Since we assume that $\psi(V')$ has codimension $1$ for every irreducible component $V'$ of $V$, it must be the case that $M$ has rank $r$, and that $\overline{\psi(V')} = \psi(\bA^{n})$ for every irreducible component, therefore the defining equation for $\overline{\psi(V)}$ is easily computed by linear algebra.

Assume now that $M$ is full rank.
Pick an invertible linear map $\psi' : \bA^{r+1} \to \bA^{r+1}$ such that the first $r$ coordinates of $\psi'$ are Noether normalizing for every $\psi(V')$.
A random linear map satisfies this property.
If we compute the defining equation of $\br{\psi' \circ \psi}(V)$, then we can use the defining linear equations of $\psi'$ to obtain the defining equation of $\psi(V)$.
Since $M$ has rank $r+1$, we can now apply a change of coordinates to $\bA^{n}$ to ensure that $\psi' \circ \psi$ is the projection to the first $r+1$ coordinates.
The map $\psi' \circ \psi$ has the two required special properties.
Therefore, in the rest of this section, we will assume that $\psi$ is the projection to the first $r+1$ coordinates, and that the projection to the first $r$ coordinates is dominant for every component $V'$ of $V$.

We start with a fact that relates the projection with the $u$-resultant of a related variety.
We do this by relating both objects to the minimal polynomial of an element of a finite dimensional algebra.
The main technical content, and the proof of the following lemma is a slight extension of \cite[Proposition~1]{schost2003computing}, the only difference being that we work with a set of polynomials that only generate $V$ set-theoretically, as opposed to ideal-theoretically.
\begin{lemma}
  \label{lemma: special projection from Chow form}
  Let $\bF$ be an arbitrary field.
  Let $V \subset \bA^{n}$ be an equidimensional variety of dimension $r$ such that projection of every irreducible component of $V$ onto the first $r$ coordinates is dense.
  Let $I \subset \bF\bs{x_{1}, \dots, x_{n}}$ denote any ideal such that $\var{I} = V$.
  Let $\bK \coloneq \overline{\bF\br{x_{1}, \dots, x_{r}}}$, and let $I' \coloneq I\bK\bs{x_{r+1}, \dots, x_{n}}$.
  Let $W \coloneq \var{I'}$.
  By assumption, $W$ has dimension zero.

  Let $\cC_{W} \in \bF\br{x_{1}, \dots, x_{r}}\bs{\vu} \subset \bK\bs{\vu}$ be a polynomial such that in $\bK\bs{\vu}$, the squarefree part of $\cC_{W}$ is a nonzero scalar multiple of the $u$-resultant of $W$.
  Let $A$ denote the specialization of $\cC_{W}$ at $u_{0} = x_{r+1}, u_{1} = -1, u_{2} = \cdots = u_{n-r} = 0$.
  Suppose $A = C / \delta$ where $C \in \bF\bs{x_{1}, \dots, x_{r}}\bs{x_{r+1}}$ is primitive and $\delta \in \bF\br{x_{1}, \dots, x_{r}}$.
  Then $\var{C} = \overline{\psi(V)}$.
\end{lemma}

\begin{proof}
  The ideal $I'$ has generators in $\bF\bs{x_{1}, \dots, x_{r}}\bs{x_{r+1}, \dots, x_{n}}$.
  Note that $W$ does not depend on the choice of ideal $I$ used to define $V$ set-theoretically.
  Let $B$ denote the algebra $\bK\bs{x_{r+1}, \dots, x_{n}} / I'$.
  By assumption, $B$ is a finite-dimensional $\bK$-algebra.

  Let $P \in \bF\bs{x_{1}, \dots, x_{r+1}}$ be a polynomial that generates the ideal $\radideal{I} \cap \bF\bs{x_{1}, \dots, x_{r+1}}$.
  Then $\var{P} = \overline{\psi(V)}$.
  By assumption, every irreducible component $V'$ of $V$ is such that the projection of $V'$ to the first $r$ coordinates is dominant.
  As a consequence, the polynomial $P$, when considered as a univariate polynomial in $x_{r+1}$ with coefficients in $\bF\bs{x_{1}, \dots, x_{r}}$, is primitive.
  Indeed, if $P$ has a factor $P'$ in $\bF\bs{x_{1}, \dots, x_{r}}$, then no minimal prime of $I$ contains $P'$, and therefore every minimal prime contains $P / P'$, which is a contradiction to the fact that $P$ generates $\radideal{I} \cap \bF\bs{x_{1}, \dots, x_{r}}$.

  Consider the endomorphism of $B$ given by multiplication by $x_{r+1}$.
  Let $\mu \in \bK\bs{T}$ denote the minimal polynomial of this endomorphism.
  Since the generators of $I'$ lie in $\bF\br{x_{1}, \dots, x_{r}}\bs{x_{r+1}, \dots, x_{n}}$, we deduce that $\mu \in \bF\br{x_{1}, \dots, x_{r}}\bs{T}$.
  By \cite[Chapter~2, Theorem 4.5]{CLO05}, the roots of $\mu$ are exactly $\alpha_{1}$ for $\valpha \in W$.
  Equivalently, $\mu$ factors as $\mu = c \prod_{\valpha \in W} \br{T - \alpha_{1}}^{n_{\valpha}}$ for some $c \in \bF\br{x_{1}, \dots, x_{r}}$ and $n_{\valpha} > 0$.
  Write $\mu = R / d$, where $R \in \bF\bs{x_{1}, \dots, x_{r}}\bs{T}$ is primitive, and $d \in \bF\br{x_{1}, \dots, x_{r}}$.

  We first show that $\overline{\psi(V)} = \var{R(x_{r+1})}$ in $\bA_{\overline{\bF}}^{r+1}$.
  Since $P$ defines $\psi(V)$, we have $P \in \radideal{I}$, and therefore $P \in \radideal{I'}$.
  Consequently, when $P$ is treated as a univariate polynomial in $\bK\bs{x_{r+1}}$, it vanishes on $\alpha_{1}$ for every $\valpha \in W$.
  The roots of $R(x_{r+1})$ are precisely $\alpha_{1}$ for each $\valpha \in W$.
  Therefore we can deduce that $R(x_{r+1})$ divides $P^{a}$ in $\bK\bs{x_{r+1}}$ for some $a \in \bN \setminus \bc{0}$.
  Since these polynomials have coefficients in $\bF\bs{x_{1}, \dots, x_{r}}\bs{x_{r+1}}$, we can write $P^{a} = R \eta$ with $\eta \in \bF\br{x_{1}, \dots, x_{r}}\bs{x_{r+1}}$.
  Since both $P$ and $R$ are primitive, we can deduce by Gauss's lemma that $\eta \in \bF\bs{x_{1}, \dots, x_{r}}\bs{x_{r+1}}$.
  This shows $\var{R(x_{r+1})} \subseteq \overline{\psi(V)}$.
  
  Conversely, we have $R(x_{r+1}) \in I' \cap \bF\bs{x_{1}, \dots, x_{r+1}}$.
  As seen above, the ideal $I'$ has generators in $\bF\bs{x_{1}, \dots, x_{r}}\bs{x_{r+1}, \dots, x_{n}}$, and is generated by a set of elements that generate $I$.
  We can therefore write $R(x_{r+1}) = \sum g_{i} h_{i}$ with $g_{i} \in \bF\bs{x_{1}, \dots, x_{n}}$ that are generators of $I$, and $h_{i} \in \bF\br{x_{1}, \dots, x_{r}}\bs{x_{r+1}, \dots, x_{n}}$.
  Clearing denominators, we get $c' R(x_{r+1}) = \sum g_{i} h'_{i}$ for some $c', h'_{i} \in \bF\bs{x_{1}, \dots, x_{n}}$ such that $h_{i} = h'_{i} / c'$ for every $i$.
  Therefore, $c' R(x_{r+1}) \in I$.
  Since no minimal prime of $I$ contains a nonzero element from $\bF\bs{x_{1}, \dots, x_{r}}$, we can deduce that $R(x_{r+1}) \in I$.
  Consequently, $\overline{\psi(V)} \subseteq \var{R(x_{r+1})}$.
  This completes the proof that $\overline{\psi(V)} = \var{R(x_{r+1})}$.

  Finally, consider now the specialization of $\cC_{W}$ at $u_{2} = \cdots = u_{n-r} = 0$.
  By definition of the $u$-resultant and the property guaranteed by $\cC_{W}$, the resulting polynomial is exactly $e \cdot \prod_{\valpha \in W} \br{u_{0} + u_{1} \alpha_{1}}^{m_{\valpha}}$, for some $e \in \bK$ and positive integers $m_{\valpha}$.
  Further setting $u_{1} = -1$ and $u_{0} = x_{r+1}$, the resulting polynomial, denoted $A$, is $e \cdot \prod_{\valpha \in W} \br{x_{r+1} - \alpha_{1}}^{m_{\valpha}}$.
  Since $\cC_{W}$ lies in $\bF\br{\vx}\bs{\vu}$, we can deduce that $e \in \bF\br{\vx}$.
  The polynomial $C$ defined in the statement of the lemma is therefore a primitive polynomial in $\bF\bs{x_{1}, \dots, x_{r}}\bs{x_{r+1}}$ with roots $\alpha_{1}$ for $\valpha \in W$.
  The polynomial $R(x_{r+1})$ also satisfies this property.
  Therefore $\var{C} = \var{R(x_{r+1})}$.
  Combined with the fact that $\var{R(x_{r+1})} = \overline{\psi(V)}$ proved above, the proof is complete.
\end{proof}

In fact by factoring an arbitrary polynomial map through its graph, we can use the above lemma to compute the image of $V$ under an arbitrary polynomial map to $\bA^{r+1}$, as long as the map preserved the dimension of every irreducible component of $V$.
Since the computation requires computing multivariate GCDs and squarefree parts, and since we have only developed the machinery for these operations in characteristic zero, the following lemma applies in a more restrictive setting than \cref{lemma: special projection from Chow form}.
Before we state and prove the lemma, we make a quick remark about Tseitin encodings, and show that we can work in the dense representation while solving this problem of computing the image.

\begin{remark}
  \label{remark: tseitin encoding}
  Suppose $C$ is a constant-free, weakly division-free arithmetic circuit of bit complexity $s$ that computes polynomials $f_{1}, \dots, f_{m} \in \bQ\bs{x_{1}, \dots, x_{n}}$.
  Let $V \subset \bA^{n}$ be the variety defined by $f_{1}, \dots, f_{m}$.
  The Tseitin encoding of $C$ is constructed as follows.
  First, we modify $C$ to ensure that it has fan-in two.
  This increases the size and bit complexity of $C$ by a polynomial factor.
  Next, we introduce one new variable $w_{v}$ for each gate $v$ in the circuit.
  We then construct the following set of equations, which we denote by $S$.
  For each addition gate $v$ in the circuit, we add the equation $w_{v} - w_{v_{1}} - w_{v_{2}} = 0$ to our set $S$, where $v_{1}, v_{2}$ are the inputs to $v$.
  For each multiplication gate $v$ in the circuit, we add the equation $w_{v} - \br{w_{v_{1}} \times w_{v_{2}}} = 0$ to our set $S$, where $v_{1}, v_{2}$ are the inputs to $v$.
  If $v$ is an input gate with input $x_{i}$, then we add $w_{v} - x_{i} = 0$ to our set $S$, and if $v$ is a constant gate with constant $c$, we add $x_{v} - c = 0$ to our set.
  For each output gate $v$, we also add $w_{v} = 0$ to our set $S$.

  Suppose the circuit $C$ had $r$ gates, and we added a total of $r$ equations to our set.
  Then $S$ defines a variety $W \subset \bA^{n+r}$, where the first $n$ coordinates correspond to the variables $x_{1}, \dots, x_{n}$.
  Consider the map $\phi: V \to W$ defined as follows.
  Given a point $\valpha \in V$, construct a point $(\valpha, \vbeta) \in \bA^{n+r}$, by setting the $w_{v}$ variables to the value obtained at gate $v$ when the circuit $C$ is evaluated at $\valpha$.
  By the construction of the equations above, we see that $(\valpha, \vbeta)$ satisfies all the equations in $S$.
  An inverse to the map $\phi$ is given by projection to the first $n$ coordinates.
  Once again, by the construction of the equations, it is easy to check that the projection of a point in $W$ to the first $n$ coordinates is a root of $f_{1}, \dots, f_{m}$.

  The main advantage of the Tseitin encoding is that the polynomials in the set $S$ are all quadratic polynomials, and therefore can be expressed in the dense representation with a total bit complexity that is only polynomially larger than $s$.
  For any questions regarding the variety $V$ where the answer does not change under isomorphism, this allows us to work in the dense representation without loss of generality.
  An example of such a problem is deciding if $V$ is empty, computing the dimension of $V$, or computing the size of $V$ (if it is promised to be finite).
  This is what enables \cite{koiran96, koiran1997randomized, AGS26} to work in the dense representation.

  However, if we are deciding some property of $V$ that is extrinsic, that is, that depends on how $V$ is embedded in $\bA^{n}$, we cannot always use the Tseitin encoding.
  For example, the Chow forms of the equidimensional components of $V$ are not invariant under isomorphisms, and it is not clear how to compute these from the Chow forms of the equidimensional components of $W$.
  Consequently, most results in \cref{section: chow form computation} worked in the circuit representation.

  Consider now the problem of computing the image of a variety under a polynomial map.
  Suppose we have two circuits $C_{1}$ and $C_{2}$ computing polynomials $f_{1}, \dots, f_{m}$ and $h_{1}, \dots, h_{p}$, respectively.
  Let $\phi_{\vh}$ be the map $\bA^{n} \to \bA^{p}$ with coordinate functions $h_{i}$.
  Suppose that $\var{f_{1}, \dots, f_{m}}$ is equidimensional of dimension $p-1$, and that the image of every irreducible component of $\var{f_{1}, \dots, f_{m}}$ under $\phi_{\vh}$ has dimension $p-1$.
  In this case, we can construct the graph $\Gamma_{\vh}$ of the map $\phi_{\vh}$ restricted to $\var{f_{1}, \dots, f_{m}}$.
  This is a variety in $\bA^{n+p}$, defined by the polynomials $f_{1}, \dots, f_{m}, y_{1} - h_{1}, \dots, y_{p} - h_{p}$.
  The image of $V$ under $\phi_{\vh}$ is the same as the projection of $\Gamma_{\vh}$ to the coordinates corresponding to the $\vy$ variables.
  
  Now the graph $\Gamma_{\vh}$ is itself defined by equations computed by the circuits $C_{1}$ and $C_{2}$.
  We can then construct the Tseitin encodings of $C_{1}$ and $C_{2}$, obtaining a variety $W$ that is isomorphic to $\Gamma_{\vh}$ such that $W$ is defined by quadratic polynomials.
  The equations for $W$ are obtained from the circuits $C_{1}, C_{2}$ as defined above.
  Since the isomorphism $W \to \Gamma_{\vh}$ is given by a projection, the image of $W$ under the projection to the coordinates corresponding to $\vy$ is the same as the image of $\Gamma_{\vh}$ under the projection to the same coordinates.
  As a consequence, for the problem of computing the equation of the image of a variety under a polynomial map with the above guarantees, we can assume without loss of generality that the input is in the dense representation.
\end{remark}

\begin{lemma}
  \label{lemma: projection images}
  There exists a family of boolean functions $\cG^{\mathcal{IM}}$ with the following properties.
  \begin{itemize}[noitemsep]
    \item The function $G^{\mathcal{IM}}$ takes as input collections of polynomials $f_{1}, \dots, f_{m} \in \bQ\bs{x_{1}, \dots, x_{n}}$ and $h_{1}, \dots, h_{p} \in \bQ\bs{x_{1}, \dots, x_{n}}$ in the dense representation.
      The function also takes as input $10(n+2)^{2}(m+p)$ integers, which we denote by $a_{1}, \dots, a_{e}$.
    \item The output of $G^{\mathcal{IM}}_{s}$ is a polynomial in the dense representation.
      If $V \coloneqq \var{f_{1}, \dots, f_{m}}$ is equidimensional of dimension $p-1$, and the image of every irreducible component of $V$ under the map $\phi_{\vh}: \bA^{n} \to \bA^{p}$ with coordinate functions $h_{1}, \dots, h_{p}$ has codimension $1$, and the inputs $a_{1}, \dots, a_{r}$ are independently and uniformly sampled from a set of size $b$, then with probability at least $1 - 100n^{10}d^{n^{10}}/b$, where $d=\max(1, \max_i \deg f_i, \max_j \deg h_j)$, the polynomial computed is a $\bZ$-multiple of the generator of the ideal of $\overline{\phi_{\vh}(V)}$
    \item The language $L_{\mathcal{IM}}$ corresponding to $\cG^{\mathcal{IM}}$ can be decided in $\CH$.
  \end{itemize}
\end{lemma}
\begin{proof}
  We start by describing the function $G^{\mathcal{IM}}$, which we denote by $G$ for brevity.
  Let $g_{1}, \dots, g_{c}$ denote the set of polynomials $f_{1}, \dots, f_{m}, y_{1} - h_{1}, \dots, y_{p} - h_{p}$.
  Our goal is to compute the image of $\var{g_{1}, \dots, g_{c}}$ under the projection to the coordinates corresponding to the $\vy$ variables.
  In the rest of this proof, all homogenizations will be done by introducing a variable $x_{0}$, and only with respect to the variables $y_{p}, x_{1}, \dots, x_{n}$.
  Let $d_{i}$ denote the degree of $g_{i}$ in the variables $y_{p}, x_{1}, \dots, x_{n}$, and let $d$ denote $\max_{i} d_{i}$.
  Let $P_{1}, \dots, P_{c'}$ denote the set of polynomials $\bc{x_{j}^{d - d_{i}} g_{i}^{h}} \cup \bc{y_{p}^{d - d_{i}} g_{i}^{h}}$.

  For $1 \leq i \leq Q$, define $Q_{i, 1} \coloneqq t \cdot x_{i}^{d} + \sum_{k} \alpha_{k, i} P_{k}$, where the coefficients $\alpha_{k, i}$ are the first few $a_{e}$'s.
  Define $Q_{n+1, 1} \coloneqq t \cdot y_{p}^{d} + \sum_{k} \alpha_{k, n+1} P_{k}$, where again the $\alpha_{k, n+1}$ are the next few $a_{e}$'s.
  Let $Q_{1, 2}, \dots, Q_{n+1, 2}$ be another set of polynomials defined the same way, using the next few $a_{e}$'s as coefficients.
  Let $L$ denote the polynomial $u_{0} x_{0} + \cdots + u_{n} x_{n} + v y_{p}$, where $\vu$ and $v$ are new variables. 
  For $i = 1, 2$, let $R_{i}$ denote $\Tt_{t} \res\br{Q_{1, i}, \dots, Q_{n+1, i}, L}$, where the resultant is computed with respect to $y_{p}, x_{0}, \dots, x_{n}$.
  Let $R$ denote the polynomial obtained by first applying \cref{lemma: multivariate GCD} to $R_{1}$ and $R_{2}$ using the next few $a_{e}$'s as the random inputs, and then applying \cref{corollary: multivariate squarefree part} to the resulting polynomial, using the next few $a_{e}$'s as the random inputs.
  By \cref{lemma: chow form without pruning} we can deduce that as long as the $a_{e}$'s are sampled uniformly, the product of those factors of $R$ that depend on the variable $u_{0}$ are the $u$-resultant of $\var{g_{1}, \dots, g_{c}}$ in $\overline{\bQ\br{y_{1}, \dots, y_{p-1}}}^{n+1}$.

  Let $S_{1}$ be the polynomial obtained by setting $u_{0}$ to the next $a_{e}$ in $R$, and similarly $S_{2}$ be the polynomial obtained by setting $u_{0}$ to the next $a_{e}$ in $R$.
  Let $S_{3}$ be the polynomial obtained by applying \cref{lemma: multivariate GCD} to $S_{1}$ and $S_{2}$, as usual using the next few $a_{e}$'s for the random inputs.
  Let $S_{4}$ be the polynomial obtained by using the multivariate polynomial division circuit \cref{theorem: integer ops in tc0} to divide $R$ by $S_{3}$, and then setting $u_{0}$ to $y_{p}$, and $v$ to $-1$, and $u_{1}, \dots, u_{n}$ to $0$.
  As long as all the $a_{e}$'s are picked uniformly and independently, by \cref{lemma: special projection from Chow form} we can deduce that with high probability, the squarefree-part of the $y_{p}$-primitive part of polynomial $S_{4}$ is the generator of the ideal of $\overline{\phi_{\vh}(V)}$.
  To compute the $y_{p}$-primitive part, we again define $S_{5}$ to be the polynomial obtained by setting $y_{p}$ to the next $a_{e}$, define $S_{6}$ to be the polynomial obtained by applying \cref{lemma: multivariate GCD} to $S_{5}$ and $S_{4}$, define $S_{7}$ to be the quotient when dividing $S_{4}$ by $S_{6}$, and finally setting $S_{8}$ to be the polynomial obtained by applying \cref{corollary: multivariate squarefree part} to $S_{7}$.

  The boolean function $G$ is defined so that $G(f_{1}, \dots, f_{m}, h_{1}, \dots, h_{p}, a_{1}, \dots, a_{r}) = S_{8}$.
  Over the course of the construction of $G$, we have also shown that it computes the image of the projection.
  That the language $L_{\mathcal{IM}}$ can be computed in $\CH$ is at this point a routine argument.
\end{proof}

\subsection{Approximate polynomial satisfiability}

In this subsection, we solve the $\APS$ problem (\cref{definition: aps}) by reducing it to a special structured instance of the projection problem defined above.
If the underlying field $\bF$ is $\bQ$, then \cref{lemma: projection images} already allows us to solve the $\APS$ problem in $\CH$.
The goal of the rest of this section is to solve $\APS$ over a larger class of fields, for example when $\bF$ is a finite field.
To this end, we will reduce an instance of $\APS$ to a very structured instance of the projection problem.
We will also crucially use the fact that solving $\APS$ only requires us to compute the constant term of the equation defining the projection.
Essentially, we want to avoid the multivariate GCD computations that are used within the proof of \cref{lemma: projection images}, since the machinery to perform these is only developed in characteristic $0$.

In the rest of this section, we will allow the input circuits to use constants, but the encoding of the circuit will include the bit complexity of the constants.
For circuits over fields of the type $\bQ\br{y_{1}, \dots, y_{k}}$ and $\bF_{p^{a}}\br{y_{1}, \dots, y_{k}}$, we will assume that the constants on the wires are encoded in the dense representation.
We use $\abs{C}$ to denote the length of encoding the circuit $C$ in binary.
The Tseitin encoding of our circuits will be appropriately modified to account for these constants: for example, if gate $w_{v}$ computes the sum of $w_{v_{1}}$ and $w_{v_{2}}$ with constants $\alpha, \beta$ on the wires, then the equation we add is $w_{v} - \alpha w_{v_{1}} - \beta w_{v_{2}} = 0$.
Before we give the reduction, we will need a subroutine to compute transcendence degrees.

\begin{lemma}
  \label{lemma: trdeg algorithm}
  Let $\bF$ be a field of the form $\bF_{p^{a}}\br{y_{1}, \dots, y_{k}}$.
  Given a circuit $C$ computing $f_{1}, \dots, f_{m} \in \bF\bs{x_{1}, \dots, x_{n}}$, the transcendence degree of $f_{1}, \dots, f_{m}$ can be computed in $\FP^{\CH}$.
\end{lemma}
\begin{proof}
  We first show that the algebraic independence of $f_{1}, \dots, f_{m}$ can be decided in $\CH$.
  Let $d$ be a bound on the maximum degrees of $f_{1}, \dots, f_{m}$, such a bound can be obtained from $C$.
  If $m > n$ then the polynomials are always algebraically dependent, so we focus on the case $m \leq n$.
  Fix a set $B$ in the coefficient subfield of $\bF$ of size $n^{3} d^{2n^{2}}$.
  If $p^{a} \leq n^{3} d^{2n^{2}}$, then we pass to a field extension of size at least $n^{3} d^{2n^{2}}$ (see \cite[Lemma~2.13]{AGS26}).
  Pick $\alpha_{1}, \dots, \alpha_{m}$ each independently and uniformly from $B$.
  We claim that with probability at least $1 - 1/n^{2}$, the polynomials $f_{1} - \alpha_{1}, \dots, f_{m} - \alpha_{m}$ are satisfiable if and only if $f_{1}, \dots, f_{m}$ are algebraically independent.

  Indeed if $f_{1}, \dots, f_{m}$ are algebraically dependent, then the closure of the image of $\bA^{n}$ under the map $\phi_{\vf} = (f_{1}, \dots, f_{m})$ is a proper subvariety of $\bA^{m}$ of degree at most $d^{m}$.
  Therefore, the probability that $(\alpha_{1}, \dots, \alpha_{m})$ lies in the image of $\phi_{\vf}$, or equivalently that $f_{1} - \alpha_{1}, \dots, f_{m} - \alpha_{m}$ is satisfiable, is at most $d^{m} / \abs{B}$.
  
  Conversely, suppose $f_{1}, \dots, f_{m}$ are algebraically independent.
  The key observation is that $\overline{\phi_{\vf}(\bA^{n})} = \bA^{m}$, therefore the set of points with nonempty fiber is dense.
  We control the degree of the bad set of the points that do not have any point in the fiber.
  To this end, consider the polynomials $f_{1} - z_{1}, \dots, f_{m} - z_{m}$, where $z_{1}, \dots, z_{m}$ are new variables.
  Consider the linear system of equations over the field $\bF\br{z}$ obtained by considering each $\vx$ monomial in the identity $1 = (f_{1} - z_{1}) g_{1} + \dots + (f_{m} - z_{m}) g_{m}$, where we treat each coefficient of each $g_{i}$ as a variable, and restrict the $g_{i}$ to have degrees at most $d^{m}$.
  Write this linear system as $M u = v$, where $M$ is the coefficient matrix, $u$ is the vector of unknowns (corresponding to the coefficients of $g$), and $b$ is the $v$ is the vector with $1$ in the entry corresponding to the constant term and $0$ everywhere else.
  Let $\Delta$ be any maximal nonzero minor of the matrix $M$.
  As a polynomial in $z_{1}, \dots, z_{m}$, it has degree at most $n d^{2n^{2}}$.
  Let $r$ be the size of the submatrix to which $\Delta$ corresponds.
  Finally, let $\Delta_{1}, \dots, \Delta_{t}$ be the minors of the augmented matrix $\bs{M|v}$ of size $r+1$.

  For a general choice of elements $\gamma_{1}, \dots, \gamma_{m} \in \bF$, the specialization of $\Delta$ at $(\gamma_{1}, \dots, \gamma_{k})$ is nonzero.
  Further, for this general choice, $f_{1} - \gamma_{1}, \dots, f_{m} - \gamma_{m}$ is satisfiable, therefore by the effective Nullstellensatz \cite{Jelonek05}, the rank of the augmented matrix $\bs{M | v}$ after specialization is the same as the rank of $M$ after specialization.
  Specifically, this implies that $\Delta_{1}(\vgamma), \dots, \Delta_{t}(\vgamma) = 0$, and since $\vgamma$ is general, we can deduce that $\Delta_{1}, \dots, \Delta_{t}$ vanish identically.
  From these two facts and another application of the effective Nullstellensatz, we can deduce that for any point $\valpha \in \bF^{m}$, as long as $\Delta(\valpha) \neq 0$, the system $f_{1} - \alpha_{1}, \dots, f_{m} - \alpha_{m}$ is satisfiable.
  If the coordinates of $\valpha$ are picked independently from $B$, then the above happens with probability at least $1 - nd^{2n^{2}} / \abs{B}$, which is at least $1 - 1/n^{2}$ as claimed.

  Once $\alpha_{1}, \dots, \alpha_{m}$ have been picked randomly, to verify if the fiber is empty or not, we can use the satisfiability algorithm of \cite{AGS26}.
  To invoke this algorithm, we must switch to the dense representation.
  For this, we switch to the Tseitin encoding of the circuit $C$ as described in \cref{remark: tseitin encoding}.
  Note that in the remark, we only described the Tseitin encoding for constant-free circuits, but the required changes to handle the case of circuits with constants are straightforward.
  We can now use the satisfiability algorithm \cite[Theorem~6.9]{AGS26} on this system, and decide if $f_{1} - \alpha_{1}, \dots, f_{m} - \alpha_{m}$ is satisfiable in $\CH$.

  Once we have the above algorithm for checking independence as a subroutine, the transcendence degree can be computed as follows.
  We start with $\cB \coloneqq \emptyset$.
  For each $i$ from $1$ to $m$, we do the following: add the polynomial $f_{i}$ to the set $\cB$, and check if the elements of $\cB$ are algebraically independent.
  If they are, then we move on to the next $i$.
  If they are not, then we remove $f_{i}$ from $\cB$ and move on to the next $i$.
  It is clear that the set $\cB$ that we end up with at the end of this algorithm is a transcendence basis for $\bc{f_{1}, \dots, f_{m}}$, and therefore the size of $\cB$ is the required transcendence degree.
\end{proof}

We now give the claimed reduction.
Our reduction requires computation of the transcendence degree of the input equations.

\begin{lemma}
  \label{lemma: aps reduction}
  Let $\bF$ be one of the fields $\bF_{p^{a}}, \bF_{p^{a}}\br{t_{1}, \dots, t_{k}}$, a number field $\bQ\bs{\alpha}$, or $\bQ\bs{\alpha}\br{t_{1}, \dots, t_{k}}$.
  There exists a randomized polynomial-time algorithm that takes as input an instance of $\APS$ over a field $\bF$ in the form of an algebraic circuit $C$, along with the transcendence degree of the input polynomials, and returns a set of polynomials $g_{1}, \dots, g_{m} \in \bK\bs{y_{1}, \dots, y_{r+1}, x_{1}, \dots, x_{m-1}}$ with the following properties.
  \begin{itemize}[noitemsep]
    \item The number of variables $r+m$, the number of polynomials $m$, and the heights of the coefficients of $g_{i}$ are bounded by $\poly\br{\abs{C}}$.
      Further, each $g_{i}$ is has degree at most $2$.
    \item If $\bF$ has characteristic $0$ then $\bK = \bQ$.
      If $\bF = \bF_{p^{a}}\br{t_{1}, \dots, t_{k}}$ then $\bK = \bF_{p^{b}}\br{t_{1}, \dots, t_{k}}$ with $b/a = \poly\br{\abs{C}}$.
    \item $V \coloneq \var{g_{1}, \dots, g_{m}} \subset \bA^{r+m}$ is an irreducible variety of dimension $r$.
    \item The projection of $V$ onto the coordinates corresponding to $y_{1}, \dots, y_{r}$ is dominant.
      The projection of $V$ onto the coordinates corresponding to $y_{1}, \dots, y_{r+1}$ has dimension $r$.
      Further, the polynomial that generates the ideal $\radideal{g_{1}, \dots, g_{m}} \cap \bF\bs{y_{1}, \dots, y_{r+1}}$, which we denote by $A$, is such that the leading coefficient of $A$ with respect to $y_{r+1}$ lies in $\bF$.
    \item The constant term of the polynomial $A$ is zero if and only if the given input to $\APS$ is a YES instance, namely if and only if the given polynomials are approximately satisfiable.
  \end{itemize}
\end{lemma}
\begin{proof}
  This reduction takes place in two steps.
  In the first step we transform the instance of $\APS$ into another instance of $\APS$ with some additional structure.
  We then reduce this to an instance of projection, and then further transform the instance of the projection problem to obtain the final structure.

  Let $f_{1}, \dots, f_{a} \in \bF\bs{u_{1}, \dots, u_{b}}$ denote the polynomials computed by $C$, and let $d$ be a bound on their maximum degree.
  Let $r$ denote the transcendence degree of $f_{1}, \dots, f_{a}$, equivalently the dimension of $\overline{\phi_{\vf}(\bA^{b})}$.
  By assumption, the integer $r$ is part of the input to the algorithm.
  For the rest of this argument, we will assume that the coefficient subfield of $\bF$ is larger than $100 b d^{b}$, by passing to a field extension if required (see for example \cite[Lemma~2.13]{AGS26}).
  Let $B$ a finite subset of the coefficient subfield of $\bF$ of size $100 b d^{b}$.

  Let $f'_{1}, \dots, f'_{r+1}$ denote random linear combinations of $f_{1}, \dots, f_{a}$, where each coefficient is picked uniformly and independently from $B$.
  By \cite[Theorem~17]{GSS19}, the following two facts hold with probability at least $1 - \br{r+1}d^{r} / \abs{B}$: the set $\bc{f'_{1}, \dots, f'_{r+1}}$ has transcendence degree exactly $r$, and further $f_{1}, \dots, f_{a}$ are approximately satisfiable if and only if $f'_{1}, \dots, f'_{r+1}$ are approximately satisfiable.
  With probability at least $1 - 2bd^{b} / \abs{B}$, the defining equation of the variety $\overline{\phi_{\vf'}(\bA^{b})}$ will also be such that its leading coefficient with respect to every variable is a scalar.

  Next, we pick $b-r$ random linear polynomials $\ell_{i} \coloneq b_{i, 0} + \sum_{j=1}^{b} b_{i, j} u_{j}$ where each $b_{i, j}$ is picked independently and uniformly from $B$.
  Let $H$ denote the linear subspace $\var{\ell_{1}, \dots, \ell_{b-r}}$.
  We claim that $\overline{\phi_{\vf'}(H)} = \overline{\phi_{\vf'}(\bA^{b})}$ holds with high probability.
  With probability at least $1 - b/\abs{B}$, the polynomials $\ell_{1}, \dots, \ell_{b-r}$ are linearly independent.
  We assume that this holds for the rest of the argument.
  Let $\valpha \in \phi_{\vf'}(\bA^{b})$ be any point such that the fiber $Y \coloneq \phi_{\vf'}^{-1}(\valpha)$ has dimension $b-r$.
  Such a point exists by the fiber dimension theorem and the fact that $\overline{\phi_{\vf'}(\bA^{b})}$ is an irreducible variety of dimension $r$.
  The variety $Y$ has degree at most $d^{r+1}$ by Bézout's theorem.
  By \cref{lemma: intersect dimension drop}, with probability at least $1 - 2bd^{r+1} / \abs{B}$, the intersection $Y \cap H$ has dimension $0$.
  We can now apply the fiber dimension theorem to the map $\phi_{\vf'}|_{H}: H \to \bA^{r+1}$.
  Since $\valpha$ is a point in the image of this restricted map whose fiber has dimension $0$, and since $H$ has dimension $r$, we can deduce that $\overline{\phi_{\vf'}(H)}$ has dimension $r$.
  Since $\overline{\phi_{\vf'}(H)} \subseteq \overline{\phi_{\vf'}(\bA^{b})}$ is a containment of irreducible varieties of the same dimension, we can deduce that these varieties are the same.

  We now compute a linear map $\psi: \bA^{r} \to \bA^{b}$ such that $\psi(\bA^{r}) = H$.
  This requires elementary linear algebra over $\bF$.
  If $\psi^{*}$ is the corresponding map of coordinate rings $\bF\bs{u_{1}, \dots, u_{b}} \to \bF\bs{v_{1}, \dots, v_{r}}$, then the polynomials $h_{1} \coloneq \psi^{*}(f'_{1}), \dots, h_{r+1} \coloneq \psi^{*}(f'_{r+1})$ are such that $\overline{\phi_{\vh}(\bA^{r})} = \overline{\phi_{\vf'}(\bA^{a})}$.
  Therefore, $(0, \dots, 0) \in \overline{\phi_{\vh}(\bA^{r})}$ if and only if $(0, \dots, 0) \in  \overline{\phi_{\vf'}(\bA^{a})}$.
  Equivalently, $h_{1}, \dots, h_{r+1}$ are approximately satisfiable if and only if $f'_{1}, \dots, f'_{r+1}$ are approximately satisfiable.
  We can thus deduce that $h_{1}, \dots, h_{r+1}$ are approximately satisfiable if and only if $f_{1}, \dots, f_{a}$ are approximately satisfiable.
  A union bound shows that if $B$ has size at least $100bd^{b}$, then the above holds with probability at least $2/3$.
  Observe that from the circuit $C$, we can easily compute a circuit $C'$ that computes the polynomials $h_{1}, \dots, h_{r+1}$.

  With this structured instance of $\APS$ at our disposal, we proceed to construct an instance of the projection problem.
  For this, we simply factor the map $\phi_{\vh}$ through its graph.
  Specifically, we consider the variety $\Gamma_{\vh} \subset \bA^{2r+1}$ defined by the equations $y_{1} - h_{1}(\vv), \dots, y_{r+1} - h_{r+1}(\vv)$.
  The graph of any morphism is isomorphic to the source, therefore $\Gamma_{\vh} \cong \bA^{r}$ is an irreducible variety of dimension $r$.
  Further, the projection of $\Gamma_{\vh}$ onto the variables $y_{1}, \dots, y_{r+1}$ is exactly $\phi_{\vh}(\bA^{r})$.
  By the properties guaranteed above, we can deduce that the projection of $\Gamma_{\vh}$ onto $y_{1}, \dots, y_{r}$ is dominant, that the projection onto $y_{1}, \dots, y_{r+1}$ is a hypersurface defined by a polynomial whose leading term with respect to $y_{r+1}$ is a scalar, and that this polynomial has zero constant term if and only if the original instance $f_{1}, \dots, f_{a}$ is approximately satisfiable.
  Therefore, we can solve the original instance of $\APS$ by computing any polynomial $Q$ such that $\radideal{Q} = \radideal{y_{1} - h_{1}(\vv), \dots, y_{r+1} - h_{r+1}(\vv)} \cap \bF\bs{y_{1}, \dots, y_{r+1}}$, and checking if the constant term of $Q$ vanishes.
  At this stage, the equations $y_{i} - h_{i}(\vv)$ are still given by algebraic circuits.
  We now further simplify this instance and reduce to the case where the defining equations are given by quadratic polynomials.
  This will allow us to work in the dense representation, as opposed to having to work in the circuit representation.

  To reduce to this dense representation, we again use the Tseitin encoding.
  We repeat the construction of the encoding here, since we want to keep track of the number of variables and equations introduced.
  Given the circuit $C'$, we first convert it into a circuit where every gate has fan-in at most $2$.
  This increases the size by a polynomial factor.
  Let $e$ denote the number of gates in the resulting circuit.
  For each gate $c$ in the circuit, we introduce a variable $x_{c}$.
  We now construct a set of polynomials in the variables $y_{1}, \dots, y_{r+1}, v_{1}, \dots, v_{r}$ and $x_{c}$.
  If a gate $c_{i}$ computes the sum of gates $c_{j}$ and $c_{j'}$ with weights $\alpha, \beta \in \bF$, we add the polynomial $x_{c_{i}} - \alpha x_{c_{j}} - \beta x_{c_{j'}}$ to our set of polynomials.
  We do the same for product gates.
  If $c_{i}$ is an input gate corresponding to variable $v_{j}$, we add $x_{c_{i}} - v_{j}$ to our set of polynomials, and similarly if $c_{i}$ is an output gate computing $\vf'_{j}$ then we add $y_{i} - x_{c_{j}}$.
  Therefore, our set consists of a total of $e + r + 1$ polynomials in $e + 2r + 1$ variables.
  Let $W$ denote the zero set of these polynomials.
  As in \cref{remark: tseitin encoding}, we have $W \cong \Gamma_{\vh}$, and also that the projection of $W$ onto $\vy$ is the same as the projection of $\Gamma_{\vh}$ onto $\vy$.
  
  This instance of projection from $W$ onto $\vy$ is the final output of our reduction.
  We rename the variables and indices and call the defining equations of $W$ constructed above $g_{1}, \dots, g_{m}$, and call the non $\vy$ variables $x_{1}, \dots, x_{m-1}$.
  The polynomials $g_{i}$ by construction have the claimed degree and height bounds, and by the arguments in the previous paragraphs, the original input equations $f_{1}, \dots, f_{a}$ are approximately satisfiable if and only if the equation defining the projection of $W$ to the $\vy$ variables has nonzero constant term.
\end{proof}

Now that we have reduced the instance of $\APS$ to a structured instance of projection, we show how this lets us solve the $\APS$ problem in $\CH$.
Note that the $\APS$ problem only requires us to decide if the constant term of the equation defining the projection is zero or not.
We do not have to compute the constant term exactly, nor do we have to compute any other coefficient.
This gives us two advantages that we will exploit.
The first is that we can perform partial substitutions throughout the computation to simplify, as long as we ensure that zero/nonzeroness of the constant term remains unchanged.
The second is that it suffices to compute the defining equation up to a radical, since this also does not affect the constant term.
Further, since the variety whose projection we are computing is a complete intersection, we also compute the $u$-resultant of the limit subvariety using a single resultant calculation, thus avoiding a multivariate GCD calculation.
Putting all these together, we obtain the following.

\begin{theorem}
  \label{theorem: aps in ch}
  Let $\bF$ be one of the fields $\bF_{p^{a}}, \bF_{p^{a}}\br{t_{1}, \dots, t_{k}}$, a number field $\bQ\bs{\alpha}$, or $\bQ\bs{\alpha}\br{t_{1}, \dots, t_{k}}$.
  Then $\APS$ over $\bF$ lies in $\CH$.
\end{theorem}
\begin{proof}
  Let $C$ be the input to the $\APS$ instance. 
  The circuit $C$ computes the polynomials whose approximate satisfiability we must decide.
  We first compute the transcendence degree of the polynomials computed by $C$.
  If $\bF$ has characteristic $0$, then the transcendence degree can be computed in $\RP$ (\cite{DGW09}).
  If $\bF$ is a finite field, then the transcendence degree can be computed in $\FP^{\PH}$ (\cite{GSS19}).
  Otherwise, the transcendence degree can be computed in $\FP^{\CH}$ (\cref{lemma: trdeg algorithm}).
  We then invoke \cref{lemma: aps reduction} on the circuit $C$, using the transcendence degree computed above.
  This gives us polynomials $g_{1}, \dots, g_{m} \in \bF\bs{y_{1}, \dots, y_{r+1}, x_{1}, \dots, x_{m-1}}$.
  If the coefficient subfield of $\bF$ has size less than $100 \br{m+r}^{2} 2^{2(m+r)}$, we pass to a field extension of size at least $100 \br{m+r}^{2} 2^{2(m+r)}$ (see \cite[Lemma~2.13]{AGS26}).
  We assume for the rest of the proof that the coefficient subfield is larger than this number, and we fix a subset $B$ of the coefficient subfield $\bF$ of size $100 \br{m+r}^{2} 2^{2(m+r)}$.

  Let $V \subset \bA^{m+r}$ denote the zero set of $g_{1}, \dots, g_{m}$.
  By \cref{lemma: aps reduction}, we know that $V$ has dimension $r$.
  Our goal is to decide if the constant term of the defining equation of the projection of $V$ onto the variables $y_{1}, \dots, y_{r+1}$ is zero or not.
  We know from the reduction also that this defining equation is monic in $y_{r+1}$.

  Following \cref{lemma: special projection from Chow form}, we must first compute the $u$-resultant of $g_{1}, \dots, g_{m}$, treated as polynomials in $\bL\bs{y_{r+1}, x_{1}, \dots, x_{m-1}}$, where $\bL \coloneq \overline{\bF\br{y_{1}, \dots, y_{r}}}$.
  In $\bA_{\bL}^{m}$, the zero set of $g_{1}, \dots, g_{m}$ is zero dimensional.
  Consequently, we can directly apply the method of generalized characteristic polynomials to $g_{1}, \dots, g_{m}$ as follows.
  We define $G_{i} \coloneq g_{i}^{h} + t \cdot x_{i}^{\deg g_{i}}$ for $1 \leq i \leq m-1$, and define $G_{m} \coloneq g_{m}^{h} + t \cdot y_{r+1}^{\deg g_{m}}$, where the degree is computed only with respect to the variables $x_{1}, \dots, x_{m-1}, y_{r+1}$, and we homogenize by introducing a new variable $z$.
  We define $L \coloneq u_{0} z + v y_{r+1} + \sum_{i=1}^{m-1} u_{i} x_{i}$, where $u_{0}, \dots, u_{m-1}, v$ are new variables.
  Define $R \coloneq \Tt_{t} \res\br{G_{1}, \dots, G_{m}, L}$.
  By \cite[Lemma~2.6]{Ierardi89}, we can deduce the following about the polynomial $R$: in the ring $\bL\bs{y_{r+1}, x_{1}, \dots, x_{m-1}}$, the squarefree part of $R$ is a $\bL$-multiple of the $u$-resultant of a limit variety $\limitvar{g_{1}^{h}, \dots, g_{m}^{h}}$.
  Specifically, if we consider the product of those factors of $R$ that do not depend on $u_{0}$, then the squarefree part of this product in $\bL\bs{y_{r+1}, x_{1}, \dots, x_{m-1}}$ is a $\bL$-multiple of the $u$-resultant of $g_{1}, \dots, g_{m}$.

  Suppose $R = R_{1} R_{2}$, where $R_{1}$ is the product of those factors of $R$ that do not depend on $u_{0}$, and $R_{2}$ is the product of the factors that depend on $u_{0}$.
  Let $R_{3}$ be the polynomial obtained by specializing $R_{2}$ at $u_{0} = y_{r+1}$ and $v = -1$ and $u_{1}, \dots, u_{m-1} = 0$.
  The polynomial $R_{3}$ lies in $\bF\bs{y_{1}, \dots, y_{r+1}}$.
  Suppose $R_{3} = R_{4} R_{5}$, where $R_{4}$ is the product of those factors of $R_{3}$ that depend on $y_{r+1}$, and $R_{5}$ is the product of factors that do not.
  By \cref{lemma: special projection from Chow form}, $\var{R_{4}}$ is the closure of the image of $V$ under the projection to the $\vy$ variables.
  Our assumptions also ensure that after rescaling $R_{4}$ and $R_{5}$, we can assume that the polynomial $R_{4}$ is monic in the variable $y_{r+1}$.

  As described above, we will apply partial substitutions when carrying out the above computation.
  For this, it is important to study the factors of $R_{2}$ more closely.
  Let $S \in \bF\bs{y_{1}, \dots, y_{r}, u_{0}, \dots, u_{m-1}, v}$ be an irreducible factor of $R_{2}$.
  We can uniquely write $S$ as $S = S_{1} + (v-1)S_{2} + S_{4} S_{5}$, with the following properties.
  The polynomial $S_{1}$ lies in the ideal generated by $u_{1}, \dots, u_{m-1}$, and none of $S_{2}, S_{4}, S_{5}$ depend on $u_{1}, \dots, u_{m-1}$.
  The polynomial $S_{2}$ lies in $\bF\bs{u_{0}, y_{1}, \dots, y_{r}, v}$, and none of $S_{4}, S_{5}$ depend on $v$.
  The polynomial $S_{5}$ lies in $\bF\bs{y_{1}, \dots, y_{r}}$, and finally, every factor of $S_{4} \in \bF\bs{y_{1}, \dots, y_{r}, u_{0}}$ depends on the variable $u_{0}$.
  It is easy to see that the polynomial $R_{4}$ described above is exactly $\prod_{S} S_{4}$, with the variable $u_{0}$ renamed to $y_{r+1}$.
  Here, the product is over all the irreducible factors of $R_{2}$.
  Since $R_{4}$ is monic in $y_{r+1}$, each of the $S_{4}$ is monic in $u_{0}$.
  This will be crucial later on in the proof.
  
  Pick $\alpha_{1}, \dots, \alpha_{m-1}$ and $\beta_{1}, \dots, \beta_{r}$ uniformly and independently from the set $B \subset \bF$.
  Define $P$ to be the specialization of $R$ at $y_{1} = \beta_{1} p, \dots, y_{r} = \beta_{r} p$ and $u_{1} = \alpha_{1} q, \dots, u_{m-1} = \alpha_{m-1} q$, where $p, q$ are new variables.
  We have $P = R_{1}(p, q, v) R_{2}(u_{0}, p, q, v)$.
  We claim that the constant term of $\Tt_{p} \Tt_{v-1} \Tt_{q} P$ is zero if and only if the constant term of $R_{4}$ is zero.

  Since both substitutions and trailing term operations commute with products, we show the above claim by considering the factors of $P$.
  First consider $R_{1}(p, q, v)$.
  Since the $\alpha_{i}, \beta_{i}$ are picked at random, with high probability $R_{1}(p, q, v) \neq 0$.
  Therefore, $\Tt_{p} \Tt_{v-1} \Tt_{q} R_{1}(p, q, v)$ is a nonzero constant, and does not affect our claim.

  We now study $R_{2}(u_{0}, p, q, v)$.
  Fix a factor $S$ of $R_{2}$.
  Writing $S = S_{1} + (v-1) S_{2} + S_{4} S_{5}$ as described above, the image of $S$ under the substitution, denoted by $S'$,  is $S' \coloneq q \cdot S'_{1}(u_{0}, p, q, v) + (v-1) S'_{2}(u_{0}, p, v) + S'_{4}(u_{0}, p) S'_{5}(p)$, where $S'_{1}, S'_{2}, S'_{4}, S'_{5}$ are obtained by specializing $S_{1}, S_{2}, S_{4}, S_{5}$ respectively.
  We thus see that $\Tt_{q} S' = (v-1) S'_{2} + S'_{4} S'_{5}$, and consequently $\Tt_{v-1} \Tt_{q} S' = S'_{4} S'_{5}$.
  Finally, with high probability, $S'_{4}$ and $S'_{5}$ are nonzero, since $S_{4}$ and $S_{5}$ are always nonzero, and $\Tt_{p} S'_{5}$ is always a nonzero constant.
  Therefore $\Tt_{p} \Tt_{v-1} \Tt_{q} S'$ is a nonzero scalar multiple of $\Tt_{p} S'_{4}$.
  Since $S_{4}$ is monic in $u_{0}$, every factor of $S_{4}'$ is bivariate polynomial that is monic in $u_{0}$.
  Consequently, $\Tt_{p} S'_{4}$ is the product of the constant terms of the factors of $S'_{4}$ written as polynomials in $p$ with coefficients in $\bF\bs{u_{0}}$.
  This is exactly the polynomial obtained by setting $y_{1}, \dots, y_{r}$ to $0$ in $S_{4}$.
  Therefore, the constant term of $\Tt_{p} S'_{4}$ is zero if and only if the constant term of $S_{4}$ is zero.
  Taking a product over all the factors, we see that the constant term of $\Tt_{p} \Tt_{v-1} \Tt_{q} R_{2}(u_{0}, p, q, v)$ is nonzero if and only if the constant term of $R_{4}$ is nonzero.
  By a union bound, all the above high probability events hold with probability at least $2/3$.

  Finally, we show how the above short-circuited computation can be carried out in $\CH$.
  Starting with the equations $g_{1}, \dots, g_{m}$, we compute the polynomials $G_{1}, \dots, G_{m}$ described above, and the polynomial $u_{0} z + v y_{r+1} + \sum_{i=1}^{m-1} u_{i} x_{i}$, all in the dense representation.
  We replace every $y_{i}$ by $\beta_{i} p$ for $1 \leq i \leq r$ and every $u_{i}$ by $\alpha_{i} q$ for $1 \leq i \leq m-1$.
  This can be done in polynomial time in the dense representation.
  By \cite[Theorem~6.7]{AGS26}, we can compute in $\CH$ the coefficients of the multivariate resultant of the above polynomials, treated as forms in the variables $x_{1}, \dots, x_{m-1}, y_{r+1}$.
  This resultant is exactly the polynomial $P$ described above.
  \footnote{
    Note that the reduction to the dense representation is necessary to invoke the results of \cite{AGS26} directly.
    The previous sections of this article discuss how the coefficients of the resultant of polynomials in the circuit representation can also be computed in $\CH$, but those results are only presented in the characteristic zero case, and do not apply (as presented) in finite fields.
  }

  Next, we can nondeterministically guess the degree of the trailing term with respect to $q$ in $P$ and verify all smaller monomials in $q$ have zero coefficient.
  This allows us to compute the coefficients of $\Tt_{q} P$ in $\CH$.
  To compute the coefficient of $\Tt_{v-1}$, we again nondeterministically guess the valuation of $\Tt_{q} P$ with respect to $v-1$.
  To check that our guess is correct, for each smaller valuation, we use the circuits for polynomial division (\cref{theorem: integer ops in tc0}) along with \cref{lemma: generalized threshold oracle ch} to divide $\Tt_{q} P$ by the appropriate power of $v-1$ and verify that the remainder is $0$.
  This task can be carried out in $\CH$.
  To compute $\Tt_{p} \Tt_{v-1} \Tt_{q} P$, we simply repeat the above procedure.
  Finally, all that remains is to check if the constant term of the resulting polynomial is $0$, which can be performed in $\CH$.
\end{proof}

\printbibliography

\end{document}